\documentclass[preprint,aps,pra, superscriptaddress, nofootinbib]{revtex4-1}
\usepackage{amsmath,amssymb}
\usepackage{dsfont,yfonts}
\usepackage{tabularx}
\usepackage{hyperref}
\usepackage{graphicx}
\usepackage{subfigure}
\usepackage{diagbox}
\DeclareMathOperator{\tr}{tr}
\DeclareMathOperator{\aut}{Aut}
\DeclareMathOperator{\inn}{Inn}
\DeclareMathOperator{\out}{Out}

\newcolumntype{C}{>{\centering\arraybackslash}X}%

\newcommand{\grp}[1]{\mathbf{#1}}

\def\B{\mathcal{B}}
\def\C{\mathcal{C}}
\def\D{\mathcal{D}}
\def\1{\mathds{1}}
\def\Z2{\mathbb{Z}_2}

\begin{document}

\title{Folding approach to topological orders enriched by mirror symmetry}
\author{Yang Qi}
\email{qiyang@fudan.edu.cn}
\affiliation{Center for Field Theory and Particle Physics, Department of Physics, Fudan University, Shanghai 200433, China}
\affiliation{State Key Laboratory of Surface Physics, Fudan University, Shanghai 200433, China}
\affiliation{Collaborative Innovation Center of Advanced Microstructures, Nanjing 210093, China}
\affiliation{Massachusetts Institute of Technology, Cambridge, Massachusetts 02139, USA}
\author{Chao-Ming Jian}
\email{cmjian@kitp.ucsb.edu}
\affiliation{Station Q, Microsoft Research, Santa Barbara, California 93106, USA}
\affiliation{Kavli Institute of Theoretical Physics, University of California, Santa Barbara, California 93106, USA}
\author{Chenjie Wang}
\email{cjwang@hku.hk}
\altaffiliation{Current address: Department of Physics and Center of Theoretical and Computational Physics, The University of Hong Kong, Pokfulam Road, Hong Kong, China}
\affiliation{Department of Physics, City University of Hong Kong, 83 Tat Chee Avenue, Kowloon, Hong Kong SAR, China}
\affiliation{Perimeter Institute for Theoretical Physics, Waterloo, ON N2L 2Y5, Canada}

\date{\today}

\begin{abstract}
We develop a folding approach to study two-dimensional symmetry-enriched topological (SET) phases with the mirror reflection symmetry. Our folding approach significantly transforms the mirror SETs, such that their properties can be conveniently studied through previously known tools: (i) it maps the nonlocal mirror symmetry to an onsite $\mathbb{Z}_2$ layer-exchange symmetry after folding the SET along the mirror axis, so that we can gauge the symmetry; (ii) it maps all mirror SET information into the boundary properties of the folded system, so that they can be studied by the anyon condensation theory---a general theory for studying gapped boundaries of topological orders; and (iii) it makes the mirror anomalies explicitly exposed in the boundary properties, i.e., strictly 2D SETs and those that can only live on the surface of a 3D system can be easily distinguished through the folding approach. With the folding approach, we derive a set of physical constraints on data that describes mirror SET, namely mirror permutation and mirror symmetry fractionalization on the anyon excitations in the topological order. We conjecture that these constraints may be complete, in the sense that all solutions are realizable in physical systems. Several examples are discussed to justify this. Previously known general results on the classification and anomalies are also reproduced through our approach.

\end{abstract}

\maketitle


\section{Introduction}
\label{sec:intro}

The interplay between topology, entanglement and symmetry has greatly broadened our understanding of gapped quantum phases of matter. First, in the absence of any symmetry, there exist gapped quantum phases of matter which hold intrinsic topological orders~\cite{WenTo1990, WenNSR2016}. Key features of topological orders include the existence of long-range entanglement~\cite{levin06,kitaev06b} and exotic excitations, known as \emph{anyons}, which obey fractional braiding statistics.  Second, for systems without intrinsic topological order but with symmetries, there are also topological phases, known as the symmetry-protected topological (SPT) phases~\cite{haldane83,gu09,pollmann10,fidkowski11,chen11a,chen11b,schuch11,XieScience2012, chen13,hasan10,qi11}. SPT phases are short-range entangled, and their topological distinction will disappear if the symmetries are absent. Well-known examples of SPT phases are topological insulators and topological superconductors. Third, when topological order and symmetries are both present, they can intertwine in various interesting ways and generate a rich family of gapped quantum phases, known as symmetry-enriched topological (SET) orders~\cite{wen12,kitaev06,mesaros13,essin13,barkeshli14, tarantino16,heinrich16,cheng16,WangLevinIndicator,BarkeshliTRSET2016X,tackikawa16a,tackikawa16b}. In particular, anyons can carry \emph{fractional} quantum numbers of the symmetries. For example, the famous fractional quantum Hall systems can be understood as SET phases with the U(1) charge conservation symmetry. There, the anyon excitations carry fractional charges of the U(1) symmetry.

While SPT and SET phases are connected in many aspects, here we would like to mention a particularly interesting connection. It is known that some two-dimensional (2D) SETs cannot be realized in a standalone 2D system. Instead, they must live on the surface of a 3D SPT state~\cite{vishwanath13,wangc-science, metlitski15,wangc13b,chen14a,bonderson13,fidkowski13,metlitski14, barkeshli14, chen14,kapustin14,cho14,wangj15,bbc,chen14,metlitski13,wangc13,qi15,HermeleFFAT,senthil15,WangLevinIndicator}. These SETs are said to be \emph{anomalous}. They realize symmetric and gapped surface terminations of a nontrivial 3D SPT system. More quantitatively, one can define an \emph{anomaly} for each SET that takes values in the Abelian group which classifies 3D SPT phases. The anomaly carries the information of which 3D SPT supports the given SET at its surface. For example, 3D bosonic time-reversal SPT phases are classified by the group $\mathbb{Z}_2\times\mathbb{Z}_2$~\cite{vishwanath13,burnell14,wangc13}. Accordingly, 2D time-reversal SETs can carry three distinct anomalies, corresponding to the three nontrivial 3D SPT phases. (The identity in $\mathbb{Z}_2\times\mathbb{Z}_2$ represents that the SET is anomaly-free and the corresponding 3D SPT state is trivial.)

A classification of 2D SETs and a comprehensive understanding of their anomalies is important for studying topological states of matter. Indeed, when the symmetry is onsite and unitary, great progress has been achieved on classification and anomalies in the last few years, both in general formalism and in physical pictures~\cite{chen14,barkeshli14}. However, many symmetries are either not onsite or not unitary, including the anti-unitary time-reversal symmetry and the mirror reflection symmetry whose action is nonlocal.  These symmetries play crucial roles in many realistic topological phases such as topological insulators (TI) and topological crystalline insulator (TCI) materials~\cite{Bernevig2006, Bernevig_PRL2006, Kane2005a, Kane2005b, Fu_PRL07, Moore_PRB07, Roy_PRB2009, kitaev2009, schnyder2008}. One of the challenges for studying SETs with these symmetries is that unlike onsite unitary symmetries, they cannot be studied using the standard approach of ``gauging the symmetry'': one promotes the global symmetry to a local gauge symmetry, so that physical properties such as symmetry fractionalization and anomalies can be deduced from the resulting gauge theories. Since the standard approach does not help much,  people turn to other approaches for  studying classification and anomalies of time-reversal and mirror-reflection SETs, including field-theoretic method and exactly-solvable models\cite{vishwanath13,wangc13,bonderson13,wangc13b,wangc14,metlitski14,chen14a,metlitski15,qi15,barkeshli14, fidkowski13,cheng16, BarkeshliTRSET2016X}, and the flux-fusion anomaly test approach~\cite{HermeleFFAT}, etc. However, these approaches are not as satisfactory as the standard gauging approach: they are neither too mathematical and physically obscure, or computationally hard, or not easy to generalize to non-Abelian topological orders.

In this work, we develop a \emph{folding approach} for studying 2D SETs with the mirror reflection symmetry. The key idea is very intuitive. Let us assume that mirror reflection maps $(x,y)$ to $(-x,y)$. Then, we fold the mirror SET along the $y$-axis, i.e., the mirror axis, after which it becomes a double-layer system and the mirror axis becomes a gapped boundary. An important goal that we have achieved simply by folding is that the reflection symmetry now becomes an \emph{onsite} layer-exchange symmetry in the double-layer system.  This makes the standard approach of gauging symmetry applicable for studying mirror SETs.
Moreover, we will see later that two additional features result immediately as well:

First, we find a way to encode all information of mirror SETs as boundary properties of the double-layer system. This encoding not only allows us to derive a universal description of the bulk of the folded system independent of the mirror enrichment, but also converts the classification of mirror SETs to the classification of layer-exchange-symmetric gapped boundaries of this universal bulk.
Following this unexpected connection between 2D SETs in the bulk and symmetric boundaries, we study the classification of symmetric gapped boundaries using a combination of techniques including gauging the symmetry and the so-called anyon-condensation theory, and apply the results to classify 2D mirror SETs.

Second, when our folding idea is further combined with the dimension reduction approach proposed by \citet{Song2017} for studying 3D mirror SPTs, we find that it is almost transparent to see the anomalies of mirror SETs after folding. More details on the idea of the folding approach will be discussed below in Sec.~\ref{sec:introidea}.

With the folding approach, we study  2D general topological orders enriched by the mirror reflection symmetry and their anomalies. The folding approach provides a clear physical picture on the differences between various mirror SETs, as well as on how the anomalies can be understood in terms of boundary properties of the double-layer system. More practically, the folding approach, together with the general anyon condensation theory (an approach for studying gapped boundaries of topological orders), allow us to derive a very strong (and perhaps complete) set of constraints on possible mirror symmetry fractionalization. The constraints are described in terms of the modular data of the topological order and hence are physical quantities. These constraints can be practically solved and lead to classification of mirror SETs, if they are complete (which we conjecture is true).
Our results are closely related to 2D time-reversal SETs, since the two symmetries are related by a Wick rotation. (A detailed discussion can be found in Sec.~\ref{sec:discussion:tr}.)
It is worth mentioning that mirror-reflection and time-reversal SETs have been studied previously\cite{WangLevinIndicator,barkeshli14,BarkeshliTRSET2016X}, and our results are consistent with those. However, our approach is completely different and is physically more transparent.

We expect that the folding approach can be generalized to study many other SETs and understand the anomalies there, for example, SETs with both mirror symmetry and onsite unitary symmetries, and fermionic SETs with mirror symmetry, etc. We shall leave them for future studies.


\begin{figure}
  \subfigure[\label{fig:mir:lut}]{\includegraphics[scale=.8]{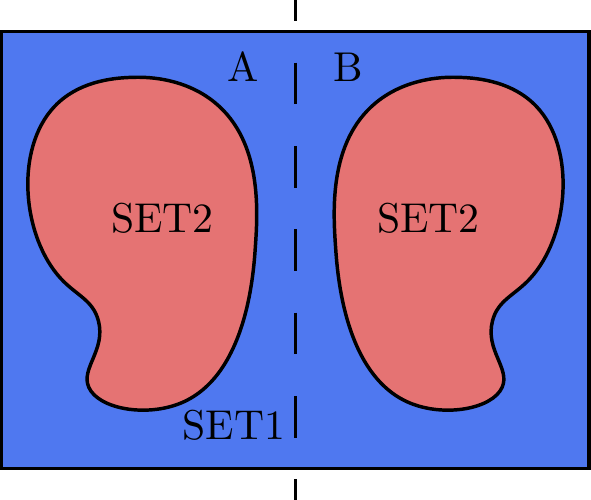}}
  \subfigure[\label{fig:mir:strip}]{\includegraphics[scale=.8]{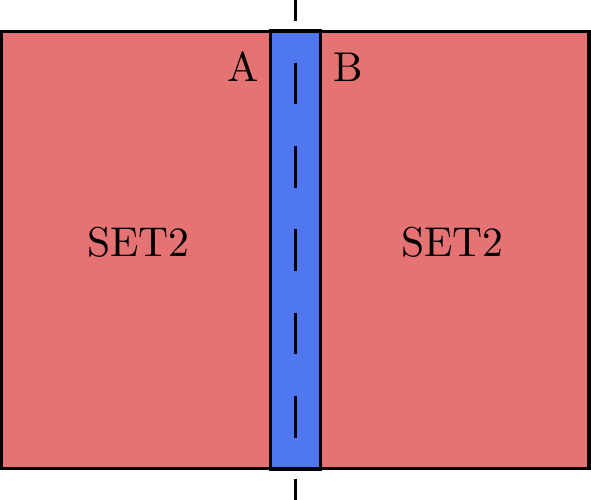}}
  \subfigure[\label{fig:mir:fold}]{\includegraphics[scale=.8]{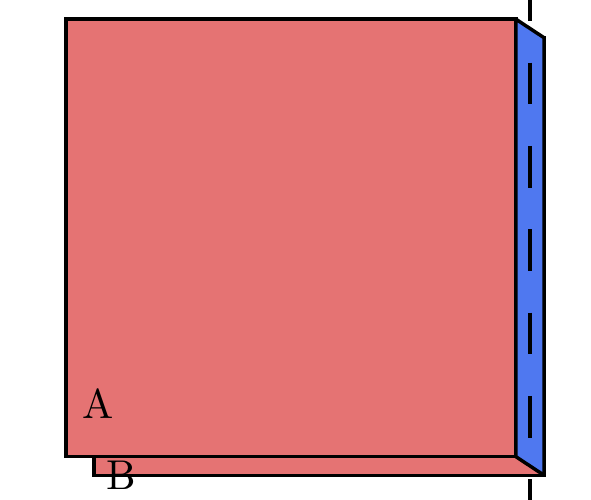}}\\
  \subfigure[\label{fig:mir:lut3d}]{\includegraphics[scale=.8]{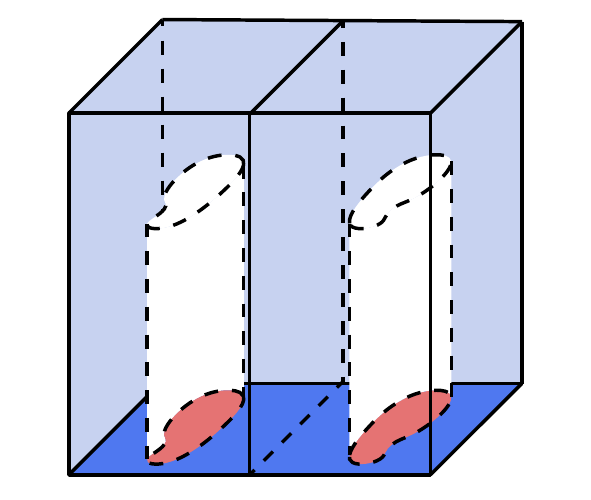}}
  \subfigure[\label{fig:mir:strip3d}]{\includegraphics[scale=.8]{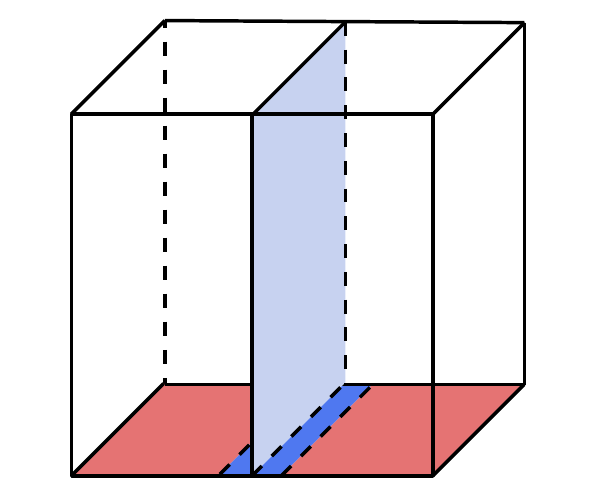}}
  \subfigure[\label{fig:mir:fold3d}]{\includegraphics[scale=.8]{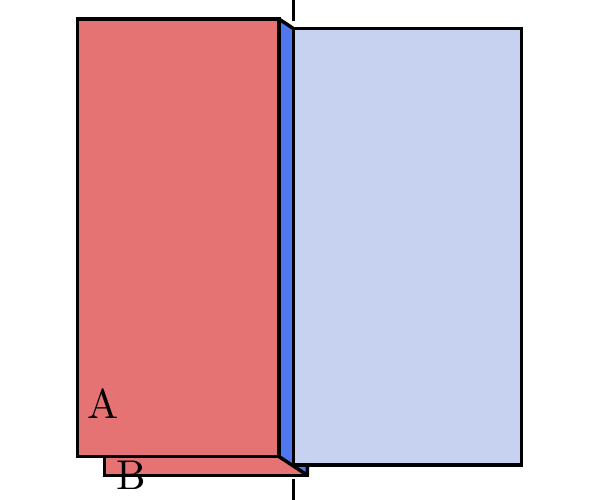}}
    \caption{Mirror symmetric local unitary transformations (LUTs) and folding of mirror SETs. [(a) and (b)]: Under mirror symmetric LUTs, the wave functions of any two strictly 2D mirror SETs can be transformed into one another, such that the difference occurs only near the mirror axis, implying that symmetry properties of mirror SETs are concentrated near the mirror axis. (c): We then fold the system, such that symmetry properties of mirror SETs become boundary properties of the folded double-layer system. [(d) and (e)]: Similar mirror symmetric LUTs can be done for anomalous SETs, which live on the surface of a 3D mirror SPT, such that the 3D bulk is transformed into a product state, with the only exception being near the mirror plane. The inverted T-like junction decouples from the rest of the system. In addition, the mirror symmetry becomes onsite and protects a $\mathbb{Z}_2$ SPT state on the mirror plane. (f) Folding the bottom surface of the inverted T junction, symmetry properties of the  anomalous SET are transformed into boundary properties between the double-layer system and the $\mathbb{Z}_2$ SPT state. } \label{fig:mir}
\end{figure}


\subsection{The general idea}
\label{sec:introidea}

Here, we give a more detailed description on the general idea of the folding approach, without referring to any technical details. As discussed above, what we want to do is simply to fold the system along the mirror axis. Then, several remarkable transformations on the problem follow.  Besides the obvious transformation that the nonlocal mirror reflection symmetry becomes an onsite layer-exchange symmetry, folding also turns the mirror SET properties into the boundary properties of the double-layer system: for a given topological order, the bulk of the double-layer system turns out to be the \emph{same} for all mirror SETs, regardless of anomalous or anomaly-free, while the information of mirror SETs is entirely encoded in the boundary properties of the double layer system. Below we explain this point.

Let us first consider strict 2D SETs and use an argument similar to the one in Ref.~\onlinecite{Song2017}, which was originally designed for mirror SPTs. Consider two different mirror SETs, based on the same intrinsic topological order. Since the difference is present only because of the mirror symmetry, we understand that the two states can be smoothly connected using local unitary transformations (LUTs) if we ignore the mirror symmetry. Let $A$ and $B$ be the left- and right-hand sides of the mirror axis respectively, as shown in Fig.~\ref{fig:mir:lut}.  Now, we apply LUTs in region $A$ such that the wave functions of the two SETs appear exactly the same in $A$. At the same time, we apply the mirror image of these LUTs onto the region $B$. It is obvious that, in region $B$, the wave fuctions of the two SETs also become the same. Overall, the combination of the LUTs are mirror symmetric. At this stage, we have smoothly connected the two SETs through mirror-symmetric LUTs in all regions except near the mirror axis.  Hence, the difference between the SETs is entirely encoded in a narrow region near mirror axis [see Fig.~\ref{fig:mir:strip}]. This argument applies for any two SETs. Then, we fold the system along the mirror axis. The bulk of the resulting double-layer system should be topologically the same for \emph{all} mirror SETs, and their distinction is solely contained in the boundary properties of the double-layer system [see Fig.~\ref{fig:mir:fold}].

The above argument can be easily adapted for anomalous mirror SETs, if we combine it with the dimensional reduction approach on 3D mirror SPTs from \citet{Song2017}. In this case, we have a mirror plane in the 3D bulk.  Then, we apply mirror-symmetric LUTs on both sides of the mirror plane [Fig.~\ref{fig:mir:lut3d}]. After that, the 3D bulk wave functions on the two sides of the mirror plane are transformed into \emph{product states}, i.e., all the entanglement is removed. Only near the mirror plane, there remains some short-range entanglement. Note that the mirror symmetry becomes onsite on the mirror plane. Hence, the remaining short-range entanglement actually describes  a 2D  SPT state with an onsite $\mathbb{Z}_2$ symmetry.  LUTs on the 2D surface work in the same way as in the strict 2D case. Hence, after the LUTs, all (long-range and short-range) entanglements are concentrated on the T-junction setup, as shown in Fig.~\ref{fig:mir:strip3d}. The rest of the system is completely decoupled with the T junction.  Such a T-junction setup was also proposed by \citet{Lake2016}. In this T junction, the perpendicular plane is a 2D $\mathbb Z_2$ SPT state, while the horizontal plane is the original surface SET and which can be turned exactly the same as those anomaly-free SETs except on the mirror axis (i.e., the intersection line of the vertical and horizontal planes).  Finally, we fold the horizontal plane of the T junction and produce a 2D system [Fig.~\ref{fig:mir:fold3d}]. One side of the 2D system is a $\mathbb{Z}_2$ symmetric double-layer topological order, while the other side is a $\mathbb Z_2$ SPT state. Again, we emphasize that the double-layer system is the same for all mirror SETs, and the distinction between mirror SETs is entirely contained on the gapped boundary, now between the double-layer system and the nontrivial $\mathbb{Z}_2$ SPT state.

In summary, both anomaly-free and anomalous mirror SETs are represented by $\mathbb Z_2$-symmetric gapped boundary conditions of the same double-layer system. The anomaly-free and anomalous SETs correspond to the trivial and nontrivial $\mathbb Z_2$ SPT states on the other side of the boundary (mirror axis), respectively. In the main text, we use the so-called anyon condensation theory to study various gapped boundaries of the double-layer system, which are eventually translated back into different mirror SETs.

One comment is that on the mirror plane of the 3D bulk, it does not have to be the $\mathbb{Z}_2$ SPT state (see Ref.~\onlinecite{Song2017}). Another possibility is the so-called $E_8$ state~\cite{e8,lu12}. The two possibilities correspond to the $\mathbb{Z}_2\times \mathbb{Z}_2$ classification of 3D mirror SPTs~\cite{vishwanath13,chen13, kapustin14a}. However, the anomaly corresponding to the $E_8$ possibility can be easily understood (see discussions in Appendix~\ref{app:umtc}). On the other hand, the anomaly corresponding to the $\mathbb{Z}_2$ SPT is much harder. We only discuss the latter in this work.

\subsection{Outline}

The rest of the paper is organized as follows. In Sec.~\ref{sec:z2}, we demonstrate the folding approach using the simplest example of mirror SETs, the mirror-symmetric toric-code states. Through this example, we demonstrate how one can read out the mirror anomaly and how some nontrivial constraints are imposed on properties of mirror SETs through anyon condensation theory. We then apply our approach to general mirror SETs in Sec.~\ref{sec:gen}, where we find very strong physical constraints on mirror symmetry fractionalization. While such constraints are more or less understood for Abelian topological orders, the ones that we find applies to general non-Abelian topological orders.  The constraints are potentially complete and can be practically solved, giving rise to a possible classification of mirror SETs.

Furthermore, in Sec.~\ref{sec:previous}, we use the constraints to derive two general results on mirror SETs. These results were previously discussed  in different languages.   In Sec.~\ref{sec:ex}, we solve the constraints for several examples, to justify that our constraints may be potentially complete. In particular, in Sec.~\ref{sec:ex:d16}, we demonstrate that our constraints are able to rule out those SETs that carry the so-called $H^3$-type obstruction. Finally, we present our conclusion and closing remarks in Sec.~\ref{sec:discuss}.


\section{Folding the toric code}
\label{sec:z2}

We start with the toric-code topological order \cite{kitaev03} as an example to demonstrate the folding approach.  The toric code is the simplest topological order that is compatible with the mirror symmetry. Through this example, we illustrate all the essential ideas, including how to read out mirror anomaly and how some nontrivial constraints on mirror symmetry fractionalization are implemented. Once it is understood, we will apply the folding approach to general 2D topological orders in Sec.~\ref{sec:gen}.

\subsection{Review on mirror-enriched toric-code states}
\label{sec:z2:known}

The classification and characterization of mirror symmetry enriched toric-code states were previously studied in Refs.~\onlinecite{qi15,Song2017}. Here, we review some of the known results, which our folding approach will later reproduce.

To describe a mirror SET state, one needs to specify two sets of data: (i) the data that describe the topological order itself and (ii) the data that describe how the mirror symmetry enriches the topological order. The first set of data includes the types of anyons, their fusion properties, and their braiding properties (see Appendix \ref{app:umtc} for a brief review and Refs.~\onlinecite{kitaev06, WenNSR2016} for the general algebraic theory of anyons). The toric-code topological order contains four types of anyons: the trivial anyon $\mathds1$, two bosons $e$ and $m$, and a fermion $\psi$.  We group them into the set $\mathcal{C}=\{\mathds1, e, m, \psi\}$. Fusing any anyon with $\mathds 1$ does not change the anyon type, i.e., $\mathds 1\times a =a$ for $a=e,m,\psi$. Other fusion rules include $e\times e= \mathds 1$, $m\times m=\mathds 1$ and $e\times m =\psi$. That is, $\psi$ is a bound state of $e$ and $m$.  Note that every anyon is its own anti-particle, i.e., $a=\bar a$., in the toric code. We denote the mutual braiding statistical phase between $a$ and $b$ by $M_{a,b}$. In the toric code topological order, we have $M_{e,m}=M_{e,\psi}=M_{m,\psi}=-1$.

Mirror-symmetry enrichment on a topological order contains two parts:  \emph{anyon permutation} and \emph{symmetry fractionalization}. First, the types of anyons can be permuted under the action of mirror symmetry. It can be described by an automorphic map $\rho_m:\mathcal{C}\rightarrow\mathcal{C}$. Fusion and braiding properties of anyons must be preserved under $\rho_m$\footnote{More precisely, the braiding phases should be complex conjugated under $\rho_m$ since the mirror reflection symmetry reverses spatial orientation. However, it does not affect the toric code, because all braiding phase factors are real.}. In the case of toric code, there are only two  kinds of consistent permutations: the trivial one, $\rho_m(a)=a$ for every anyon $a$, and the nontrivial one that exchanges $e$ and $m$, described by the following map:
\begin{equation}
\rho_m(\mathds 1)= \mathds 1, \quad \rho_m(e)=m, \quad \rho_m(m)= e, \quad \rho_m(\psi) =\psi.
\label{em-exchange}
\end{equation}

\begin{figure}
  \subfigure[\label{fig:mir:anyons:act}]{\includegraphics{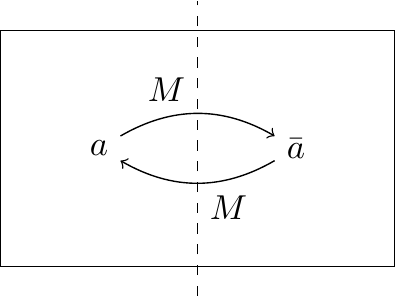}}
  \subfigure[\label{fig:mir:anyons:ev}]{\includegraphics{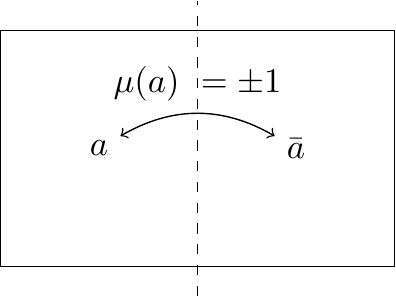}}
  \caption{Definition of mirror-symmetry fractionalization. (a) A pair of anyons, including an anyon $a$ and its antiparticle $\bar a$, located symmetrically on the two sides of the mirror axis. Mirror symmetry $\mathbf{M}$ maps $a$ to $\rho_m(a)$, such that the wave function respects the mirror symmetry only if $\bar a = \rho_m(a)$. (b) The mirror eigenvalue of the two-anyon wave function defines the fractional quantum number $\mu(a)=\pm1$.}
  \label{fig:mir:anyons}
\end{figure}

Second, for those anyons satisfying $\rho_m(a)=\bar a$, we can further define how the mirror symmetry is ``fractionalized'' on $a$. To define symmetry fractionalization, imagine an excited state containing an anyon pair, $a$ and its antiparticle $\bar a$, located symmetrically on the two sides of the mirror axis (Fig.~\ref{fig:mir:anyons}). The pair $a$ and $\bar a$ is created from the ground state through a string-like operator. Requiring that the state is mirror symmetric, we are led to the condition that $\rho_m(a)=\bar a$ (note that in the toric code, $a=\bar a$ for every $a$). Now, we can ask that what is the mirror eigenvalue, denoted by $\mu(a)$,  of this state, $+1$ or $-1$? Interestingly, as pointed out by Ref.~\onlinecite{zaletel15b} , the eigenvalue $\mu(a)$ is ``topologically robust'' in the sense that any mirror symmetric local perturbations around $a$ and $\bar a$ cannot change it. It follows from that a state containing two mirror symmetric local excitations must have mirror eigenvalue $1$. That is,  $\mu(\mathds 1) \equiv 1$. Hence, composing local excitations onto $a$ and $\bar{a}$ cannot change $\mu(a)$. Different sets of mirror eigenvalues $\{\mu(a)\}|_{\rho_m(a)=\bar a}$ are refereed to as different symmetry fractionalization classes of the topological order with a given permutation $\rho_m$. (It might look a bit unnatural to call $\mu(a)$ symmetry fractionalization. However, if $\mu(a)=-1$, the mirror charge ``$-1$'' is split between $a$ and $\bar{a}$, with each carrying a part. In this sense, it is indeed a kind of fractionalization.)


Like anyon permutation $\rho_m$,  symmetry fractionalizations  $\{\mu(a)\}$ also satisfy certain constraints.  For general non-Abelian topological orders, the complete constraints on physical symmetry fractionalizations $\{\mu(a)\}$ are not known. One of the purposes of this work is to  find these constraints through the folding approach. However, the complete constraints for the toric code are known, which we list below.  When the permutation $\rho_m$ is trivial, $\mu(\psi)$ is determined by $\mu(e)$ and $\mu(m)$.  They satisfy the constraint
\begin{equation}
  \label{eq:piemf}
  \mu(\psi) = \mu(e)\mu(m).
\end{equation}
It follows from the observation that a two-$\psi$ state can be viewed as a state with two anyons $e$ and $m$ on one side, and with two other anyons $e$ and $m$ on the other side. Hence, there exist four possible mirror SET states, corresponding to the two independent assignments, $\mu(e)=\pm1$ and $\mu(m)=\pm1$. Following the notations of Wang and Senthil \cite{wangc13}, we denote these four states as $e1m1$, $eMm1$, $e1mM$ and $eMmM$, where $a1$ and $aM$ denotes $\mu(a)=+1$ and $-1$, respectively. For the nontrivial anyon permutation given by Eq.~\eqref{em-exchange},  $\psi$ is the only anyon that satisfies the condition $\rho_m(a)=\bar a$ besides the trivial anyon $\mathds 1$.  In this case, $\mu(\psi)$ satisfies another constraint:
\begin{equation}
\mu(\psi)= M_{e,m}=-1
\label{eq:constraint2}
\end{equation}
where $M_{e,m}$ is the mutual braiding statistics between $e$ and $m$. According to Refs.~\onlinecite{bonderson13,chen14a,metlitski15,zaletel15,barkeshli14}, this constraint (and its variant for time-reversal symmetry) is related to the fact that $\psi=e\times m = e\times \rho_m(e)$. Therefore, there is only a single SET state for the nontrivial anyon permutation \eqref{em-exchange}. To sum up, there are in total five mirror-enriched toric code states, four associated with the trivial permutation and one for the nontrivial permutation.



Finally, a particularly interesting phenomenon is that among the five mirror-enriched toric code states, the $eMmM$ is \emph{anomalous}, meaning that it cannot be realized in a standalone 2D system. Instead, it can only be realized on the surface of a 3D mirror SPT state~\cite{vishwanath13,wangc13}. On the contrary, the other four states are anomaly-free, and can be realized in strictly 2D systems, such as exactly solvable models, tensor-product states and $\mathbb Z_2$ spin liquids~\cite{SJiangTPS2015X,Song2015}. In what follows, we will show that our folding approach can reproduce this classification of mirror-enriched toric states and reveal the mirror anomaly in the $eMmM$ state.


\subsection{Qualitative description of the folding approach}

\label{sec:z2qualitative}

What makes the mirror symmetry difficult to deal with is its nonlocal nature: it maps one side of the mirror axis to the other side. Here, we develop a folding approach following the general idea illustrated in the introduction (Sec.~\ref{sec:introidea}) for studying mirror symmetric SETs.  For simplicity, in this and the next subsection, we describe the folding method in the absence of anyon permutation. The case where the mirror symmetry permutes $e$ and $m$ will be discussed in Sec.~\ref{sec:z2:nma}.

Consider a mirror-symmetric toric-code state, where the mirror symmetry does a trivial permutation on anyons. The mirror axis divides the system into two regions $A$ and $B$, as shown in Fig.~\ref{fig:mir}. The mirror symmetry maps an anyon located in region $A$ to an anyon of the same type in region $B$, and vice versa. We now fold region $B$ along the mirror axis, such that it overlaps with region $A$. After folding, it becomes a double-layer system, where each layer hosts a copy of the toric-code topological order.\footnote{In general,  folding also reverses the spatial orientation of the region $B$ and correspondingly changes the nature of the topological order therein. However, this does not occur for the simple example of toric code, where all the self and mutual statistics among the anyons are real. This issue will be dealt more carefully in Sec.~\ref{sec:folding}, where we generalize our folding method to a general topological order.} In this double-layer system, the mirror symmetry acts  as an interlayer exchange symmetry, which can be treated as an onsite unitary $\mathbb{Z}_2$ symmetry. The anomalous mirror SET that lives on the surface of a 3D SPT state (i.e., the $eMmM$ state) can be folded in a similar way, as illustrated in Fig.~\ref{fig:mir}.

In this section, we give a qualitative description on the bulk and boundary properties of the doubler-layer system, and on how these properties encode the information of the original mirror-enriched toric-code states. In Sec.~\ref{sec:z2cl}, we will apply the standard method for studying onsite unitary symmetries, the method of gauging global symmetry \cite{levin12}, to give a more quantitative analysis.

\subsubsection{Bulk of the double-layer system}
\label{sec:z2bulk}

As discussed in the introduction (Sec.~\ref{sec:introidea}), information about mirror SETs is encoded only near the mirror axis. Away from the mirror axis, all mirror SETs look alike after appropriate mirror-symmetric local unitary transformations. Accordingly, after folding, the bulk of the double-layer system is the \emph{same} for all mirror SETs, including both anomaly-free and anomalous ones.

Let us describe the bulk of the double-layer system. To begin, we introduce some notations. The double-layer system hosts a topological order of two copies of the toric code. We denote an anyon of type $a$ on the top layer and the bottom layer as $(a,\mathds1)$ and $(\mathds1, a)$, respectively. More generally, we denote a composite anyon, with charge $a$ on the top layer and $b$ on the bottom layer, as $(a, b)$. Since the two layers are decoupled, the fusion and braiding properties follow immediately as a direct sum of those in each layer.


Generally speaking, the onsite $\mathbb{Z}_2$ symmetry enriched bulk topological order are characterized at three levels: anyon permutation by the symmetry, symmetry fractionalization, and stacking of a $\mathbb{Z}_2$ SPT state~\cite{barkeshli14,SET2}. With a trivial mirror permutation in the original system, the $\mathbb Z_2$ interlayer symmetry permutes the anyons in the double-layer system in the following form,
\begin{equation}
  \label{eq:m}
  \mathbf m:(a,b)\mapsto(b,a).
\end{equation}
Here, we use $\grp m$ to denote the interlayer $\mathbb Z_2$ symmetry in the folded system, to distinguish it from the original mirror symmetry which we denote as $\grp M$. Next, according to Ref.~\onlinecite{barkeshli14}, any double-layer system with a unitary $\mathbb{Z}_2$ interlayer exchange symmetry that permutes the anyons in the form of Eq.~\eqref{eq:m} has only a unique symmetry fractionalization class. Finally, the remaining flexibility is to attach a $\mathbb{Z}_2$ SPT state. For onsite $\mathbb{Z}_2$ symmetry, it is known that besides the trivial state, there is only one nontrivial SPT state~\cite{levin12,chen13}.

We conclude that given the anyon permutation in Eq.~\eqref{eq:m}, there are two possible $\mathbb{Z}_2$ SETs.  The bulk $\mathbb{Z}_2$ SET order of the double-layer system, obtained from folding the mirror-enriched toric-code states, should be one of the two possibilities. To specify which one is the actual bulk SET order, it is more convenient to use braiding statistics of gauge flux excitations after we gauge the $\mathbb{Z}_2$ symmetry. Once the $\mathbb{Z}_2$ symmetry is gauged, the two possible SETs can be easily distinguished by braiding statistics of gauge flux excitations. So, we postpone this discussion to Sec.~\ref{sec:z2cl}.

We remark that since the two $\mathbb{Z}_2$  SETs differ by stacking an SPT state, readers may expect that they can be distinguished by the edge mode stability: one has symmetry protected gapless edge modes, and the other does not. However, the fact is that both SETs have no symmetry protected gapless edges modes. We will come back to this observation below when discuss the boundary properties.


\subsubsection{Boundary of the double-layer system}

\label{sec:z2boundary}

After folding, the mirror axis becomes a boundary between the double-layer system and the vacuum (i.e., the trivial $\mathbb{Z}_2$ SPT state) for anomaly-free mirror SETs, or a boundary between the double-layer system and the nontrivial $\mathbb{Z}_2$ SPT state for the anomalous mirror SET. In either case, the boundary is gapped and symmetric under the onsite $\mathbb{Z}_2$ symmetry\footnote{The phenomenon that the double-layer system admits a gapped and symmetric boundary with \emph{both} trivial and nontrival $\mathbb{Z}_2$ SPT states is not unusual for SETs; similar discussions can be found in Refs.~\onlinecite{wang13,lu14}. In addition, it implies that there is no stable gapless edge modes for both possible bulk  $\mathbb{Z}_2$ SETs: one can further fold the trivial/nontrivial $\mathbb{Z}_2$ SPT state onto the double-layer system, leading to the two bulk $\mathbb{Z}_2$ SETs, the edges of which are gapped.}. Since the information of mirror SETs is encoded near the mirror axis, we expect that different mirror symmetry fractionalization classes ($e1m1$, $eMm1$, $e1mM$ or $eMmM$) should correspond to different types of gapped and $\mathbb{Z}_2$ symmetric boundaries. Below, we explain how to resolve this correspondence.

Gapped boundaries or domain walls (i.e., boundaries between two topological orders) have been widely studied in the literature~\cite{bravyi98,beigi11,levin09,levin12b,levin13,kitaev11,barkeshli13,barkeshli13b,Fuchs2013, TLanAnyCon2015,hung15, wan17,hu17}. One way to describe them is to use the language of ``anyon condensation''~\cite{bais2009, eliens14,TLanAnyCon2015,NeupertAnyCon2016}: a set of self-bosonic anyons on one side of the boundary/domain wall ``condense'' into the trivial anyon on the other side. In other words, this set of self-bosons can be annihilated by \emph{local operators} once they move to the boundary/domain wall. In the current example, the one side of the gapped boundary is the double-layer system, which has a topological order of two copies of the toric code, and the other side is a trivial topological order that has no nontrivial anyons. The gapped boundary corresponds to condensing these self-bosons:  $(e,e)$, $(m,m)$, $(\psi,\psi)$, and the trivial anyon $(\mathds 1, \mathds 1)$ from the double-layer system. To see this,  we recall that in the unfolded picture, two anyons of the same type $a$ near the mirror axis can annihilate each other (note that $a=\bar a$ for the toric code). In the folded picture, such an anyon pair is described as a composite anyon $(a,a)$. The fact that such two-anyon pair can be created or annihilated at the mirror axis in the unfolded picture then translates to that $(a,a)$ can be created or annihilated at the boundary of the folded system. That is, the anyons $(e,e)$, $(m, m)$, and $(\psi,\psi)$ can be condensed at the boundary of the double-layer system. An consequence of such a condensate is that it confines all other anyon charges that have nontrivial braiding statistics with any of the anyons in the condensate. Eventually, the condensation gives rise to a trivial vacuum state on the other side of the boundary.


Furthermore, the gapped boundary is symmetric under the $\mathbb{Z}_2$ interlayer exchange symmetry. Mirror symmetry fractionalization will be translated into symmetry properties of the gapped boundary. Recall that the symmetry fractionalization $\mu(a)$ is defined as the mirror eigenvalue of the wave function containing a pair of the same anyon $a$ located symmetrically on the two sides of the mirror axis.  After folding, $\mu(a)$ becomes the $\mathbb Z_2$ eigenvalue of the double-layer wave function, which has a \emph{single} anyon $(a,a)$ in the bulk.  Accordingly, we can view $\mu(a)=\pm1$ as the ``charge'' carried by $(a,a)$ under the $\mathbb Z_2$ symmetry: if $\mu(a)=+1$, $(a,a)$ is neutral; otherwise, $(a,a)$ carries a $\mathbb{Z}_2$ charge. Generally speaking, \emph{integer} charge of an anyon cannot be well defined, since there is no canonical way to name what is $+1$ or $-1$ charge on an anyon; only \emph{fractional} charge (modulo integer charge) is well defined. However, the current double-layer system is special. It has a ``memory'' of the unfolded system --- one can simply define the $\mathbb{Z}_2$ charge $\mu(a)$ through the original mirror symmetry by unfolding the system. Unfolding cannot be generally done if interaction is introduced between the two layers, since local interaction will be mapped to nonlocal one by unfolding.

It is important to note that the wave function containing a single anyon $(a,a)$ in the bulk is permitted only because of the fact that $(a,a)$ can be annihilated on or created out of the gapped boundary. Hence,  $\mu(a)$ should really be viewed as a property of both the gapped boundary and the anyon $(a,a)$. Now, if we insist that the whole system to be even under $\mathbb{Z}_2$ (which will be strictly imposed once we gauge the $\mathbb{Z}_2$ symmetry), then permitting a single anyon $(a,a)$ carrying a charge $\mu(a)$ in the bulk translates to the property that the anyon $(a,a)$ carrying a charge $\mu(a)$ can be annihilated/condensed on the boundary. Accordingly, the boundary should be viewed as a condensate of $(\mathds 1, \mathds 1)$, $(e,e)$, $(m, m)$ and $(\psi,\psi)$ which carry $\mathbb{Z}_2$ charges $\mu(\mathds 1)$, $\mu(e)$, $\mu(m)$ and $\mu(\psi)$, respectively. Hence, we have mapped the mirror symmetry fractionalization patterns into boundary properties through the folding trick.


In the above discussion, we have not touched the $\mathbb{Z}_2$ symmetry properties of the system living on the other side of the gapped boundary, i.e., the ``anyon-condensed system''. It supports no anyons, but it can either be a trivial or nontrivial $\mathbb{Z}_2$ SPT state~\cite{levin12,chen13}. According to the picture discussed in Sec.~\ref{sec:introidea}, the anomalous $eMmM$ mirror SET leads to a nontrivial $\mathbb{Z}_2$ SPT state on the other side of the gapped boundary after folding. Therefore, if we condense $\{(\mathds 1, \mathds 1),(e,e), (m,m), (\psi,\psi)\}$ with $\mu(e)=-1$ and $\mu(m)=-1$, it should be able to argue that, after anyon condensation, the system must be a nontrivial $\mathbb{Z}_2$ SPT state. Similarly, other anyon condensation patterns should generate a trivial $\mathbb{Z}_2$ SPT state on the other side of the boundary/mirror axis. Indeed, it can be derived through a more quantitative analysis on anyon condensation, which will be discussed shortly in the next subsection.

In summary, different mirror symmetry fractionalization patterns are described by different anyon-condensation patterns on the mirror axis that preserve the $\mathbb Z_2$ symmetry in the folded double-layer system. Depending on the symmetry properties of the anyon-condensation pattern, the other side of the mirror axis can either be trivial or nontrivial $\mathbb Z_2$ SPT states, which correspond to anomaly-free and anomalous mirror SETs, respectively.

\subsection{Gauging the $\mathbb Z_2$ symmetry}
\label{sec:z2cl}

In this subsection, we employ the method of ``gauging the symmetry'' to study the double-layer system and its boundary properties in more detail. We give a more quantitative analysis of the anyon condensation patterns on the boundary. In particular, based on general principles of anyon condensation, we are able to show that condensing $(e,e)$ and $(m,m)$ with $\mu(e)=\mu(m)=-1$ indeed leads to a nontrivial $\mathbb{Z}_2$ SPT state, while other anyon condensation patterns lead to a trivial $\mathbb{Z}_2$ SPT state.

To gauge the $\mathbb{Z}_2$ symmetry, we minimaly couple the $\mathbb{Z}_2$-symmetric double-layer system to a dynamical $\mathbb{Z}_2$ gauge field. General gauging procedure on lattice systems can be found in Refs.~\onlinecite{levin12, wangcj15}, which however is not important for our discussions here. Once the $\mathbb{Z}_2$ symmetry is gauged, the bulk topological order will be enlarged. It contains those anyons that have origins from the ungauged topological order, as well as new anyons that carry $\mathbb{Z}_2$ gauge flux. Here, we provide a heuristic description of the nature of the gauged topological order, while a detailed derivation is given in Appendixes~\ref{app:d8} and \ref{app:dl}.

The bulk topological order in the gauged double-layer system contains $22$ anyons in total, where eight are Abelian and $14$ are non-Abelian. (For experts, it is the same as the quantum double of the group $\mathbb D_8$.)  First, let us describe those anyons that have counterparts in the ungauged system.   The original diagonal anyons $(a,a)$, carrying $\mathbb Z_2$ eigenvalues $\pm1$, become Abelian anyons $(a,a)^+$ and $(a,a)^-$ in the gauged double-layer system, respectively. Taking $a=\mathds 1, e, m,\psi$, we obtain eight Abelian anyons. In particular, the anyon $(\mathds1,\mathds1)^+$ is the trivial anyon in the new topological order, while $(\mathds1,\mathds1)^-$ represents the $\mathbb Z_2$ gauge charge. In gauge theories, gauge charge excitations have to be created in pairs and thereby become anyons. The off-diagonal anyons, $(a,b)$ and $(b,a)$ (with $a\neq b$), merge into one non-Abelian anyon with quantum dimension $2$, which we denote by $[a,b]$. Since $(a,b)$ and $(b,a)$ transform into one another under the $\mathbb{Z}_2$ symmetry, gauging the symmetry enforces them to be merged. There are six of these anyons. The braiding statistics of all these 14 anyons are the same as their counterparts in the ungauged system, e.g., the topological spins are $\theta_{(a,a)^\pm} = \theta_a^2 = 1$ and $\theta_{[a,b]} = \theta_a\theta_b$ (see Appendix~\ref{app:umtc} for the meaning of topological spins).

In addition, there are eight anyons that carry $\mathbb{Z}_2$ gauge flux and that do not have counterparts in the ungauged system, which we call $\mathbb Z_2$ defects\footnote{In this manuscript, we use ``defects'' to name the excitations that carry $\mathbb{Z}_2$ gauge flux. It is important to note that they are not extrinsic defects, but actual excitations in the gauged system, since we treat the $\mathbb{Z}_2$ gauge field as a dynamical field.}. These defects are non-Abelian and have quantum dimension 2. We denote them by $X_{\mathds1}^\pm$, $X_e^\pm$, $X_m^\pm$ and $X_\psi^\pm$. The meaning of this notation is as follows.  The subscript ``$b$'' of the defects $X_b^\pm$ is defined by the mutual braiding statistics between the defects and the anyons $(a,a)^\pm$. In particular, there exists a defect, which we name $X_{\mathds 1}^+$, that detects the $\mathbb Z_2$ charge carried by the diagonal anyons $(a,a)^\pm$, through the mutual braiding statistics,
\begin{equation}
  \label{eq:M_X1aa}
  M_{X_{\mathds1}^+, (a,a)^\pm} = \pm1
\end{equation}
where the ``$\pm$'' signs are correlated on the two sides of the equation. Instead, the other three flavors of defects, $X_e^+$, $X_m^+$ and $X_\psi^+$, have the following mutual braiding statistics with respect to $(a,a)^\pm$:
\begin{equation}
  \label{eq:M_Xbaa}
  M_{X_b^+,(a,a)^\pm} = \pm M_{b,a}.
\end{equation}
where $M_{b,a}$ is the mutual braiding statistics between $a$ and $b$ in the original toric code.  The defects $X_e^+$, $X_m^+$ and $X_\psi^+$ can be obtained by attaching an anyon charge to $X_{\mathds1}^+$, as indicated by the fusion rules,
\begin{equation}
  \label{eq:X1xa1}
  X_{\mathds1}^+ \times[a,\mathds1] = X_a^++X_a^-.
\end{equation}
On the other hand, the superscript ``$\pm$'' of the defect $X_b^\pm$ is conventional, in contrast to that of $(a,a)^\pm$. The sign $\pm$ only denotes a \emph{relative} $\mathbb Z_2$ gauge charge difference between $X_{b}^+$ and $X_b^-$, reflected in the fusion rule $X_b^+\times (\mathds 1, \mathds 1)^-= X_b^-$. The mutual braiding between $X_b^-$ and $(a,a)^\pm$ is the same as in Eqs.~\eqref{eq:M_X1aa} and \eqref{eq:M_Xbaa} for $X_b^+$.

As discussed in Sec.~\ref{sec:z2bulk}, there are two possible bulk $\mathbb{Z}_2$ SETs in the double-layer system. The two SETs give rise to two different bulk topological orders. The above topological properties of anyons in the gauged system do not distinguish them. To distinguish the two SETs, we need to look at the topological spins of the defects. We find that one of the SETs gives rise to the topological spins
\begin{align}
\theta_{X_{\mathds 1}^\pm} = \theta_{X_{e}^\pm}= \theta_{X_{m}^\pm} = \pm 1, \quad  \theta_{X_{\psi}^\pm }= \pm i   \label{eq:topospin}
\end{align}
while the other SET has all the topological spins multiplied by a factor $i$ from those in Eq.~\eqref{eq:topospin} (see Appendix~\ref{app:dl} for a derivation). We claim that the bulk topological order obtained from folding the original mirror-enriched toric code states correspond to the one described by Eq.~\eqref{eq:topospin}. As discussed below, this bulk SET indeed reproduces the anomaly of all symmetry fractionalization classes. Instead, the other SET produces a completely opposite conclusion on the anomaly. [The opposite conclusion is expected, because one can imagine attaching a $\mathbb Z_2$ SPT state to both sides of the mirror axis, in the folded setup in Fig.~\ref{fig:mir:fold3d}. This attachment changes the bulk SET state on the left-hand side to the other SET and flips the SPT state on the right-hand side, while keeps the gapped boundary unchanged.]


Next, we move on to the boundary properties of the gauged double-layer system. As discussed in Sec.~\ref{sec:z2boundary}, before gauging the symmetry, the symmetry fractionalization patterns correspond to condensing $(e,e),(m,m),(\psi,\psi)$ with $\mathbb{Z}_2$ eigenvalues $\mu(e),\mu(m),\mu(\psi)$, respectively. In the gauged system, this translates to condensing the anyons $(e,e)^{\mu(e)}$, $(m, m)^{\mu(m)}$, and $(\psi,\psi)^{\mu(\psi)}$ at the boundary. The four symmetry fractionalization classes $e1m1, e1mM, eMm1$, and $eMmM$ correspond to the four independent choices of $\mu(e), \mu(m), \mu(\psi)$ that satisfy the constraint Eq.~\eqref{eq:piemf}. One important question that the qualitative discussion in Sec.~\ref{sec:z2qualitative} does not address is that: why the choice that $\mu(e)=\mu(m)=-1$ leads to a nontrivial $\mathbb{Z}_2$ SPT state after condensation, while other choices lead to a trivial $\mathbb{Z}_2$ SPT state?

Here we answer this question. First, let us describe the topological order $\mathcal{U}$ on the other side of the mirror axis, i.e., the topological order after anyon condensation. The trivial (nontrivial) $\mathbb{Z}_2$ SPT state becomes the untwisted (twisted) $\mathbb{Z}_2$ topological order after gauging the $\mathbb{Z}_2$ symmetry~\cite{levin12}, respectively. Both $\mathbb{Z}_2$ topological orders contain four anyons: the trivial anyon, a $\mathbb{Z}_2$ gauge charge, and two $\mathbb{Z}_2$ defects that carry gauge flux and that are related by attaching a $\mathbb{Z}_2$ gauge charge. The difference is that the $\mathbb{Z}_2$ defects are fermion/bosons in the untwisted $\mathbb{Z}_2$ topological order, while  they are semions/antisemions (topological spin being $\pm i$) in the twisted $\mathbb{Z}_2$ topological order. The untwisted $\mathbb{Z}_2$ topological order is actually the same as the toric code, where $e$ is identified as the gauge charge and $m,\psi$ can be identified as the gauge flux excitations. The twisted $\mathbb{Z}_2$ topological order is also called the double-semion topological order.

\begin{table}
  \caption{Anyon condensation patterns. In the columns of $X_{\mathds1}^\pm$, $X_{e}^\pm$ , $X_{m}^\pm$ and $X_{\psi}^\pm$, the symbol ``$\times$'' means that the defect is confined and ``$\bigcirc$'' means that the defect is deconfined after anyon condensation. The $\theta_{X_w^\pm}$ column lists the topological spins of the deconfined defects, and the last column denotes the nature of the condensed phase $\mathcal{U}$, with ``TC'' being the toric code topological order and ``DS'' being the double-semion topological order.}
  \label{tab:condz2}
  \begin{tabularx}{.9\textwidth}{C CCC CCCC CC}
    \hline\hline
    SET & $\mu(e)$ & $\mu(m)$ & $\mu(\psi)$ & $X_{\mathds1}^\pm$ & $X_e^\pm$ & $X_m^\pm$
    & $X_\psi^\pm$ & $\theta_{X_w^\pm}$ & $\mathcal U$ \\
    \hline
    $e1m1$ & + & + & +
    & $\bigcirc$ & $\times$ & $\times$ & $\times$
    & $\pm1$ &TC\\
    $eMm1$ & $-$ & + & $-$
    & $\times$ & $\times$ & $\bigcirc$ & $\times$
    & $\pm1$ & TC\\
    $e1mM$ & + & $-$ & $-$
    & $\times$ & $\bigcirc$ & $\times$ & $\times$
    & $\pm1$ &TC\\
    $eMmM$ & $-$ & $-$ & +
    & $\times$ & $\times$ & $\times$ & $\bigcirc$
    & $\pm i$ & DS\\
    \hline\hline
  \end{tabularx}
\end{table}

Accordingly, we expect that the remaining topological order $\mathcal{U}$ is the double-semion topological order if $(e,e)^-$ and $(m,m)^-$ are condensed, and $\mathcal{U}$ is the toric code topological order for other anyon condensation patterns. Indeed, this expectation follows from the general principles of anyon condensation. One consequence of anyon condensation is \emph{confinement}. Anyons which have nontrivial mutual braiding statistics with respect to any of the anyons in the condensate will be confined.  In particular, some $\mathbb Z_2$ defects will be confined after condensing  $(e,e)^{\mu(e)}$, $(m, m)^{\mu(m)}$ and $(\psi,\psi)^{\mu(\psi)}$. Using Eq.~\eqref{eq:M_Xbaa}, one can work out the confinement of $\mathbb Z_2$ defects in different anyon-condensation patterns, as listed in Table~\ref{tab:condz2}. For each anyon-condensation pattern, three of the four types of defects are confined, and one type of defect $X_w^\pm$ remain deconfined, where $w$ is an anyon charge. The defects $X_w^\pm$ become the $\mathbb Z_2$ gauge flux excitation in the $\mathbb Z_2$ topological order on the other side of the boundary. According to the values of topological spins in Eq.~\eqref{eq:topospin}, we find that for the first three cases  in Table~\ref{tab:condz2}, the deconfined $\mathbb{Z}_2$ defects are bosons and fermions, indicating that $\mathcal{U}$ is an untwisted $\mathbb Z_2$ gauge theory (i.e., the toric-code topological order). The last row in Table~\ref{tab:condz2}, however, has semionic defects that remain deconfined, indicating that $\mathcal{U}$ is the twisted $\mathbb{Z}_2$ gauge theory (i.e., double-semion topological order).


Hence, we conclude that general principles of anyon condensation indeed reproduce the fact that the $eMmM$ symmetry fractionalization class generates a nontrivial $\mathbb{Z}_2$ SPT state through the folding approach. This indicates that it is an anomalous mirror SET, following the argument given in Sec.~\ref{sec:introidea}.

\subsection{Nontrivial anyon permutation}
\label{sec:z2:nma}

We have learned from  Sec.~\ref{sec:z2cl} that the folding method and general principles of anyon condensation allow us to identify anomaly-free and anomalous mirror SETs. Here, we demonstrate another aspect of the power of this method, through the mirror SET where the mirror symmetry interchanges $e$ and $m$ anyons, as described by Eq.~\eqref{em-exchange}. We show that nontrivial constraints on mirror symmetry fractionalization, e.g. Eq.~\eqref{eq:constraint2}, are secretly encoded in the principles of anyon condensation. [The constraint Eq.~\eqref{eq:piemf} is also encoded. However, we do not emphasize it in Sec.~\ref{sec:z2cl}, since that constraint can be easily understood without referring to anyon condensation.]

When the mirror symmetry permutes $e$ and $m$,  folding translates it into a $\mathbb{Z}_2$ interlayer symmetry which permutes the anyons in the double-layer system in the following way:
\begin{equation}
\mathbf{m}: (a,b) \mapsto (\rho_m(b),\rho_m(a))
\label{eq:mrho}
\end{equation}
where $\rho_m$ is given by Eq.~\eqref{em-exchange}. The permutation in Eq.~\eqref{eq:mrho} is apparently  different from the one in Eq.~\eqref{eq:m}. However, they are actually equivalent, up to relabeling of anyon types. The equivalence can be revealed by relabeling the $e$ anyon as $m$, and vise versa, on the second layer. After this relabeling, the $\mathbb Z_2$ symmetry permutation Eq.~\eqref{eq:mrho} becomes the simple interlayer exchange in Eq.~\eqref{eq:m}. Accordingly, the bulk SET of the double-layer system is the same as that discussed in Sec.~\ref{sec:z2bulk}, and the gauged topological order becomes the same as the one discussed in Sec.~\ref{sec:z2cl}.

What is changed by the relabeling $e\leftrightarrow m$ on the second layer is the description of anyon condensation pattern on the boundary. Before the relabeling, the gapped boundary corresponds to a condensation of anyons $(\mathds 1, \mathds 1)$, $(e,e)$, $(m,m)$ and $(\psi, \psi)$ with the bulk SET equipped with the permutation Eq.~\eqref{eq:mrho}. After relabeling $e$ and $m$ anyons on the second layer, the boundary corresponds to a condensation of anyons $(\mathds 1, \mathds 1)$, $(e,m)$, $(m,e)$ and $(\psi,\psi)$ with the bulk SET equipped with the permutation Eq.~\eqref{eq:m}. Here, we take the latter notation so that all discussions in Sec.~\ref{sec:z2qualitative} and Sec.~\ref{sec:z2cl} about the bulk SET are still valid.

The condensation on the boundary is $\mathbb{Z}_2$ symmetric, since $(\mathds 1, \mathds 1)$ and $(\psi, \psi)$ are invariant, and $(e,m)$, $(m,e)$ transform into one another. In addition, $(\psi,\psi)$ may carry a $\mathbb Z_2$ eigenvalue $\mu(\psi)=\pm 1$.  As reviewed in Sec.~\ref{sec:z2:known}, $\mu(\psi)$ can only take $-1$ according to the constraint Eq.~\eqref{eq:constraint2}. Below, we would like to argue that $\mu(\psi)=+1$ indeed violates general principles of anyon condensation, and only $\mu(\psi)=-1$ is a valid choice.

To do that, we again gauge the $\mathbb{Z}_2$ symmetry. According to Sec.~\ref{sec:z2cl}, the anyons $(e,m)$ and $(m,e)$ merge into the non-Abelian anyon $[e,m]$, which has a quantum dimension 2. Hence, in the gauged double-layer system, the condensed anyons are $(\mathds 1,\mathds 1)^+$, $[e,m]$ and $(\psi,\psi)^{\mu(\psi)}$. This type of anyon condensation, where non-Abelian anyons are condensed, requires more complicated mathematical descriptions, comparing to the case discussed in Sec.~\ref{sec:z2cl}. Here, we only give an intuitive and simplified argument, while leaving the full details to Sec.~\ref{sec:anycond}, where we discuss such anyon condensations in a more general setting.

The argument is as follows. Similar to the examples in Sec.~\ref{sec:z2cl}, we are interested in finding which $\mathbb Z_2$ defects remain deconfined after the condensation. First, since $(\psi,\psi)^{\mu(\psi)}$ is condensed, the deconfined defects must have trivial mutual braiding statistics with respect to $(\psi,\psi)^{\mu(\psi)}$. According to Eq.~\eqref{eq:M_Xbaa}, if $(\psi,\psi)^+$ is condensed, only $X_{\mathds1}^\pm$ and $X_\psi^\pm$ are possibly deconfined;  if $(\psi,\psi)^-$ is condensed, only $X_{e}^\pm$ and $X_m^\pm$ are possibly deconfined. Second, besides confinement, another important consequence of anyon condensation that we do not mention in the Abelian case is \emph{identification}. Intuitively, a condensed anyon $a$ is regarded as the trivial anyon after the condensation.  Accordingly, two distinct defects $X^\pm_w$ and $X^\pm_w\times a$ is identified as the same defect after the condensation. Here,  since we are condensing $[e,m]$, the defects $X^\pm_w$ and $X_w^\pm\times[e,m]$ become the same $\mathbb Z_2$ defect after condensation. For any $w$, we show in Appendix~\ref{app:dl} that the fusion rule is $X^\pm_w\times[e,m] =X^+_{w\times\psi} + X^-_{w\times\psi}$. Therefore, $X_{e}^\pm$ and $X_m^\pm$ are identified, and  $X_{\mathds1}^\pm$ and $X_\psi^\pm$ are also identified. Finally, we expect that any defects that are \emph{identified} and \emph{deconfined} should have the same topological spin (in contrast, a confined defect cannot be associated with a well-defined topological spin). Combining the first and second points, we see that condensing $(\psi,\psi)^+$ leads to a contradiction: $X_{\mathds1}^\pm$ and $X_\psi^\pm$ are deconfined and identified, but they have different topological spins according to Eq.~\eqref{eq:topospin}. Hence, it is not a valid anyon condensation. This leaves condensing $(\psi,\psi)^-$ the only option. After condensing $(\psi,\psi)^-$, the defects $X_{e}^\pm$ and  $X_{m}^\pm$  become the deconfined defects. Since they are fermions/bosons, the topological order on the other side of the boundary (mirror axis) is the toric-code topological order, indicating that the original mirror SET is anomaly-free.

\section{General formulation}
\label{sec:gen}

We now apply the folding approach to general 2D topological orders that are enriched by the mirror symmetry. As we learn from Sec.~\ref{sec:z2},  folding turns a mirror SET into a $\mathbb{Z}_2$ symmetric anyon condensation pattern on the mirror axis (i.e., the boundary of the folded system).  With this mapping, we use the consistency conditions on anyon condensation to understand constraints on the data that describe mirror SETs. Ideally, when the constraints are complete in the sense that all solutions can be realized in physical systems, we obtain a classification of mirror SETs from the solutions. We conjecture that the constraints that we find in this section are complete. This conjecture is tested in many examples to be discussed in Sec.~\ref{sec:ex}. Based on these constraints, we describe practical algorithms to find possible mirror SETs for a given topological order. In addition, it is very easy to distinguish anomaly-free and anomalous SETs in our formulation.

\subsection{Mirror SET states}

We begin with the data that describe topological and symmetry properties of general mirror SET states.

\subsubsection{Topological data}

A general 2D topological order can be described by a unitary modular tensor category (UMTC)~\cite{kitaev06, WenNSR2016}. The basic properties of a UMTC, which we will use in this section, is reviewed in Appendix~\ref{app:umtc}. Here, we give a brief summary. The content of a UMTC $\mathcal{C}$ includes the anyon types, their fusion properties and braiding properties. We denote anyons in $\mathcal C$ using letters $a$, $b$, $c$, $\cdots$, and in particular, the trivial anyon is denoted by $\mathds 1$. We will abuse the notation $a\in\mathcal C$ to indicate that $a$ is an anyon in $\mathcal C$. Two anyons can fuse into other anyons, according to the fusion rules,
\begin{equation}
  \label{eq:fusion}
  a\times b = \sum_{c\in\mathcal C}N^{ab}_cc,
\end{equation}
where  the fusion multiplicity $N^{ab}_c$ is a non-negative integer. There is an antiparticle $\bar{a}$ of each anyon $a$, with $N^{a\bar{a}}_{\mathds 1}=1$. Each anyon $a$ is associated with a quantum dimension $d_a$, which is the largest eigenvalue of the matrix $\hat N^a$: $(\hat N^a)_{bc} = N^{ab}_c$. Abelian anyons have $d_a=1$, and non-Abelian anyons have $d_a > 1$. The quantum dimensions satisfy the following relation
\begin{equation}
  \label{eq:ddNd}
  d_ad_b = \sum_{c\in\mathcal C}N^{ab}_cd_c.
\end{equation}
One can define the total quantum dimension $D_{\mathcal C}$ of the topological order $\mathcal{C}$:
\begin{equation}
  \label{eq:DC}
  D_{\mathcal C} = \sqrt{\sum_{a\in\mathcal C}d_a^2}.
\end{equation}
Each anyon $a$ is also associated with a topological spin $\theta_a$, which is a unitary phase factor. It is a non-Abelian generalization of the self-statistics of Abelian anyons. The modular $S$ and $T$ matrices are defined as follows
\begin{align}
T_{a,b} & = \theta_a \delta_{a,b} \nonumber\\
S_{a,b} & = \frac{1}{D_{\mathcal{C}}} \sum_c N^{a\bar b}_c \frac{\theta_c}{\theta_a\theta_b} d_c
\end{align}
The $S$ and $T$ matrices essentially summarize the braiding properties of anyons.

Throughout our discussion, we assume that $N^{ab}_c$, $d_a$, $\theta_a$, $S_{a,b}$ and $T_{a,b}$ are all known. In addition,  we assume that the chiral central charge of the 1D conformal field theory living on the boundary of the topological order  $\mathcal C$ is zero (see Appendix~\ref{app:umtc}), and that  $\mathcal{C}$ is compatible with a mirror symmetry.

\subsubsection{Symmetry data}
\label{sec:mirror-set}

Symmetry properties of general SETs are described by three layers of data~\cite{barkeshli14,SET2,tarantino16}: (1) how symmetry permutes anyon types, (2) symmetry fractionalization, and (3) stacking of a possible SPT state. For the mirror symmetry, the last layer of data does not exist, because there is no nontrivial mirror SPT state in 2D. Below we discuss the first two layers of data.

First, the mirror symmetry may permute anyon types. Anyon permutation can be described by a group homomorphism $\rho:\mathbb Z_2^M\rightarrow\aut^\ast(\mathcal C)$, where $\mathbb{Z}_2^M=\{\grp{1},\grp{M}\}$ is the mirror symmetry group and $\aut^\ast(\mathcal C)$ is the group that contains all autoequivalences and anti-autoequivalences of $\mathcal{C}$.  An autoequivalence is a one-to-one map from $\mathcal{C}$ to itself that preserves all the fusion and braiding properties. An antiautoequivalence is a one-to-one map from $\mathcal{C}$ to itself that that preserves the fusion rules but put a complex conjugation on the braiding phase factors.  Specifically, the image $\rho(\grp{1})$ of the identity $\grp{1}$ is a (trivial) autoequivalence, while the image $\rho(\grp M)$ of the mirror symmetry $\grp{M}$ is an antiautoequivalence of $\mathcal C$. The latter follows from the fact that mirror symmetry reverses the spatial orientation. We will use the shorthand notation $\rho_m \equiv \rho(\grp{M})$ from now on. Under the action of the mirror symmetry, an anyon $a$ is sent to another anyon $\rho_m(a)$. It is required that $\rho_m(\mathds{1})=\mathds{1}$. To be compatible with the $\mathbb{Z}_2^M$ group structure, we have $\rho_m(\rho_m(a))=a$. As an antiautoequivalence, $\rho_m$ also satisfies the constraints $N^{\rho_m(a)\rho_m(b)}_{\rho_m(c)}=N^{ab}_c$ and $\theta_{\rho_m(a)} = \theta_a^\ast$. From these constraints, one deduces that $\rho_{m}(\bar a) = \overline{\rho_m(a)}$. It is convenient to define an antiautoequivalence $\bar\rho_m$, with $\bar{\rho}_m(a) = \overline{\rho_m(a)}$, for later discussions.

Second, some anyons may carry mirror-symmetry fractionalization. As a nonlocal symmetry, the fractionalization of the mirror symmetry on an anyon $a$ is defined through the mirror eigenvalue of the wave function that contains two anyons, $a$ and its antiparticle $\bar a$, located symmetrically on the two sides of the mirror axis, shown in Fig.~\ref{fig:mir:anyons}.~\cite{LuBFU, QiCSF, CincioCSL2016}. Since $N^{a\bar a}_{\mathds 1}=1$, this is a non-degenerate state. For the wave function to be mirror symmetric, the anyons satisfy
\begin{equation}
  \label{eq:ra=bara}
  \rho_m(a) = \bar a
\end{equation}
That is, $a$ is \emph{invariant} under the antiautoequivalence $\bar\rho_m$, with $\bar\rho_m(a)=a$. Then, the wave function has a well-defined mirror eigenvalue $\pm1$, which we denote as $\mu(a)$. The collection $\{\mu(a)\}$ defines the mirror symmetry fractionalization class of the SET under a given anyon permutation $\rho_m$. These quantum numbers $\{\mu(a)\}$ should satisfy various constraints. For example, they should be consistent with anyon fusions,
\begin{equation}
  \label{eq:pifusion}
  \mu(a)\mu(b)=\mu(c),
\end{equation}
if $N^{ab}_c\neq 0$ and all $a,b,c$ satisfy Eq.~\eqref{eq:ra=bara}. For another example, if $a=\bar\rho_m(a)$ and $N^a_{b\bar\rho_m(b)}=1$, then\cite{barkeshli14}
\begin{equation}
\label{eq:bbarrhob}
\mu(a) = \theta_a
\end{equation}
In the case of Abelian topological order, it reduces to Eq.~\eqref{eq:constraint2}, since $\theta_a = \theta_b\theta_{\bar\rho_m(b)}M_{b,\bar\rho_m(b)}=M_{b,\bar\rho_m(b)}$, where $M_{b,\bar\rho_m(b)}$ is the mutual braiding statistics between $b$ and $\bar\rho_m(b)$. For Abelian topological orders, it is generally believed that the constraints Eqs.~\eqref{eq:pifusion} and \eqref{eq:bbarrhob} are complete. However, for general non-Abelian topological orders, these constraints are known to be incomplete.

One of the main purposes of this paper is to find a stronger and hopefully complete set of constraints on the symmetry fractionalization $\{\mu(a)\}$ for a given anyon permutation $\rho_m$, through the folding method and principles of anyon condensation. In most of the following discussions, we will assume that a valid antiautoequivalence $\rho_m$ is known (however, see an example with an \emph{invalid} $\rho_m$ in Sec.~\ref{sec:ex:d16}, which carries the so-called $H^3$-type obstruction).



\subsection{Folded double-layer system}
\label{sec:folding}


We now fold a general mirror SET with the topological order $\mathcal C$ into a double-layer system, as shown in Fig.~\ref{fig:mir}.  The mirror axis divides the system into region $A$ and $B$. We fold $B$ such that it overlaps with $A$. When the region $B$ is folded, its spatial orientation is reversed, and the topological order becomes the reverse of the original one, which we denote as $\mathcal C^{\text{rev}}$. The topological spins of all anyons in $\mathcal C^{\text{rev}}$ and their mutual statistics are reversed, i.e., get complex conjugated. More specifically, let $a^{\text{rev}}\in\mathcal{C}^{\text{rev}}$ be the ``reverse'' of the anyon $a\in\mathcal{C}$. Then, $\theta_{a^{\text{rev}}} = \theta_a^*$. The bulk topological order of the double-layer system should be described by $\mathcal{C}\otimes \mathcal{C}^{\text{rev}}$.

For a general UMTC $\mathcal C$, its reverse $\mathcal C^{\text{rev}}$ could be a different topological order.
However, if $\mathcal C$ preserves a mirror symmetry, it must be equivalent to its reverse $\mathcal C\simeq\mathcal C^{\text{rev}}$. The equivalence is naturally defined by the mirror action $\rho_m$, as an antiautoequivalence of $\mathcal{C}$, through the one-to-one mapping
\begin{equation}
a^{\text{rev}}\leftrightarrow \rho_m(a)
\label{eq:relabel}
\end{equation}
where $a^{\text{rev}}\in \mathcal{C}^{\text{rev}}$ and $\rho_m(a)\in\mathcal{C}$. Using this mapping, we generalize the relabeling trick first introduced in Sec.~\ref{sec:z2:nma}: the anyon $a^{\text{rev}}$ in $\mathcal{C}^{\rm rev}$ is relabeled as $\rho_m(a)$. After this relabeling, we can view the double-layer system as $\mathcal{C}\otimes\mathcal{C}$. In this notation, anyons of the double-layer system are labeled by the doublets $(a, b)$, whose topological spin is given by $\theta_{(a,b)} = \theta_a\theta_b$ and quantum dimension is given by $d_{(a,b)}=d_ad_b$. We will use this relabeled notation in the rest of the paper.

After folding, the mirror symmetry $\mathbf{M}$ becomes an onsite unitary $\mathbb{Z}_2$ symmetry $\mathbf{m}$. In the relabeled notation, it  permutes the anyons as follows
\begin{equation}
\mathbf{m}: (a,b)\rightarrow(b,a)
\label{eq:permute}
\end{equation}
As discussed in Sec.~\ref{sec:z2bulk}, there are only two possible $\mathbb{Z}_2$ SETs for this type of anyon permutation, which differ from each other by stacking a $\mathbb{Z}_2$ SPT state. The bulk SET order of the double-layer system is one of the two possibilities, which we will specify after we gauge the $\mathbb{Z}_2$ symmetry.

As discussed in Sec.~\ref{sec:introidea}, information of mirror SET states is encoded only near the mirror axis. Different mirror SETs are described by different anyon-condensation boundary conditions of the same bulk.
An anyon $a$ in region $A$ can move across the mirror axis and become an anyon $a$ in region $B$, which is relabeled as the anyon $\rho_m(a)$ on the second layer in the folded picture.
Hence, on the boundary, we should identify $(a,0)$ with $(0,\rho_m(a))$, or equivalently, condense the anyon $(a,\bar\rho_m(a))$. That is, the gapped boundary correspond to a condensate of $\{(a,\bar\rho_m(a)) \}_{a\in\mathcal{C}}$.

Furthermore, the gapped boundary is symmetric under $\mathbb{Z}_2$ symmetry. We need to consider symmetry properties of the condensate $\{(a,\bar\rho_m(a)) \}_{a\in\mathcal{C}}$. Under the $\mathbb{Z}_2$ symmetry $\grp{m}$,  $(a,\bar\rho_m(a))\rightarrow (\bar\rho_m(a),a)$. It is obvious that both $(a,\bar\rho_m(a))$ and $(\bar\rho_m(a),a)$ are contained in the condensate $\{(a,\bar\rho_m(a)) \}_{a\in\mathcal{C}}$. When $a$ satisfies $a=\bar\rho_m(a)$, i.e., when $(a,\bar\rho_m(a)) = (\bar\rho_m(a),a)$, we can further define a $\mathbb Z_2$ symmetry eigenvalue $\mu(a)$. The eigenvalue $\mu(a)$ is inherited from the mirror symmetry fractionalization through folding. Similarly to the discussion in Sec.~\ref{sec:z2boundary}, the gapped boundary should be understood as a condensate of anyons $(a,\bar\rho_m(a))$ that carry $\mathbb Z_2$ eigenvalue $\mu(a)$. In this way, both the symmetry permutation $\rho_m$ and the symmetry fractionalization factors $\{\mu(a)\}$ are encoded in the pattern of anyon condensation on the boundary.



\subsection{Bulk of gauged double-layer system}
\label{sec:math}

To better understand the anyon condensation pattern and in particular to derive possible constraints on $\{\mu(a)\}$ from principles of anyon condensation, we promote the $\mathbb{Z}_2$ interlayer exchange symmetry to a gauge symmetry~\cite{levin12}. In this subsection, we describe the bulk of the double-layer system after gauging the $\mathbb{Z}_2$ symmetry. The gauged double-layer system is described by another UMTC, which we denote $\mathcal D$. The modular data of $\mathcal D$, including $S$ and $T$ matrices as well as topological spins and fusion rules, can be deduced from $\mathcal C$. Details of the derivation is discussed in Appendix~\ref{app:dl}. Here, we give a summary of topological order $\mathcal{D}$.

The topological order $\mathcal D$ consists of those anyon charges that have counterparts in the ungauged double-layer system and the $\mathbb Z_2$ defects that do not have counterparts. Let us first list their quantum dimensions and topological spins. (1) For each pair of different anyons $a\neq b$ from $\mathcal C$, there is an anyon in $\mathcal D$, denoted by $[a,b]$, with the quantum dimension $d_{[a,b]}=2d_ad_b$ and the topological spin $\theta_{[a,b]}=\theta_a\theta_b$. Physically, this anyon represents one $a$ anyon and one $b$ anyon on each layer, respectively, and it does not carry a well-defined $\mathbb Z_2$ charge. (2) For each pair of identical anyon charges $a=b\in\mathcal C$, there are two anyons in $\mathcal D$, denoted by $(a,a)^\pm$, respectively. They have the same quantum dimension $d_{(a,a)^\pm}=d_a^2$ and the same topological spin $\theta_{(a,a)^\pm}=\theta_a^2$.  (3) Finally, the $\mathbb Z_2$ symmetry defects are labeled by $X_a^\pm$, where $a\in\mathcal C$ denotes an anyon charge that can be attached to the defect.  The quantum dimension of the defect is given by $d_{X_a^\pm}=d_aD_{\mathcal C}$, and its topological spin is given by
\begin{equation}
  \label{eq:thetaXa}
  \theta_{X_a^\pm}=\pm\sqrt{\theta_a}.
\end{equation}
where the signs $\pm$ are correlated on the two sides.  Equation \eqref{eq:thetaXa} only describes one of the two possible $\mathbb{Z}_2$ SETs; the other possible SET is associated with the topological spin $\theta_{X_a^\pm}=\pm i\sqrt{\theta_a}$. We claim that the bulk SET order of the double-layer system is associated with the topological spin in Eq.~\eqref{eq:thetaXa}, which will later produce the desired mirror anomaly (see more discussions in Sec.~\ref{sec:anomaly}). Summing over all anyons listed above, one can compute the total quantum dimension of $\mathcal D$,
\begin{equation}
  \label{eq:DD}
  D_{\mathcal D}=2D_{\mathcal C}^2.
\end{equation}


Here we make a comment on the notational signs ``$\pm$'' on $(a,a)^\pm$ and $X_a^\pm$. First, the sign $\pm$ on $X_a^\pm$ is conventional and has no absolute meaning. It only represents that $X^+_a$ and $X_a^-$ differ relatively by a $\mathbb{Z}_2$ charge $(\mathds{1},\mathds{1})^-$. We use a convention such that Eq.~\eqref{eq:thetaXa} holds. Second, the $\pm$ sign on $(a,a)^\pm$ can be roughly understood as the integer $\mathbb Z_2$ symmetry charge that the anyon carries. However, strictly speaking, the $\pm$ sign corresponds to the absolute $\mathbb{Z}_2$ charge on $(a,a)$ only if $\bar\rho_m(a)=a$. Under this condition, one can define the $\mathbb{Z}_2$/mirror eigenvalue of $(a,a)$ simply through unfolding. The $\pm$ sign on $(a,a)^\pm$ will finally be translated to the symmetry fractionalization $\mu(a)$ through anyon condensation on the boundary, to be discussed in the next subsection. On the other hand, if $\bar\rho_m(a) \neq a$, there is no physically transparent way to define the $\mathbb{Z}_2$ charge on $(a,a)$. However, regardless the existence of physical definitions, we will abuse the language and call the $\pm$ sign  ``$\mathbb{Z}_2$ charge'' for both $(a,a)^\pm$ and $X_a^{\pm}$.

Next, we give a partial list of the fusion coefficients of $\mathcal D$, which we will use later. First, the fusion among the anyons $[a,b]$ and $(a,a)^\pm$ naturally follows the fusion in $\mathcal C$. The $\mathbb Z_2$ charges in $(a,a)^\pm$ and $(b,b)^\pm$ can be simply multiplied. Second, the fusion rule between the anyon $[b,c]$ and the defect $X_a^\pm$ is as follows,
\begin{equation}
  \label{eq:Xabc}
  X_a^+\times[b,c]=X_a^-\times[b,c]=\sum_{d,e}N^{bc}_dN^{da}_e(X_e^++X_e^-).
\end{equation}
The rest of the fusion coefficients, however, are harder to compute; in particular, it is not easy to determine the $\mathbb Z_2$ charges of the fusion outcomes. Fortunately, they are not needed for our later discussion. As pointed out in Appendix~\ref{app:dl}, they can in principle be computed using the Verlinde formula in Eq.~\eqref{eq:verlinde}, from the $S$ matrix of $\mathcal{D}$ given in Appendix~\ref{app:dl}.

Finally, the $S$ matrix of $\mathcal D$ can also be computed, using the $S$ and $T$ matrices of $\mathcal C$. In particular, using the fusion rule in Eq.~\eqref{eq:Xabc} and the topological spins in Eq.~\eqref{eq:thetaXa}, one immediately sees that
\begin{equation}
  \label{eq:SXa[bc]}
  S(X_a^\pm,[b,c])=0.
\end{equation}
[Here, for clarity, we denote entries of the $S$ matrix of $\mathcal D$ by $S(a,b)$ instead of $S_{a,b}$.] Also, as shown in Appendix \ref{app:dl}, the $S$ matrix entry between $X_a^\pm$ and $(b,b)^\pm$ is determined by $S_{a,b}$ in $\mathcal C$ and the $\mathbb Z_2$ charge of $(b,b)^\pm$,
\begin{equation}
  \label{eq:SXa(bb)}
  S(X_a^+,(b,b)^\pm) = S(X_a^-,(b,b)^\pm) = \pm\frac12S_{a,b}.
\end{equation}

\subsection{Anyon condensation on the boundary}
\label{sec:anycond}


The folding approach turns the mirror SETs, described by $\rho_m$ and $\{\mu(a)\}$, into anyon condensation patterns that happen on the boundary. We now derive a (potentially complete) set of  the constraints on the $\{\mu(a)\}$ for a given $\rho_m$ through various consistency conditions of anyon condensation in the gauged double-layer system.

\subsubsection{Brief review on anyon condensation}


We first give a brief review on the general anyon condensation theory (see also Refs.~\onlinecite{bais2009, eliens14} for general theories, however, we will use the formulation from Refs.~\onlinecite{TLanAnyCon2015, NeupertAnyCon2016} ). In particular, we review a set of consistency conditions that constrain possible anyon condensation patterns.

\begin{figure}
  \includegraphics{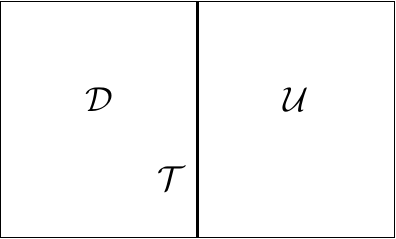}
  \caption{Anyon condensation theory describing a gapped boundary $\mathcal T$, between topological orders $\mathcal D$ and $\mathcal U$.}
  \label{fig:dtu}
\end{figure}

Despite of its name, anyon condensation theory describes a gapped boundary between two topological orders. An anyon condensation pattern is associated with three categories (Fig.~\ref{fig:dtu}): a parent UMTC $\mathcal{D}$ and a child UMTC $\mathcal{U}$ that live on the two sides of the boundary respectively, and  a unitary fusion category (UFC) $\mathcal{T}$ that lives on the boundary and that does not have a well-defined braiding structure. The UFC $\mathcal{T}$ should be considered as an intermediate stage of the ``condensation transition'' from $\mathcal{D}$ to $\mathcal{U}$, while $\mathcal{U}$ is the physical outcome of the transition. So, people often do not care much about properties of $\mathcal{T}$ and mainly focus on the connection between $\mathcal{D}$ and $\mathcal{U}$.

An anyon condensation pattern can be described in two steps. In the first step, one needs to specify a restriction map $r:\mathcal D\rightarrow\mathcal T$.  It can be encoded in a $|\mathcal D|\times|\mathcal T|$ matrix $n_{\alpha,t}$, where $\alpha\in \mathcal D$ and $t\in\mathcal T$, respectively, as the following
\begin{equation}
  \label{eq:rant}
  r(\alpha)=\sum_tn_{\alpha,t}t.
\end{equation}
Each entry of this matrix $n_{\alpha,t}$ must be a nonnegative integer. Physically, the restriction map means that when it moves to the boundary, $\alpha$ \emph{splits} into those anyons  $t\in\mathcal{T}$ with $n_{\alpha,t}\neq 0$. When $n_{\alpha,t}>1$, $\alpha$ can split into multiple copies of $t$. At the same time, one can define a ``reverse'' map of Eq.~\eqref{eq:rant}, called the lifting map, which is also encoded by $n_{\alpha,t}$ as follows
\begin{equation}
  \label{eq:ltna}
  l(t)=\sum_\alpha n_{\alpha,t}\alpha.
\end{equation}
Physically, it means that for a fixed $t$, all $\alpha$'s with $n_{\alpha,t}\neq 0$ will be \emph{identified} into the same anyon $t$ on the boundary. We will use the notation that  $\alpha\in l(t)$ to indicate that $n_{\alpha,t}\neq 0$. A special lifting is the collection of anyons in $l(\mathds{1})$ with $n_{\alpha,\mathds{1}}\neq 0$, which will be referred to as  ``the condensed anyons'' or ``the anyons in the condensate''. These anyons all become the  vacuum anyon in $\mathcal{T}$.

In the second step, a subset of anyons in $\mathcal{T}$ will be associated with a well-defined braiding structure. These anyons are called \emph{deconfined}, and they form the UMTC $\mathcal{U}$. Other anyons in $\mathcal{T}$ are said to be \emph{confined}. A simple criterion to distinguish confined and deconfined anyons is that: if all anyons in the lift $l(t)$ have the same topological spin, $t$ is deconfined; otherwise, $t$ is confined. For deconfined anyons, $\theta_t=\theta_\alpha$ for any $\alpha\in l(t)$; for confined anyons, there is no physical way to define their topological spins.  It is required  that the vacuum anyon $\mathds{1}\in \mathcal{T}$ must be deconfined, therefore all anyons in the condensate $l(\mathds 1)$ are self-bosons.

We see that the matrix $n_{\alpha,t}$ specifies an anyon condensation pattern.\footnote{In certain complicated cases, the matrix $n_{\alpha,t}$ cannot uniquely specify an anyon condensation pattern, i.e., the type of the gapped domain wall, under proper definition of the latter (e.g., see Ref. \onlinecite{davydov13}). However, in this work we neglect this subtlety and treat $n_{\alpha,t}$ equivalent to an ``anyon condensation pattern''.} In fact, for most of the discussions below, we are mainly interested in the part of the matrix $n_{\alpha,t}$ when $t\in \mathcal{U}$, which describes the connection between $\mathcal{D}$ and $\mathcal{U}$. For an anyon condensation to be well defined, the matrix $n_{\alpha,t}$ should satisfy various constraints. Below we list two main constraints that will be used later. Generally speaking, there may exist more independent constraints other than the following ones.

First, the restriction map $r$ commutes with the fusion process: $r(\alpha)\times r(\beta) = r(\alpha\times \beta)$. Expanding the fusion rules on both sides, we are led to the following constraint on $n_{\alpha,t}$:
\begin{equation}
  \label{eq:nnN=Nn}
  \sum_{r,s\in\mathcal T}n_{\alpha,r}n_{\beta,s}N^{rs}_t
  = \sum_{\gamma\in\mathcal D}N^{\alpha\beta}_\gamma n_{\gamma,t}
\end{equation}
where $N^{rs}_t$ is a fusion coefficient in $\mathcal{T}$ and $N^{\alpha\beta}_\gamma$ is a fusion coefficient in $\mathcal{D}$.  Two corollaries of this constraint are
\begin{equation}
  \label{eq:rcons:qdr}
  d_\alpha = \sum_tn_{\alpha,t}d_t.
\end{equation}
and
\begin{equation}
  \label{eq:rcons:qdl}
  d_t = \frac{D_{\mathcal U}}{D_{\mathcal D}}\sum_an_{\alpha,t}d_\alpha.
\end{equation}
See e.g. Ref.~\onlinecite{NeupertAnyCon2016} for a proof of the corollaries based on Eq.~\eqref{eq:nnN=Nn}.


Second, the matrix $n_{a,t}$ commutes the modular matrices of $\mathcal D$ and $\mathcal U$ in the following sense
\begin{equation}
  \label{eq:STn=nST}
  S^{\mathcal D}n = nS^{\mathcal U},\quad
  T^{\mathcal D}n = nT^{\mathcal U},
\end{equation}
where the superscripts of $S$ and $T$ denote the topological orders that the matrices are associated with. The multiplications appear in these equations are matrix multiplications.  The first equation in \eqref{eq:STn=nST} can be more explicitly written as
\begin{equation}
  \label{eq:Sn=nS:app}
  \sum_{\beta\in\mathcal D}S^{\mathcal D}_{\alpha,\beta}n_{\beta,t}
  = \sum_{s\in\mathcal U}n_{\alpha,s}S^{\mathcal U}_{s,t}, \quad t\in \mathcal{U}
\end{equation}
which will be an important constraint below. The second equation in Eq.~\eqref{eq:STn=nST} is equivalent to the statement that for a given $t\in\mathcal{U}$, all $\alpha$'s with $n_{\alpha,t}\neq 0$ share the same topological spin and $\theta_\alpha=\theta_t$.

\subsubsection{Application to the double-layer system}
\label{sec:app2dl}

We now specialize to our double-layer system. The parent UMTC  $\mathcal{D}$ is the bulk topological order discussed in Sec.~\ref{sec:math}. Also, we understand that the child UMTC $\mathcal{U}$ is either the toric code or double semion topological order, i.e., the untwisted or twisted $\mathbb{Z}_2$ topological order respectively.

Let us understand some aspects of the unitary fusion category (UFC) $\mathcal T$ that lives on the boundary. Before gauging the $\mathbb{Z}_2$ symmetry, it is easy to understand that those anyons that live on the boundary are exactly those in $\mathcal{C}$: when anyons in $\mathcal{C}\otimes\mathcal{C}$ move to the boundary, the anyon from the top and bottom layers fuse together, the outcomes of which are exactly the anyons in $\mathcal{C}$ (although their braiding structure is lost when we confine them on the boundary). When the $\mathbb{Z}_2$ symmetry is gauged, $\mathcal T$ should contain anyons in $\mathcal U$, which are two $\mathbb Z_2$ symmetry charges and two $\mathbb Z_2$ symmetry defects. Therefore,  it is not hard to see that anyons in $\mathcal T$ should be labeled as follows
\begin{equation}
\mathcal T = \{ a^\pm, x_a^\pm | a\in\mathcal{C} \}
\end{equation}
where the anyons $a^\pm$ originate from the anyon $a$ before gauging the symmetry and are further decorated with the $\mathbb{Z}_2$ charge after gauging,  and $x_a^\pm$ are understood as symmetry defects (we use the little $x$ to distinguish it from the symmetry defects in $\mathcal{D}$).  Among all anyons in $\mathcal{T}$, only the vacuum anyon $\mathds1^+\equiv \mathds 1$, the $\mathbb{Z}_2$ charge $\mathds1^-$, and two defects which we conventionally denote as $x_{\mathds1}^\pm$ are deconfined, i.e., they are free to move to the other side of the boundary and become anyons in $\mathcal{U}$. All other anyons are confined to the boundary. One may compute the total quantum dimension of $\mathcal{T}$ using the general relation $ D_{\mathcal T} = \sqrt{D_{\mathcal D}D_{\mathcal U}} $, and find that $D_{\mathcal{T}}=2 D_{\mathcal{C}}$.


To describe the anyon condensation patterns, we need to specify the restriction/lifting maps. Let us begin with some understanding of the restriction map. From physical picture in the ungauged system discussed in Sec.~\ref{sec:folding}, we can understand that before gauging the $\mathbb{Z}_2$ symmetry, at the boundary, an anyon $a$ on one layer is identified with the anyon $\rho_m(a)$ on the other layer. Hence, two anyons $a$ and $b$ from each layer are merged into a single anyon in the fusion product $a\times\rho_m(b)$. This understanding makes us to  assume that after gauging the $\mathbb{Z}_2$ symmetry, the anyon charges $[a,b]$ and $(a,a)^\pm$ are restricted in the following way,
\begin{align}
  \label{eq:rab}
  r([a,b])&=\sum_c N^{a\rho_m(b)}_c(c^++c^-),\\
  \label{eq:raa}
  r((a,a)^+)&=\sum_c N^{a\rho_m(a)}_cc^{\mu(a,\rho_m(a);c)},\\
  \label{eq:raa-}
  r((a,a)^-)&=\sum_c N^{a\rho_m(a)}_cc^{-\mu(a,\rho_m(a);c)},
\end{align}
where $\mu(a,\rho_m(a);c)=\pm$ is a $\mathbb Z_2$ symmetry charge, which differs from different anyon condensation patterns. In particular, when $a=\bar\rho_m(a)$, $\mu(a,\rho_m(a);\mathds1)=\mu(a)$ which encodes the mirror-symmetry fractionalization and is the main quantity that we are interested in. When writing down Eqs.~\eqref{eq:rab}-\eqref{eq:raa-}, we have used the observation that $x^\pm_a$ is not contained in $ r([a,b])$ and $r((a,a)^\pm)$. On the other hand, we expect that the restriction $r(X_a^\pm)$ should only contain defects $x_a^\pm$, but none of the anyons $a^\pm$. Due to the fact that restriction map commute with fusion, we have that
\begin{equation}
r(X_a^-)=r((\mathds1,\mathds 1)^-)\times r(X_a^+)
\end{equation}
which follows from the fusion rule that $X_a^-=X_a^+\times (\mathds 1,\mathds 1)^-$.

Next, we describe the lifting map. In fact, we are only interested in the lifting of $\mathcal{U}$ to $\mathcal{D}$, i.e. $l(\mathds 1^\pm)$ and $l(x_{\mathds 1}^\pm)$. The lifting of the confined anyons in $\mathcal{T}$ are not relevant for us to understand mirror symmetry fractionalizations. From the above understanding of the restriction map, we immediately find that
\begin{equation}
  \label{eq:l1}
  l(\mathds1)= \sideset{}{'}\sum_{a\neq\bar\rho_m(a)}[a,\bar\rho_m(a)]
  +\sum_{a=\bar\rho_m(a)}(a,a)^{\mu(a)},
\end{equation}
where $\sideset{}{'}\sum$ means that only one among the two anyons, $a$ and $ \bar\rho_m(a)$, are taken in the summation (note that $[a,\bar\rho_m(a)]=[\bar\rho_m(a),a]$). The condensate $l(\mathds{1})$ in Eq.~\eqref{eq:l1} is what we expect from the discussion in Sec.~\ref{sec:folding}. Similarly, we have
\begin{equation}
  \label{eq:l1-}
  l(\mathds1^-)= \sideset{}{'}\sum_{a\neq\bar\rho_m(a)}[a,\bar\rho_m(a)]
  +\sum_{a=\bar\rho_m(a)}(a,a)^{-\mu(a)},
\end{equation}
One the other hand, it is not apparent to us what are the liftings $l(x_{\mathds 1}^\pm)$. However, we do expect that the lifting $l(x_{\mathds1}^+)$ has the following form
\begin{equation}
  \label{eq:lx1}
  l(x_{\mathds1}^+) = \sum_{a\in\mathcal C}w_a X_a^+,
\end{equation}
where $w_a\equiv n_{X_a^+,x_{\mathds1}^+}$ are nonnegative integers, and $w_a\neq0$ indicates that $X_a^+$ can become the deconfined $\mathbb Z_2$ defect $x_{\mathds 1}^+$ on the other side of the mirror axis. The fact that the lifting of $x_{\mathds1}^+$ only contains defects with the $+$ charge is due to two reasons: (i) all defects in $l(x_{\mathds1}^+)$ should have the same topological spin, and this is possible only if the defects carry the same sign due to Eq.~\eqref{eq:thetaXa}; and (ii) the sign on $x_{\mathds 1}^\pm$ is also conventional, so we just choose the sign on $x_{\mathds 1}^\pm$ to match that of the $X$'s that are contained in the lifting. The fact that all $X_a^+$'s with $w_a\neq0$ share the same topological spin implies that all $a$'s also share the same topological spin. That is,
\begin{equation}
\label{eq:spinconstraint}
\theta_a = \theta_{a'}, \quad \text{for any $w_a,w_{a'}\neq0$}
\end{equation}
Since $x_{\mathds 1}^+$ is its own antiparticle, we always have $w_a=w_{\bar a}$. It is not hard to see that
\begin{equation}
  \label{eq:lx1-}
  l(x_{\mathds1}^-) = \sum_{a\in\mathcal C}w_a X_a^-.
\end{equation}
One compact way of expressing the coefficients $w_a$ is to define the following superposition of anyons,
\begin{equation}
  \label{eq:w=wa}
  \textswab w = \sum_{a\in\mathcal C}w_aa.
\end{equation}
We will abuse the notation $a\in\textswab w$ to indicate that $w_a\neq0$. We summarize all the matrix elements $n_{\alpha,t}$ for $\alpha\in\mathcal{D}$ and $t\in\mathcal{U}$ in Table \ref{tab:nat}.

\begin{table}
  \caption{The matrix elements $n_{\alpha,t}$ for $\alpha\in\mathcal{D}$ and $t\in\mathcal{U}$ in the gauged double-layer system. }
  \label{tab:nat}
  \begin{tabularx}{0.7\textwidth}{C|CCC CCCC CC}
    \hline\hline
    \diagbox[width=4em]{$\alpha$}{$t$} & $\mathds1$  & $\mathds1^-$ & $x^+_{\mathds1}$ &  $x_{\mathds1}^-$\\
    \hline
    $(a,a)^{\mu(a)}$  & $\delta_{a,\bar\rho_m(a)}$ & 0 & 0 & 0\\
    $(a,a)^{-\mu(a)}$ & 0 & $\delta_{a,\bar\rho_m(a)}$ & 0 & 0\\
    $[a,b]$      & $\delta_{b,\bar\rho_m(a)}$ & $\delta_{b,\bar\rho_m(a)}$  & 0 & 0 \\
    $X_a^+$           & 0 & 0 & $w_a$ & 0 \\
    $X_a^-$           & 0 & 0 & 0 & $w_a$ \\
    \hline\hline
  \end{tabularx}
\end{table}

For a given $\rho_m$, we see that the liftings $l(\mathds 1^\pm)$ and $l( x_{\mathds 1}^\pm)$ are specified by $\{\mu(a)\}$ and $\{w_a\}$, where $\mu(a)=\pm 1$ and $w_a$ is a nonnegative integer.  Intuitively, we expect that once the condensate $l(\mathds1)$ in Eq.~\eqref{eq:l1} is specified, i.e., $\rho_m$ and $\mu(a)$ are given, the anyon condensation pattern should be fully determined. That is, $w_a$ is not independent. Indeed, we show that $\mu(a)$ and $w_a$ are mutually determined through the following relations:
\begin{equation}
  \label{eq:wa=Sp}
  w_a=\sum_{b=\bar\rho_m(b)}S_{a,b}\mu(b),
\end{equation}
and
\begin{equation}
  \label{eq:pa=Sw}
  \mu(a)=\sum_{b}S_{a,b}w_b,
\end{equation}
where $S_{a,b}$ is the $S$ matrix of $\mathcal{C}$, and the $a$ in Eq.~\eqref{eq:pa=Sw} satisfies $a=\bar\rho_m(a)$. (We will show shortly that the summation in Eq.~\eqref{eq:wa=Sp} can be extended to all $b$'s, and Eq.~\eqref{eq:pa=Sw} can be extended to arbitrary $a$, if we define $\mu(a)=0$ for $a\neq\bar\rho_m(a)$.) Since the $S$ matrix is unitary, different $\{w_a\}$'s lead to different symmetry fractionalization $\{\mu(a)\}$. To prove the relations, we make us of the consistency condition Eq.~\eqref{eq:Sn=nS:app}, which in the current notation is
\begin{equation}
  \label{eq:snns}
  \sum_{\beta\in\mathcal D}S(\alpha,\beta)n_{\beta,t}=\sum_{s\in\mathcal U}n_{\alpha,s}S_{s,t},
\end{equation}
where $\alpha$ and $t$ are anyons in $\mathcal D$ and $\mathcal U$ respectively, and $S(\alpha,\beta)$ and $S_{s,t}$ are $S$ matrices of $\mathcal{D}$ and $\mathcal{U}$ respectively.  Taking $\alpha=X_a^+$ and $t=\mathds1$ in Eq.~\eqref{eq:snns}, and further using $n_{\alpha,t}$ listed in Table \ref{tab:nat} and the $S$ matrices in Eqs.~\eqref{eq:SXa[bc]} and \eqref{eq:SXa(bb)}, we immediately obtain Eq.~\eqref{eq:wa=Sp}. At the same time, taking $\alpha=(a,a)^+$ that satisfies $a=\bar\rho_m(a)$ and $t=x_{\mathds1}^+$ in  Eq.~\eqref{eq:snns}, and further using Table \ref{tab:nat} and $S$-matrix entries, we immediately obtain Eq.~\eqref{eq:pa=Sw}.

In fact, Eq.~\eqref{eq:pa=Sw} can be extended to arbitrary $a$, not necessarily for those satisfying $a=\bar\rho_m(a)$. This is possible if we define $\mu(a)=0$ for $a\neq \bar\rho_m(a)$.  To see that, we take $\alpha=(a,a)^+$ with $a\neq\bar\rho_m(a)$ and $t=x_{\mathds 1}^+$ in Eq.~\eqref{eq:snns}. Since $(a,a)^+$ does not restrict to any anyon in $\mathcal U$, the right-hand side of Eq.~\eqref{eq:snns} equals 0. This implies that Eq.~\eqref{eq:pa=Sw} holds if we define $\mu(a)=0$ for $a\neq \bar\rho_m(a)$.  With this extension on $\mu(a)$, the summation in Eq.~\eqref{eq:wa=Sp} can also be extended to all anyons $b$. Then, we see that Eq.~\eqref{eq:wa=Sp} and Eq.~\eqref{eq:pa=Sw} are actually equivalent. Indeed, from Eq.~\eqref{eq:pa=Sw}, we have
\begin{align}
w_a = \sum_{b} S^{-1}_{a,b}\mu(b) = \sum_{b} S^*_{b,a}\mu(b) = \sum_{b} S_{a,b}\mu(b)
\end{align}
where the first equality follows from that $S$ is invertible,  the second equality follows from that $S$ is unitary, and the last equality follows from that $S$ is symmetric and that both $w_a$ and $\mu(a)$ are real. Therefore, in the rest of our paper, when we refer to Eqs.~\eqref{eq:wa=Sp} and \eqref{eq:pa=Sw}, we will implicitly assume that $\mu(a)=0$ for $a\neq\bar\rho_m(a)$ and the summations are over all anyons in $\mathcal{C}$.

Now that $\{\mu(a)\}$ and $\{w_a\}$ are equivalent, any constraint on $\{w_a\}$ can be translated into that on the symmetry fractionalization $\{\mu(a)\}$. The consistency condition Eq.~\eqref{eq:nnN=Nn} of anyon condensation can be used to derive an important constraint on $w_a$. The constraint is
\begin{equation}
  \label{eq:wxw}
  \textswab w\times\textswab w=\sum_{a\in\mathcal C}a\times\bar\rho_m(a).
\end{equation}
The derivation is a little involved, so we separately give it in Appendix \ref{app:derive_wxw}. It can be more explicitly written as
\begin{equation}
\label{eq:wxw2}
\sum_{ab}w_aw_b N^{ab}_c = \sum_a N^{a\bar\rho_m(a)}_c
\end{equation}
which holds for any $c\in \mathcal{C}$. Applying Eq.~\eqref{eq:ddNd} to this result, we can show that the total quantum dimension of $\textswab w$ is $D_{\mathcal C}$,
\begin{equation}
  \label{eq:Dw}
  \sum_{a\in\textswab w}w_ad_a=D_{\mathcal C}.
\end{equation}

\subsubsection{Summary of main results}
\label{sec:constraints}

We have translated the data $\rho_m$ and $\{\mu(a)\}$ of mirror SETs into a description of anyon condensation through the folding approach. From the principles of anyon condensation, we have defined a new set of data $\{w_a\}$, which is equivalent to the symmetry fractionalization $\{\mu(a)\}$ for a given $\rho_m$. The equivalence follows from Eqs.~\eqref{eq:wa=Sp} and \eqref{eq:pa=Sw}. Using consistency conditions of anyon condensation, we find that $\{w_a\}$ satisfies the following constraints:
\begin{enumerate}
\item $w_a$ is a non-negetative integer;
\item $\theta_a$ is the same for all $w_a\neq 0$, i.e., satisfying Eq.~\eqref{eq:spinconstraint};
\item $\textswab w$ satisfies Eq.~\eqref{eq:wxw}, or explicitly $\{w_a\}$ satisfy Eq.~\eqref{eq:wxw2}.
\end{enumerate}
At the same time, we summarize the constraints on the original symmetry fractionalization data $\{\mu(a)\}$ here:
\begin{enumerate}
\item $\mu(a)=\pm 1$ for $a=\bar\rho_m(a)$;
\item $\mu(a)$ satisfies Eq.~\eqref{eq:pifusion};
\item $\mu(a)$ satisfies Eq.~\eqref{eq:bbarrhob}.
\end{enumerate}
We show in Appendix \ref{app:concon} that some parts of the constraints on $\{\mu(a)\}$ can be derived from those on $\{w_a\}$. In the case of Abelian topological orders,  we can show that the two sets of constraints turn out to be equivalent.


\subsection{Finding mirror-SETs}
\label{sec:alg}




In this section, we describe how to find possible mirror symmetry fractionalization $\{\mu(a)\}$ for a given $\rho_m$ by solving the constraints summarized in Sec.~\ref{sec:constraints}, which are either directly on $\{\mu(a)\}$ or indirectly defined through $\{w_a\}$. Ideally when the constraints are complete, every solution should correspond to a physical mirror SET. While we cannot prove the completeness, all examples that we work out in Sec.~\ref{sec:ex} seem to support it.

We describe two algorithms to solve the constraints. The two algorithms treat the data $\{\mu(a)\}$ and $\{w_a\}$ in different priorities: the first algorithm uses $\{\mu(a)\}$ as the primary data and then it is supplemented with constraints of $\{w_a\}$;  on the other hand, the second algorithm uses $\{w_a\}$ as the primary data and then it is supplemented with constraints of $\{\mu(a)\}$. Both algorithms assume that an anyon permutation $\rho_m$ is given.

In the first algorithm, we start with $\{\mu(a)=\pm 1\}$ for $a=\bar\rho_m(a)$. Without affecting any physics, we set $\mu(a)=0$ if $a\neq \bar\rho_m(a)$. There are finitely many possible choices of the values of $\{\mu(a)\}$. Next, one picks out the combinations $\{\mu(a)\}$ that satisfy  Eq.~\eqref{eq:pifusion}. In addition, one can use Eq.~\eqref{eq:bbarrhob} to fix the value of $\mu(a)$ for certain anyons.  After that, we insert each picked combination $\{\mu(a)\}$ into Eq.~\eqref{eq:wa=Sp}, and obtain a set of $\{w_a\}$. Then, we test (i) whether $w_a$ are nonnegative integers, (ii) whether $\theta_a$ are equal for all anyons with $w_a\neq0$ and (iii) whether Eq.~\eqref{eq:wxw} holds. Only those combinations that satisfy all three constraints are kept and potentially describe physical mirror SETs.  This algorithm has a computational cost that scales exponentially with numbers of independent  $\mu(a)$.  

The second algorithm starts with the non-negative integers $\{w_a\}$. The first step is to solve Eq.~\eqref{eq:wxw} or Eq.~\eqref{eq:wxw2}. This is a set of $|\mathcal C|$ quadratic equations with $|\mathcal{C}|$ unknowns. We do not know a way to systematically search for solutions to Eq.~\eqref{eq:wxw} or Eq.~\eqref{eq:wxw2}, but answers can usually be guessed when the topological order $\mathcal C$ is not too big. Moreover, Eq.~\eqref{eq:Dw} can often be used to narrow down possible solutions. Among the solutions, we only keep those in which all $a$'s with $w_a\neq 0$ have the same topological spin. Next, we insert these $\{w_a\}$ into Eq.~\eqref{eq:pa=Sw} and obtain a set $\{\mu(a)\}$. Surprisingly,  we find that $\{\mu(a)\}$ obtained this way automatically satisfy that (i) $\mu(a)=\pm 1$ if $a=\bar\rho_m(a)$ and $\mu(a)=0$ otherwise,  (ii) the Abelian case of Eq.~\eqref{eq:bbarrhob}, and (iii) the Abelian case of Eq.~\eqref{eq:pifusion} (see Appendix \ref{app:concon} for proofs). Accordingly, we only need to test the non-Abelian case of Eqs.~\eqref{eq:pifusion} and \eqref{eq:bbarrhob}. Since it tests less constraints, most examples in Sec.~\ref{sec:ex} are studied under this algorithm.

Several comments are in order.  First, the two algorithms are essentially the same. One just uses the same constraints in different orders. In fact, one does not have to stick with the orders discussed above.  Applying the constraints in a more flexible order can often help to find the final solutions quicker. Second, not all the constraints are independent. In particular, as we see from Appendix \ref{app:concon}, the constraints that we have on $w_a$ determine many of those constraints on $\mu(a)$ through the relation Eq.~\eqref{eq:pa=Sw} (see Appendix \ref{app:concon}).  Third, when the anyon $a$ is Abelian, computing $\mu(a)$ through Eq.~\eqref{eq:pa=Sw} can be simplified. We prove in Appendix~\ref{app:simplification} that
\begin{equation}
  \label{eq:pa=Mab}
  \mu(a) = M_{a,b}^\ast,\quad\forall b\in\textswab w.
\end{equation}
where $a$ is an Abelian anyon and $M_{a,b}$ is the mutual statistical phase between $a$ and $b$. Finally, it is possible that the constraints that we have are incomplete. That is, the data $\{\mu(a)\}$ or $\{w_a\}$ found through our algorithms might  not desdribe physical mirror SETs. For Abelian topological orders, the constraints on $\{\mu(a)\}$ are known to be complete. However, we conjecture that the constraints on $\{w_a\}$ and $\{\mu(a)\}$ are complete for general non-Abelian topological orders. (In fact, a stronger conjecture is that the constraints on $\{w_a\}$ alone are complete.) Evidence of the conjecture follows from the examples discussed in Sec.~\ref{sec:ex}. In particular, the $H^3$-type obstruction\cite{BarkeshliTRSET2016X}  can be ruled out by the constraints, as illustrated in the example in Sec.~\ref{sec:ex:d16}.


\subsection{Determining mirror anomaly}
\label{sec:anomaly}

After finding mirror SETs for a given topological order, we need to determine which ones are anomalous. It can be easily done in our formulation.  As discussed previously, the topological spin of the deconfined defect $x_{\mathds1}^+$ in $\mathcal{U}$ determines the anomaly: if $\theta_{x_{\mathds 1}^+}=1$, $\mathcal{U}$ is the toric code topological order and it corresponds to an anomaly-free mirror SET; if $\theta_{x_{\mathds 1}^+}=i$, $\mathcal{U}$ is the double semion topological order and it corresponds to an anomalous mirror SET.  The principles of anyon condensation asserts that $\theta_{x_{\mathds 1}^+} = \theta_{X_{a}^+} = \sqrt{\theta_a}$, for any $a\in \textswab w$. Accordingly, $\theta_a=\pm 1$. We define
\begin{equation}
  \label{eq:eta}
  \eta = \theta_a,\quad\forall a\in\textswab w
\end{equation}
which we will call \emph{anomaly indicator}. Then, $\eta=+1$ indicates that the  mirror SET is anomaly-free, and $\eta=-1$ indicates that the  mirror SET is anomalous.

In all the above discussion, we have assumed that the bulk SET order of the double-layer system is associated with the topological spins given in Eq.~\eqref{eq:thetaXa}. Instead, if we choose the other bulk SET order where all the topological spins of the defects $X_a^\pm$ differ from Eq.~\eqref{eq:thetaXa} by a factor $i$,  the anomaly prediction will be opposite and accordingly incorrect.


\section{Two general results}
\label{sec:previous}

In this section, we prove two general results on mirror SETs in our formulation, one on the mirror anomaly and the other on solutions to the constraints summarized in Sec.~\eqref{sec:constraints}. These results were previously discussed in Refs.~\onlinecite{WangLevinIndicator,BarkeshliTRSET2016X, barkeshli14} in different setups or languages.

\subsection{Anomaly indicator}
\label{sec:prev:wl}

The quantity $\eta$ in Eq.~\eqref{eq:eta}, which we name \emph{anomaly indicator}, is defined through anyons in $\textswab w$.  There exists an explicit expression of $\eta$ in terms of the symmetry fractionalization $\{\mu(a)\}$:
\begin{equation}
  \label{eq:eta2}
  \eta = \frac1{D_{\mathcal C}}\sum_{a\in\mathcal{C}}\mu(a)d_a\theta_a.
\end{equation}
where we have set $\mu(a)=0$ if $a\neq \bar\rho_m(a)$. The expression was first proposed by \citet{WangLevinIndicator} in the context of time-reversal SETs and later proved in Ref.~\onlinecite{BarkeshliTRSET2016X} by putting the SETs on nonorientable manifolds.  In fermionic systems, a similar anomaly indicator was discussed in Refs.~\onlinecite{WangLevinIndicator,tackikawa16b}. More discussions about the connection between time-reversal SETs and mirror SETs will be given in Sec.~\ref{sec:discussion:tr}.

To show Eq.~\eqref{eq:eta2}, we first notice that the topological order $\mathcal{C}$ is assumed to have a vanishing chiral central charge. Accordingly, we have
\begin{equation}
\label{eq:zeroc}
1= \frac1{D_{\mathcal C}}\sum_{a} d_a^2\theta_a.
\end{equation}
Multiplying Eqs.~\eqref{eq:eta2} with the complex conjugate of \eqref{eq:zeroc}, we rewrite $\eta$ as follows
\begin{align}
\eta & =\frac1{D_{\mathcal C}^2}\sum_{b,c}\mu(b)d_b\theta_b  d_c^2\theta_c^*\nonumber\\
& =\frac1{D_{\mathcal C}^2}\sum_{b,c} \mu(b) \theta_b d_c\theta_c^* \sum_a N^{bc}_a d_a\nonumber \\
& = \frac1{D_{\mathcal C}} \sum_{a,b} \theta_a^*d_a\mu(b) \left(\frac1{D_{\mathcal C}} \sum_c  \frac{\theta_b\theta_a}{\theta_c}d_cN^{a\bar b}_c  \right)\nonumber\\
& = \frac1{D_{\mathcal C}}\sum_{a} w_ad_a\theta_a
\end{align}
where in the second line we have used $d_bd_c=\sum_{a}N^{bc}_ad_a$, the term in the  parenthesis in the third line equals $S_{a,b}^*$, and in the forth line we have used Eq.~\eqref{eq:wa=Sp} and the fact that $\theta_a$ is real for $w_a\neq0$. With this rewriting, the deviration of \eqref{eq:eta2} from Eq.~\eqref{eq:eta} is straightforward with the help of Eq.~\eqref{eq:Dw}. (This connection was hinted in Ref.~\onlinecite{WangLevinIndicator}. However, no clear physical interpretation of $w_a$ was given there.)




\subsection{Symmetry fractionalization}
\label{sec:prev:sf}

It was shown in Ref.~\cite{BarkeshliTRSET2016X} that for an unobstructed symmetry permutation $\rho_m$, symmetry fractionalization is classified by a torsor group $H^2_{\bar\rho}[\mathbb Z_2^P,\mathcal A]$, where $\mathcal A$ is the Abelian group formed by the Abelian anyons in $\mathcal C$ under fusion. More precisely, in the language of Ref.~\onlinecite{BarkeshliTRSET2016X}, a symmetry fractionalization pattern is described by a set of phase factors, $\pi(a)$, defined for each $a\in\mathcal{C}$. When $a=\bar\rho_m(a)$, $\pi(a)$ is believed to be the same as $\mu(a)$ defined in this manuscript. Given a symmetry fractionalization pattern $\pi^{(1)}(a)$, a representative cocycle $\textswab u\in H^2_{\bar\rho}[\mathbb Z_2^P, \mathcal A]$ generates another pattern as follows:
\begin{equation}
  \label{eq:pi+u}
  \pi^{(2)}(a) = \pi^{(1)}(a) M_{a,\textswab u(P, P)}^\ast.
\end{equation}
In other words, the cocycle $\textswab u$ encodes the difference between two possible symmetry-fractionalization patterns.


Here, we show that cocycles in the torsor group $\textswab u\in H^2_{\bar\rho}[\mathbb Z_2^P, \mathcal A]$ can also be used to generate new symmetry fractionalization patterns in our formulation. 
Below we  use the data $\textswab w$ to show this. To do that, we show two claims: First, given a solution $\textswab w^{(1)}$ satisfying the constraints Eqs.~\eqref{eq:wxw} and \eqref{eq:spinconstraint}, the following quantity $\textswab w^{(2)}$ also satisfies these constraints:
\begin{equation}
  \label{eq:w2=w1xu}
  \textswab w^{(2)}=\textswab w^{(1)}\times\textswab u(P, P),
\end{equation}
where $\textswab u$ is a 2-cocycle in $H^2_{\bar\rho}(\mathbb{Z}_2^P,\mathcal A)$. Second, the two solutions $\textswab w^{(1)}$ and $\textswab w^{(2)}$ are identical, if $\textswab u$ is a 2-coboundary.

Let us review a bit on the definitions of cocycles and coboundaries. We notice that, a 2-cocycle $\textswab u$ in $H^2_{\bar\rho}(\mathbb Z_2^P, \mathcal A)$ is a function $\mathbb Z_2^P\times\mathbb Z_2^P \rightarrow \mathcal{A}$ that satisfies the so-called \emph{cocycle condition}
\begin{equation}
\bar\rho_{g}(\textswab w(h,k))\times \textswab w(g,hk) = \textswab w(gh,k)\times \textswab w(g,h)
\end{equation}
where $g,h,k=1, P$ are group elements of $\mathbb{Z}_2^P$, and $\bar\rho_g$ is a permutation associated with $g$. Two cocycles are equivalent if they are related by a coboundary
\begin{equation}
\textswab w(g,h) \sim \textswab w(g,h)\times \textswab{v}(g)\times\bar\rho_g(\textswab v(h)) \times \overline{\textswab v (gh)}
\end{equation}
where the function $\textswab{v}(g)\times\bar\rho_g(\textswab v(h)) \times \overline{\textswab v (gh)}$ is called a 2-coboundary. It is always possible to choose a representative cocycle such that only $\textswab u(P, P)$ is nontrivial, while $\textswab u(1, 1)=\textswab u(1, P)=\textswab u(P, 1)=\mathds 1$. For simplicity, we denote $u=\textswab u(P, P)$, which is an Abelian anyon. In this convention, the only nontrivial  cocycle condition is given by
\begin{equation}
  \label{eq:cocycle}
  u=\bar\rho_m(u),
\end{equation}
At the same time, the coboundary equivalence becomes
\begin{equation}
  \label{eq:cobdry}
  u\sim u\times v\times \bar\rho_m(v),\quad\forall v\in\mathcal A.
\end{equation}

First, we show that $\textswab w^{(2)}=\textswab w^{(1)}\times u$ also satisfies Eq.~\eqref{eq:wxw} if $\textswab w^{(1)}$ does. We compute the square of $\textswab w^{(2)}$ as follows
\[\textswab w^{(2)}\times \textswab w^{(2)}=u\times u\times\sum_{a}a\times \bar\rho_m(a)=\sum_{a}u\times a\times \bar\rho_m(u\times a),\]
where we have used the cocycle condition in Eq.~\eqref{eq:cocycle} in the last step. Since $u$ is Abelian, $u\times a$ is a unique anyon in $\mathcal{C}$. Moreover, when $a$ goes through all anyons in $\mathcal{C}$, so does $u\times a$. Thus, the above equation can be rewritten as Eq.~\eqref{eq:wxw}.

Next, we show that $\textswab w^{(2)}$ also satisfies Eq.~\eqref{eq:spinconstraint}, i.e., all anyons in $\textswab w^{(2)}$ have the same topological spin, if $\textswab w^{(1)}$ does. To see that, take any anyon $a\in\textswab w^{(1)}$. The topological spin of  $a\times u$ is given by
\begin{equation}
  \label{eq:tau=}
  \theta_{a\times u}=\theta_a\theta_uM_{a,u}^\ast.
\end{equation}
Equation~\eqref{eq:eta} asserts that $\theta_a=\eta^{(1)}=\pm1$. Also, we notice that the cocycle equation in Eq.~\eqref{eq:cocycle} ensures that $u$ has a well-defined mirror symmetry fractionalization $\mu^{(1)}(u)=\pm1$. Hence, the relation in Eq.~\eqref{eq:pa=Mab} reveals that the braiding phase $M_{a,u}^\ast=\mu^{(1)}(u)$, independent of $a$. Therefore, Eq.~\eqref{eq:tau=} becomes
\begin{equation}
  \label{eq:tau=tatupu}
  \theta_{a\times u}=\eta^{(1)}\theta_u\mu^{(1)}(u).
\end{equation}
which is independent of $a$. This implies that all anyons in $\textswab w^{(2)}=\textswab w^{(1)}\times u$ have the same topological spin. In fact, Eq.~\eqref{eq:tau=tatupu} also gives the anomaly indicator of the mirror SET state represented by $\textswab w^{(2)}$,
\begin{equation}
  \label{eq:eta1xu=2}
  \eta^{(2)}=\eta^{(1)}\theta_u\mu^{(1)}(u).
\end{equation}
This result provides a quick way to compare the anomaly indicators of two mirror-symmetry-fractionalization patterns.

Lastly, we show that $\textswab w^{(1)}=\textswab w^{(1)}\times u$, if $u=v\times\bar\rho_m(v)$ is a coboundary. More explicitly, we need to show
\begin{equation}
\sum_{a}w_a a = \sum_a w_a a\times u =\sum_a w_{a\times \bar u} a
\end{equation}
That is, $w_a=w_{a\times\bar\mu}$ for each $a$, if $u$ is a coboundary. This is indeed true, as we show in Appendix \eqref{app:concon} that any $\textswab{w}$ satisfying the constraint Eq.~\eqref{eq:wxw} holds the property that $w_a=w_{a\times v\times\bar\rho_m(v)}$ for any Abelian anyon $v$. Hence, we prove the claim.

We make several comments here. First, a question that one can ask is whether all the solutions constructed from inequivalent cocycles of $H^2_{\bar\rho}(\mathbb{Z}_2^P, \mathcal{A})$ are distinct. That is, whether the equality $\textswab w^{(1)}=\textswab w^{(1)}\times u$ \emph{necessarily} results that $u$ is a coboundary (what we have shown above is that $u$ being a coboundary is a \emph{sufficient} condition). From  $\textswab w^{(1)}=\textswab w^{(1)}\times u$, we have $w_{a}=w_{a\times\bar u}$ for every $a$. Combining this with Eq.~\eqref{eq:wa=Sp} and using the fact that $u$ is Abelian, we are led to
\begin{equation}
\label{eq:trivialbraiding}
M_{u,b}=1, \quad \text{for every $b=\bar\rho_m(b)$}
\end{equation}
Therefore, it is equivalent to ask whether every $u$ satisfying Eq.~\eqref{eq:trivialbraiding} has to be a coboundary. It is generally believed to be true, but we do not have a proof. For Abelian topological orders, there is indeed a proof, which is given in Ref.~\onlinecite{WangLevinIndicator}. In fact, for Abelian topological orders, solutions constructed from $H^2_{\bar\rho}(\mathbb{Z}_2^P, \mathcal{A})$  are complete. Our second comment is that in principle, there may exist a solution $\textswab{w}'$ which has no connection to a given solution $\textswab{w}$ through fusion of Abelian anyons. However, we are not sure if this situation can occur or not. Finally, in certain cases, there is even no solution to the constraints for a given $\rho_m$. Such an example is given in Sec.~\ref{sec:ex:d16}.


\section{Examples}
\label{sec:ex}

In this section, we find possible mirror SETs for several specific topological orders, using the algorithms outlined in Sec.~\ref{sec:alg}.

\subsection{Toric code}
\label{sec:ex:z2}

We start with revisiting the example of toric code studied in Sec.~\ref{sec:z2}, using the general framework developed in Sec.~\ref{sec:gen}. First, we consider the trivial symmetry permutation on anyons. With this permutation, the constraint in Eq.~\eqref{eq:wxw} becomes
\begin{equation}
  \label{eq:Ptc0}
  \textswab w\times\textswab w
  = \mathds1\times\mathds1+e\times e+m\times m+\psi\times \psi = 4\mathds1.
\end{equation}
It is not hard to see that there exist and only exist four solutions to this equation: $\textswab w_1=2\mathds1$, $\textswab w_2=2e$, $\textswab w_3=2m$ and $\textswab w_4=2\psi$. They correspond to the $e1m1$, $e1mM$, $eMm1$ and $eMmM$ mirror SETs, respectively. Judging from the topological spins of anyons in $\textswab w$, we see that the first three mirror SETs are anomaly-free and the last one is anomalous.

Next, consider the symmetry permutation that exchanges $e$ and $m$: $\rho_m(e)=m$.
With this action, the constraint in Eq.~\eqref{eq:wxw} becomes
\begin{equation}
  \label{eq:Ptc1}
  \textswab w\times\textswab w
  =\mathds1\times\mathds1+e\times m+m\times e+\psi\times \psi = 2\mathds1+2\psi.
\end{equation}
In this case, there are two possible solutions $\textswab w_1=e+m$ and $\textswab w_2=1+\psi$. However, not all anyons in $\textswab w_2$ have the same topological spin, thereby violating the second constraint on $\textswab w$ in Sec.~\ref{sec:constraints}.  Accordingly, the only solution is $\textswab w_1$. It is anomaly-free, since both $e$ and $m$ are bosons. Also, in this mirror SET, we see that the $\psi$ anyon carries $\mu(\psi)=-1$, according to Eq.~\eqref{eq:pa=Sw}.

We comment that in fact, the constraints on $\{w_a\}$ can formally be solved for general Abelian topological orders. A general solution can be found in Ref.~\onlinecite{WangLevinIndicator}, which was presented in the context of time-reversal SETs.

\subsection{$S_3$ gauge theory}
\label{sec:ex:s3}

Now, we study our first example of non-Abelian topological orders. It is the topological order of the deconfined gauge theories associated with the smallest non-Abelian group $S_3$. The UMTC describing this topological order is the quantum double of $S_3$, which is often denoted as $\mathcal C=D(S_3)$.

We first review how to compute the topological data of a quantum double $\mathcal C=D(G)$\cite{Bakalov2001}. Besides this example, such a review is also helpful for understanding a later example in Sec.~\ref{sec:ex:d16} and Appendix~\ref{app:d8}. Anyons in the quantum double $D(G)$ are labeled as $([g], \phi)$, where $[g]=\{hgh^{-1}|h\in G\}$ denotes conjugacy classes of $G$ represented by $g$, and $\phi$ is an irreducible representation of the centralizer group $Z([g])=Z_g=\{h|hg=gh, h\in G\}$. These anyons are usually named \emph{dyons}. The topological spin of $([g], \phi)$ is given by
\begin{equation}
  \label{eq:theta-dyon}
  \theta_{([g],\phi)} = \frac{\tr\phi(g)}{\tr\phi(1)},
\end{equation}
and the entries of the $S$-matrix is given by
\begin{equation}
  \label{eq:S-dyon}
  S_{([g],\phi),([h],\nu)} = \frac1{|Z_g||Z_h|}
  \sum_{r\in G,rhr^{-1}\in Z_g}\tr\phi(rh^{-1}r^{-1})
  \tr\nu(r^{-1}g^{-1}r).
\end{equation}
For a given group $G$, the traces appeared in Eqs.~\eqref{eq:theta-dyon} and \eqref{eq:S-dyon} can be found in the character tables of the centralizer groups. From the $S$ matrix, the fusion coefficients $N^{ab}_c$ can be computed, using the Verlinde formula Eq.~\eqref{eq:verlinde}.

We now specialize to $G=S_3$. The symmetric group $S_3$ contains six elements, and can be presented by two generators $a$ and $r$, satisfying $a^3=r^2=1$ and $rar=a^{-1}$. The six elements are divided into three conjugacy classes: $[1]=\{1\}$, $[a]=\{a, a^2\}$ and $[r]=\{r, ra, ra^2\}$. The centralizer of $[1]$ is $S_3$ itself, which has three irreducible representations, including a trivial representation, a nontrivial 1D representation, and a 2D representation. We follow Refs.~\cite{beigi11,BarkeshliTRSET2016X} and denote these representations by $A$, $B$, and $C$, respectively. The centralizer of $[a]$ is a $\mathbb Z_3$ group, generated by $a$. It has three 1D irreducible representations, which can be labeled by integers $n=0,1,2$. The representations are $\phi(a)=e^{n\frac{2\pi i}{3}}$. The centralizer of $[r]$ is a $\mathbb Z_2$ group generated by $r$. It has two 1D irreducible representations, which can be labeled by the sign $\pm$. The representations are associated with $\phi(r)=\pm1$ respectively. Hence, there are in total eight anyons. Following Refs.~\cite{beigi11,BarkeshliTRSET2016X}, we label them by letters from $A$ through $H$, as the following:
\begin{equation}
  \label{eq:s3label}
  \begin{split}
    ([1], A)=A,\quad ([1], B)=B,\quad ([1], C)=C, \\
    ([r],+)=D,\quad ([r],-)=E, \\
    ([a],0)=F,\quad ([a],1)=G,  \quad ([a],2)=H.
  \end{split}
\end{equation}
In particular, $A=\mathds1$ is the trivial anyon. Using the character tables, one can compute the $S$ matrix, the topological spins and the fusion rules.  We summarize the quantum dimensions, topological spins, and fusion rules in Table~\ref{tab:fusion:s3}. The $S$ matrix can be found in Refs.~\cite{beigi11,BarkeshliTRSET2016X}.

\begin{table}
  \caption{Quantum dimensions and topological spins (top) and fusion rules (bottom) of the quantum double $D(S_3)$. The fusion rules between the trivial anyon $A\equiv \mathds 1$ and other anyons are not listed.}
  \label{tab:fusion:s3}
\begin{tabularx}{0.7\textwidth}{C|C|C|C|C|C|C|C|C}
\hline\hline & $A$ & $B$ & $C$ & $D$ & $E$ & $F$ & $G$ & $H$ \\
\hline $d_a$ & 1 & 1 & 2 & 3 & 3 & 2 & 2 & 2 \\
$\theta_a$ & 1 & 1 & 1 & 1 & $-1$ & 1 & $e^{i2\pi/3}$ & $e^{-i2\pi/3}$\\
\hline\hline
\end{tabularx}
  \begin{tabular}{c|c|c|c|c|c|c|c}
    \hline\hline
    $\times$ & $B$ & $C$ & $D$ & $E$ & $F$ & $G$ & $H$\\
    \hline
    $B$ & $A$ & $C$ & $E$ & $D$ & $F$ & $G$ & $H$\\
    $C$ & $C$ & $A+B+C$ & $D+E$ & $D+E$ & $G+H$ & $F+H$ & $F+G$\\
    $D$ & $E$ & $D+E$ & $A+C+F+G+H$ & $B+C+F+G+H$ & $D+E$ & $D+E$ & $D+E$\\
    $E$ & $D$ & $D+E$ & $B+C+F+G+H$ & $A+C+F+G+H$ & $D+E$ & $D+E$ & $D+E$\\
    $F$ & $F$ & $G+H$ & $D+E$ & $D+E$ & $A+B+F$ & $C+H$ & $C+G$\\
    $G$ & $G$ & $F+H$ & $D+E$ & $D+E$ & $C+H$ & $A+B+G$ & $C+F$\\
    $H$ & $H$ & $F+G$ & $D+E$ & $D+E$ & $C+G$ & $C+F$ & $A+B+H$\\
    \hline\hline
  \end{tabular}
\end{table}

With these topological data, we now consider mirror-symmetry enriched quantum double $D(S_3)$. First,  we consider a simple mirror-symmetry permutation, which can be canonically defined for $D(G)$ of any $G$:
\begin{equation}
  \label{eq:rho1}
  \rho_m^1:([a], \phi)\mapsto([a^{-1}], \phi).
\end{equation}
The permutation $\rho_m^1$, mapping a conjugacy class to its inverse, is an antilinear autoequavilence of $D(G)$. We note that $\rho^1$ defined here can be realized trivially in the Levin-Wen models~\cite{kitaev03,levin05}. Using the fact that the dual particle of a dyon in $D(G)$ is
\begin{equation}
  \label{eq:dual-DG}
  \overline{([a],\phi)}=([a^{-1}],\phi^\ast),
\end{equation}
we conclude that the composite symmetry action $\overline{\rho_m^1}$ has the following form,
\begin{equation}
  \label{eq:barrho1}
  \overline{\rho_m^1}:([a],\phi)\mapsto([a],\phi^\ast).
\end{equation}
As discussed in Sec.~\ref{sec:prev:wl}, the action $\overline{\rho^1}$ can also be interpreted as a time-reversal symmetry action, and when interpreted that way, it is also realized trivially by the Levin-Wen model.

With realizations in the Levin-Wen models, $\rho^1$ and $\overline{\rho^1}$ represents canonical definitions of mirror and time-reversal symmetry actions on $D(G)$, respectively.  We denote $\rho^1$ and $\overline{\rho^1}$ with the superscript $1$, to distinguish them from other symmetry permutations which we will discuss in the example in Sec.~\ref{sec:ex:d16}.

Now we use the second algorithm discussed in Sec.~\ref{sec:alg} to find possible mirror symmetry fractionalization. Applying the symmetry action in Eq.~\eqref{eq:barrho1} to $D(S_3)$, we see that $\overline{\rho_m^1}$ permutes $G$ and $H$, while other anyons are invariant. With the fusion rules in Table~\ref{tab:fusion:s3}, Eq.~\eqref{eq:wxw} reduces to
\begin{equation}
  \label{eq:s3:wxw}
  \textswab w\times\textswab w = 6A+2B+5C+5F+2G+2H.
\end{equation}
It might be hard to spot a solution of this equation at the first glance, but we can easily construct a solution from the fact that the trivial symmetry fractionalization of $\rho^1$ is realized by the Levin-Wen model. Because of this, the symmetry fractionalization $\mu(A)=\mu(B)=\mu(C)=\mu(D)=\mu(E)=\mu(F)=+1$ should be a solution. Plugging this to Eq.~\eqref{eq:wa=Sp}, we get $\textswab w_1=2A+C+F$. One can easily check that this indeed is a solution of Eq.~\eqref{eq:s3:wxw}. Using the result of Sec.~\ref{sec:prev:sf}, another solution can be obtained by fusing $B$ to $\textswab w_1$: $\textswab w_2=\textswab w_1\times B=2B+C+F$. Using Eq.~\eqref{eq:pa=Sw}, we see that this solution gives the symmetry fractionalization $\mu(A)=\mu(B)=\mu(C)=\mu(F)=+1$ and $\mu(D)=\mu(E)=-1$. Indeed, these two solutions exhaust all possible symmetry fractionalization, because the fusion rules $C\times C=A+B+C$ and $D\times D=A+C+F+G+H$ fixes $\mu(A)=\mu(B)=\mu(C)=\mu(F)=+1$, while $B\times D=E$ dictates that $\mu(D)=\mu(E)$. Both of the two symmetry-fractionalization patterns are not anomalous, because $A$, $B$, $C$, and $F$ are all bosons.

Next, we consider a more complicated mirror permutation $\rho_m^\prime$, obtained by composing $\rho_m^1$ and a linear autoequivalence of $D(G)$ that exchanges $C$ and $F$, while leaves other anyons invariant. Accordingly, the dual mirror permutation $\bar{\rho}_m^\prime$ exchanges $C$ with $F$, and exchanges $G$ and $H$, and keeps anyons $A$, $B$, $D$, and $E$ invariant. With this mirror action, Eq.~\eqref{eq:wxw} becomes
\begin{equation}
  \label{eq:s3:wxw:2}
  \textswab w\times\textswab w=4(A+C+F+G+H).
\end{equation}
From the fusion rules in Table~\ref{tab:fusion:s3}, we immediately find two solutions to Eq.~\eqref{eq:s3:wxw:2}: $\textswab w_1=2D$ and $\textswab w_2=2E$. Through Eq.~\eqref{eq:pa=Sw}, the first solution $\textswab w_1=2D$ generates the symmetry fractionalization $\mu(B)=\mu(E)=-1$ and $\mu(D)=+1$. This is not anomalous, because $D$ is a boson. On the other hand, the second solution $\textswab w_2=2E$ generates $\mu(B)=\mu(D)=-1$ and $\mu(E)=+1$, and this state is anomalous, because $E$ is a fermion. We have checked that there is no other valid solutions of symmetry fractionalization associated with $\rho_m^\prime$.

\subsection{Gauged T-Pfaffian}
\label{sec:ex:tp}

In this subsection, we consider the gauged T-Pfaffian topological order, obtained by gauging the fermion parity of the T-Pfaffian state, which is a fermionic topological order. The T-Pfaffian topological order was first proposed in Refs.~\onlinecite{chen14a, bonderson13} as a realization of gapped and symmetric surface state of 3D topological insulators. For our formulation to be applicable, we gauge the fermion parity symmetry, so that the resulting gauged T-Pfaffian topological order is described by a UMTC. Moreover, it is originally an SET state with time-reversal symmetry. Here, we adapt it into a mirror SET state, as we have seen that the time-reversal and mirror symmetries behave quite similarly.

\begin{table}
\caption{List of anyons in the gauged T-Pfaffian topological order, and their quantum dimensions and topological spins. The mirror permutation $\bar\rho_m$ on anyons are also listed.}\label{tab:tp}
\begin{tabular}{c|c|c|c|c|c|c|c|c|c|c|c|c|c|c|c|c|c|c}
\hline\hline & $I_0$ & $I_2$ & $I_4$ & $I_6$ & $\psi_0$ & $\psi_2$ & $\psi_4$ & $\psi_6$ & $\sigma_1$ & $\sigma_3$ & $\sigma_5$ & $\sigma_7$ & $s_1$ & $s_3$ & $s_5$ & $s_7$ & $s\sigma_0$ & $s\sigma_2$ \\
\hline $d_a$ & 1 & 1 & 1 & 1 & 1 & 1 & 1 & 1 & $\sqrt{2}$ &   $\sqrt{2}$ &  $\sqrt{2}$&  $\sqrt{2}$ &$\sqrt{2}$ & $\sqrt{2}$ & $\sqrt{2}$ & $\sqrt{2}$ &2 & 2 \\
$\theta_a$ & $1$ & $-i$ & $1$ & $-i$ & $-1$ & $i$ & $-1$ & $i$ & $1$ & $-1$ & $-1$ & $1$ & $1$ & $-1$ & $-1$ & $1$ & $e^{i\frac{\pi}{4}}$ & $e^{-i\frac{\pi}{4}}$\\
$\bar\rho_m(a)$ & $I_0$ & $\psi_2$ & $I_4$ & $\psi_6$ & $\psi_0$ & $I_2$ & $\psi_4$ & $I_6$ & $\sigma_1$ & $\sigma_3$ & $\sigma_5$ & $\sigma_7$ & $s_7$ & $s_5$ & $s_3$ & $s_1$ & $s\sigma_2$ & $s\sigma_0$ \\
\hline\hline
\end{tabular}
\end{table}

The topological data of the gauged T-Pfaffian topological order can be found in Ref.~\onlinecite{chen14a}. It contains 18 anyons in total. The quantum dimension and topological spin of each anyon is listed in Table \ref{tab:tp}  (we follow the notation of Ref.~\onlinecite{BarkeshliTRSET2016X}). We assume that the mirror symmetry permutes the anyons in a way such that $\bar\rho_m$ is the same as the time-reversal permutation discussed in Refs.~\onlinecite{chen14a,BarkeshliTRSET2016X}. The permutation action $\bar\rho_m$ is also listed in Table \ref{tab:tp}. To compute the fusion product $a\times\bar\rho_m(a)$ for Eq.~\eqref{eq:wxw},  we list the relevant fusion rules as follows:
\begin{align}
I_k\times I_l  = I_{k+l}, \quad I_k\times\psi_l =\psi_{k+l}, \quad \psi_k\times\psi_l = I_{k+l}\nonumber\\
\sigma_k\times\sigma_l = I_{k+l} + \psi_{k+l}, \quad s_k \times s_l = I_{k+l}+\psi_{k+l+4} \nonumber \\
 s\sigma_0\times s\sigma_2 = I_2+I_6 + \psi_2+\psi_6
\end{align}
where the sums over the subscripts are defined modulo $8$. With these information, Eq.~\eqref{eq:wxw} becomes
\begin{equation}
  \label{eq:PTP}
  \textswab w\times\textswab w
  =\sum_aa\times\bar\rho_m(a)=8I_0+8\psi_4+4I_2+4\psi_6+4I_6+4\psi_2.
\end{equation}
From Eq.~\eqref{eq:Dw}, we know that  $\sum_aw_a d_a = 4\sqrt{2}$.  With this irrational number, we understand that $\textswab{w}$ should only contain non-Abelian anyon with quantum dimension $\sqrt2$. In fact, it is easy to check that there are only two solutions,
\begin{equation}
  \label{eq:wTP}
  \textswab w_1=2s_1+2s_7,\quad \textswab w_2=2s_3+2s_5.
\end{equation}
We see that $\textswab w_1$ describes an anomaly-free mirror SET, since $s_1$ and $s_7$ both have topological spins $\theta_{s_1}=\theta_{s_7}=+1$. On the other hand, $\textswab w_2$ corresponds to an anomalous SET since $s_3$ and $s_5$ both have topological spins $\theta_{s_3}=\theta_{s_5}=-1$. Using Eq.~\eqref{eq:pa=Sw}, we can compute the mirror symmetry fractionalization $\mu(a)$. It turns out that $\textswab w_{1}$ corresponds to the so-called gauged (T-Pfaffian)$_+$ state, which is known to be anomaly-free, and $\textswab w_{2}$ corresponds to the so-called gauged (T-Pfaffian)$_-$ state, which is known to be anomalous~\cite{metlitski15b}.

\subsection{$\mathbb D_{16}$-gauge theory: an example with an $H^3$ obstruction}
\label{sec:ex:d16}

Finally, we present an example of a symmetry permutation $\rho_m$ with an $H^3$ obstruction~\cite{barkeshli14,fidkowski15,barkeshli17}. In our language, it means that for a seemingly valid $\rho_m$, when we apply the algorithms in Sec.~\ref{sec:alg}, we find no valid mirror symmetry fractionalization at all. Accordingly, this $\rho_m$ is actually invalid, i.e., obstructed. This makes us conjecture that our algorithms can correctly detect the $H^3$ obstruction in general cases.

This example is constructed based on a gauge theory with gauge group $\mathbb D_{16}$. In other words, the intrinsic topological order is described by the quantum double $\mathcal C=D(\mathbb D_{16})$. As described in Sec.~\ref{sec:ex:s3}, anyons in $D(\mathbb D_{16})$ are dyons in the form $([g], \phi)$. Below we describe the idea behind this example.

The group $\mathbb D_{16}$ contains 16 group elements, generated by $a$ and $r$ satisfying $a^8=1$, $r^2=1$ and $rar=a^{-1}$.
The 16 group elements belong to seven conjugacy classes: $[1]$, $[r]$, $[a]$, $[a^2]$, $[a^4]$, $[ra]$ and $[a^3]$. Pairing them with the irreducible representations of the corresponding centralizer group, we get 46 anyons in total. The modular data of $\mathbb D_{16}$, including the $S$ and $T$ matrices, the topological spins and the fusion coefficients, can be computed from Eqs.~\eqref{eq:S-dyon} and \eqref{eq:theta-dyon}, using the character tables of the centralizer groups.

Next, we consider mirror actions on the topological order $D(\mathbb D_{16})$.  We consider a mirror permutation $\rho_m^f$, induced by a group automorphism $f: G\rightarrow G$.  For a topological order $\mathcal C=D(G)$ and given an automorphism $f$, we can define  a linear autoequivalence $\xi^f$ on $D(G)$ through the following permutation on the dyons,
\begin{equation}
  \label{eq:xif}
  \xi^f:([g],\phi)\mapsto([f(g)], \phi\circ f^{-1}),
\end{equation}
where $\phi\circ f^{-1}$ denotes a representation of the centralizer $Z_{f(g)}$: $\phi\circ f^{-1}(b)=\phi(f^{-1}(b))$. We notice that, if $f\in \inn G$ is an inner automorphism, $\xi^f$ actually does not permute the anyons, because an inner automorphism $f$ maps $g$ to another group element within the same conjugacy class of $g$. Furthermore, if two automorphisms $f$ and $f^\prime$ differ only by an inner automorphism, $\xi^f$ and $\xi^{f^\prime}$ permute anyons in the same way. Hence, $\xi^f$ is determined by an element $f$ of the outer automorphism group, which is the quotient $\out G=\aut G/\inn G$.

Furthermore, composing the linear autoequivalence $\xi^f$ and the antilinear autoequivalence $\rho_m^1$ in Eq.~\eqref{eq:rho1} gives the demanded antilinear autoequivalence $\rho_m^f\equiv \rho_m^1\circ\xi^f$, given by
\begin{equation}
  \label{eq:rhof}
  \overline{\rho^f_m}:([g],\phi)\mapsto([f(g)], (\phi\circ f^{-1})^\ast).
\end{equation}

In this example, we choose the following automorphism of $\mathbb D_{16}$:
\begin{equation}
  \label{eq:d16:f}
  f(a) = a^5,\quad f(r)=ra.
\end{equation}
which is taken from Ref.~\onlinecite{fidkowski15}. This is an order-2 element of $\out G$, since $f$ acting twice on $r$ gives $ra^6$, which is in the same conjugacy class of $r$ as $ra^6=a^{-3}ra^3$. Hence, $\xi^f$ and $\rho_m^f$ induced by $f$ are order-2 linear and antilinear autoequivalences, respectively. It is known that the autoequivalence $\xi^f$ induced by this outer automorphism $f$ carries a nonvanishing $H^3$ obstruction associated with an unitary $\mathbb{Z}_2$ symmetry. On the other hand, it is clear that the symmetry permutation $\rho^1_m$ is obstruction-free, because it can be implemented in the Levin-Wen models. Therefore, the mirror permutation $\rho^f_m$, defined as the composition of $\rho^1_m$ and $\xi^f$, is expected to exhibit an $H^3$ obstruction.

Indeed, using $\rho^f_m$ as the input, the algorithm described in Sec.~\ref{sec:alg} generates no consistent symmetry fractionalization. Due to the complexity of $D(\mathbb D_{16})$, it is easier to use the first algorithm in Sec.~\ref{sec:alg} and Eq.~\eqref{eq:wa=Sp} to search for solutions to the constraints. Using the explicit form of $\rho^f_m$ in Eq.~\eqref{eq:rhof}, we find that there are only eight anyons satisfying $a=\bar\rho_m^f(a)$, which have a well-defined $\mu(a)$. Eq.~\eqref{eq:pifusion} further eliminates independent choices of $\mu(a)$ to just three. Hence, there are in total $2^3=8$ combinations to try. Feeding them to Eq.~\eqref{eq:wa=Sp}, only two combinations generate nonnegative integral results of $\{w_a\}$. However, none of the two candidate sets of $\{w_a\}$ satisfies the condition that all $a$'s with nonzero $w_a$ have topological spins. Therefore, we conclude that our algorithm finds no solution for the mirror permutation $\rho^f_m$, consistent with the expectation that it has an $H^3$ obstruction.

\section{Discussion and Conclusion}
\label{sec:discuss}

\subsection{Time-reversal SETs}
\label{sec:discussion:tr}

We have mentioned in several places that time-reversal SETs are very similar to mirror SETs. Here, we summarize their connection. The usual argument of the similarity is based on the assumption that the lower-energy topological quantum field theory description of SETs is compatible with Lorentz invariance. In the presence of Lorentz symmetry, the time-reversal symmetry (reflection in the time direction) and mirror symmetry (reflection in one of the spatial directions) are indeed of no difference. While no violation  has been discovered, there is either no rigorous argument for the assumption.

Similarly to mirror SETs, time-reversal SETs are also described by a time-reversal permutation $\rho_t$ and a set of time-reversal symmetry fractionalization $\{\mathcal{T}^2_a\}$. The permutation $\rho_t$ is an antiautoequivalence, satisfying $\rho_t^2=1$. The symmetry fractionalization $\mathcal{T}^2_a$ is defined only for those anyons that are invariant under permutation, i.e., $\rho_t(a)=a$, and take values $+1$ or $-1$. More precisely, if $a$ carries a Kramers singlet, $\mathcal{T}^2_a=1$; if $a$ carries a Kramers doublet, $\mathcal{T}^2_a=-1$ (see Ref.~\onlinecite{levin12b} for a precise definition). Then, descriptions of mirror SETs and time-reversal SETs admit the same mathematical structure, if we make the identifications $\rho_t\leftrightarrow\bar\rho_m$ and $\mathcal{T}^2_a\rightarrow\mu(a)$.  With these identifications, the formula for mirror anomaly indicator \eqref{eq:eta2} matches exactly the time-reversal anomaly indicator proposed in Ref.~\onlinecite{WangLevinIndicator}. In addition, all the constraints on mirror SETs discovered in this work can be translated to time-reversal SETs.


\subsection{Interpretation of $\{w_a\}$}

We define $\{w_a\}$ as the coefficients in the restriction map $r(X^+_{\mathds 1})$ which describes the anyon condensation patterns [see Eq.~\eqref{eq:lx1}]. It is required  that $w_a$ is a nonnegative integer by the principles of anyon condensation. An alternative definition of $w_a$ was given in Ref.~\onlinecite{BarkeshliTRSET2016X}: $w_a$ is defined as the ground state degeneracy of a mirror SET when it is put on a M\"{o}bius strip, whose boundary carries a topological charge $a$. In this definition, $w_a$ must also be  an integer. It is important to note that the M\"{o}bius strip cannot host all anyons on its boundary. For those anyons that cannot live on a M\"{o}bius strip, $w_a$ is defined to be 0.

At the physical level, we do not know why the two definitions should give rise to the same quantity. However, at the mathematical level, from both definitions, one can show that $\{w_a\}$ is related to the symmetry fractionalization $\{\mu(a)\}$ through Eq.~\eqref{eq:wa=Sp}. Moreover, in both cases, the mirror anomaly is detected by the topological spin $\theta_a$ with $w_a\neq 0$. Hence, we believe the two definitions indeed describe the same physical quantity. Compared to Ref.~\onlinecite{BarkeshliTRSET2016X}, we derive the important constraint Eq.~\eqref{eq:wxw} on $\{w_a\}$ through the anyon condensation picture. We do not know how to derive the same constraint in the language of Ref.~\onlinecite{BarkeshliTRSET2016X}.

\subsection{Summary and Outlook}
To sum up, we develop a folding approach to study the classification and anomaly of 2D SET states with the mirror symmetry. Folding does dramatic transformations on mirror SET systems and eventually makes it possible to tackle the problem through previously available tools. More specifically, folding does the following reductions on the problem: (i) It turns an SET with the nonlocal mirror symmetry into a double-layer system with an onsite layer-exchange $\mathbb{Z}_2$ symmetry, thereby allowing us to study the system through the method of gauging symmetry; (ii) It turns the properties of 2D mirror SETs  into properties of the 1D gapped boundary of the double-layer system, thereby allowing us to study the boundary properties through the anyon condensation theory; and (iii) Combining folding with the dimension reduction idea of \citet{Song2017}, it allows us to easily read out the mirror anomaly/non-anomaly associated with  each SET from the anyon condensation pattern of the gapped boundary.

Using the folding approach and anyon condensation theory, we define a new set of data $\{w_a\}$ to describe mirror SETs. It complements the original mirror SET data, an anyon permutation $\rho_m$ and a set of quantities $\{\mu(a)\}$ that describes mirror symmetry fractionalization. The data $\{w_a\}$ and $\{\mu(a)\}$ are equivalent and related to each other through the $S$ matrix of the topological order. With the help of $\{w_a\}$, we find very strong constraints on physical mirror SETs through our approach  (summarized in Sec.~\ref{sec:constraints}). We conjecture that these constraints are complete, which is justified by our examples in Sec.\ref{sec:ex}. If they are indeed complete, then we have a classification of 2D mirror SETs.\footnote{Here, we mean all mirror SETs that are free from the $H^3$-type obstruction.} These constraints allow us to  establish practical algorithms (see Sec.\ref{sec:alg}) to find possible mirror SET states --- at least rule out unphysical ones if the constraints are incomplete --- as well as to detect the mirror anomaly.

We expect several generalizations and applications of the folding approach to other system, including fermionic SET states with the mirror symmetry, and SET states with both mirror and other onsite unitary symmetries. Folding these systems will result a double-layer system with the $\mathbb{Z}_2$ interlayer-exchange symmetry and additional symmetries (such as fermion parity conservation) on each layer. We leave these interesting generalizations to future works.

\begin{acknowledgements}
We gratefully acknowledge Meng Cheng,  Liang Kong, and Michael Levin for very helpful discussions. Y.Q. and C.W. thank The Chinese University of Hong Kong for hospitality, where this work was initiated. Y.Q. acknowledges support from Minstry of Science and Technology of China under grant numbers 2015CB921700, and from National Science Foundation of China under grant number 11874115. Chao-Ming Jian is partly supported by the Gordon and Betty Moore Foundation’s EPiQS Initiative through Grant GBMF4304. C.W. is partially supported by the Research Grant Council of Hong Kong, ECS HKU21301018.  This research was supported in part by Perimeter Institute for Theoretical Physics. Research at Perimeter Institute is supported by the Government of Canada through the Department of Innovation, Science and Economic Development Canada and by the Province of Ontario through the Ministry of Research, Innovation and Science.

\end{acknowledgements}

\appendix

\section{Review of UMTC describing a topological order}
\label{app:umtc}

In this appendix, we briefly review some of the basic properties of a UMTC $\mathcal C$, a mathematical object that describes a 2D intrinsic topological order. For a more thorough introducion on this topic, we refer the readers to Ref.~\onlinecite{kitaev06} and the review article Ref.~\onlinecite{WenNSR2016}.

One defining feature of a 2D intrinsic topological order is the existence of fractionalized quasiparticle excitations, known as anyons. We denote the total number of anyons in $\mathcal C$ by $|\mathcal C|$, which is assumed to be finite. We label the anyons using letters $a$, $b$, $\ldots$, and we abuse the notation $a\in\mathcal C$ to denote that $a$ is an anyon in $\mathcal C$. Among all anyons in $\mathcal C$, the anyon $\mathds1$ is special, which represents the trivial (unfractionalized) local quasiparticles.

Two anyons can be fused together and form a linear superposition of other anyons, as described by the fusion rules,
\begin{equation}
  \label{eq:app:fusion}
  a\times b = \sum_{c\in\mathcal C}N^{ab}_cc,
\end{equation}
where $N^{ab}_c$ are nonnegative integers, and are known as the fusion coefficients. In a UMTC, the fusion is commutative: $a\times b = b\times a$, or equivalently $N^{ab}_c=N^{ba}_c$. The trivial anyon $\mathds1$ serves as the identity element of the fusion operation: $a\times\mathds 1=a$ for any anyon $a$. Furthermore, for any anyon $a\in\mathcal C$, there exist another anyon, known as the dual or the antiparticle of $a$ and denoted by $\bar a$, such that $N^{a\bar a}_\mathds1=1$, or $a\times\bar a=\mathds1+\cdots$. We notice that the trivial particle $\mathds1$ can only appear once in any fusion outcomes. We always have $\bar{\bar a}=a$.

In general, a system containing several anyon excitations can have a nontrivial topologically protected degeneracy. The information of the degeneracy is encoded in the quantum dimension $d_a$ for each anyon $a$. Roughly speaking,  $d_a$ describes the degeneracy contributed by the anyon $a$. More precisely, $d_a$ is the largest eigenvalue of the matrix $\hat N^a$: $(\hat N^a)_{bc} = N^{ab}_c$. We notice that $d_a$ in general are not integers, and can even be irrational numbers. The quantum dimension $d_a$ satisfies the following relation,
\begin{equation}
  \label{eq:app:dd=nd}
  d_ad_b=\sum_{c\in\mathcal C}N^{ab}_cd_c.
\end{equation}
Using quantum dimensions $d_a$, one can define the total quantum dimension $D_{\mathcal C}$,
\begin{equation}
 \label{eq:app:Dc}
 D_{\mathcal C}=\sqrt{\sum_{a\in\mathcal C}d_a^2}.
\end{equation}
If $d_a=1$, $a$ is called an Abelian anyon. An Abelian anyon does not support any topologically protected degeneracy. It can be shown that the fusion outcome of an Abelian anyon $a$ and another anyon $b$ (can be non-Abelian in general) is always unique, $a\times b=c$. In particular, the trivial anyon $\mathds1$ is Abelian. When all anyons in $\mathcal C$ are Abelian, we say $\mathcal C$ describes an Abelian topological order. In that case, we have $D_{\mathcal C}=\sqrt{|\mathcal C|}$. However, in general, $D_{\mathcal C}\geq\sqrt{|\mathcal C|}$.

The fractionalized topological excitations are called anyons, because they have anyonic statistics. The self-statistics of an anyon $a$, defined as the Berry phase accumulated by rotating $a$ by $2\pi$, is called the topological spin of $a$, and denoted by $\theta_a$. Notice that, in our notation, $\theta_a$ is a unimodular complex number, not the phase angle of the self-statistics. For example, $a$ is a fermion if $\theta_a=-1$ (not $\pi$). Using $\theta_a$, one can compute the so-called chiral central charge  $c$ modulo 8:
\begin{equation}
  \label{eq:c}
  e^{i2\pi\frac c8} = \frac1{D_{\mathcal C}}\sum_{a\in\mathcal C}d_a^2\theta_a.
\end{equation}
The chiral central charge $c$ is defined through the 1D conformal field theory that lives on the boundary of the 2D topological order. More precisely,  $c$ is the deference of the central charges associated with the holomorphic and antiholomorphic parts of the conformal field theory.  It is obvious that time-reversal or mirror-reflection symmetry maps $c$ to $-c$. Therefore, a 2D mirror SET must have $c=0$. We notice that there is a class of anomalous 2D mirror or time-reversal SETs which are equipped with $c=4\mod8$. They can be realized on the surface of a 3D mirror/time-reversal SPT. The latter is an SPT beyond the group cohomology classification and can be realized by decorating the so-called $E_8$ bosonic SPT state on the mirror plane. This type of anomaly is easy to describe, and is not the subject of our paper. Therefore, in the main text of our paper, we always assume $c=0$.

Next, we introduce the modular matrices $S$ and $T$, which describe how wave functions on a torus transform under modular transformations. The sizes of the $S$ and $T$ matrices are $|\mathcal C|$ by $|\mathcal C|$, with each row and column labeled by one anyon. The $T$ matrix is diagonal and directly related to the topological spins of the anyons, through
\begin{equation}
  \label{eq:Tmat}
  T_{a,b}=\delta_{a,b}\theta_a.
\end{equation}
The entries of the $S$ matrix can also be computed from the topological spins and the fusion coefficients,
\begin{equation}
  \label{eq:Smat}
  S_{a,b} = \frac1{D_{\mathcal C}}\sum_{c\in\mathcal C}N^{ab}_c
  d_c\frac{\theta_a\theta_b}{\theta_c}.
\end{equation}
Reversely, given the $S$ matrix, the fusion coefficients can be computed using the Verlinde formula,
\begin{equation}
  \label{eq:verlinde}
  N^{ab}_c=\sum_{x\in\mathcal C}\frac{S_{a,x}S_{b,x}S_{c,x}^\ast}
  {S_{\mathds 1,x}}.
\end{equation}
The $S$ matrix is unitary, i.e., $S^\dag S=1$. In addition, it satisfies $S_{a,b}=S_{b,a}=S_{\bar a,\bar b} = S_{\bar a ,b}^*$, and $S^4=1$.

The mutual braiding statistics between two anyons $a$ and $b$ can be read out from the $S$ matrix. To be precise, the monodromy scalar component is defined as
\begin{equation}
  \label{eq:Mab}
  M_{ab}^\ast=\frac{S_{a,b} S_{\mathds1,\mathds1}}
  {S_{\mathds1,a}S_{\mathds1,b}}=\frac{D_{\mathcal C}}{d_ad_b}S_{a,b}.
\end{equation}
When $a$ is an Abelian anyon, this equation can be simplified as
\begin{equation}
  \label{eq:Mab-da=1}
  M_{ab}^\ast=\frac{\theta_a\theta_b}{\theta_{a\times b}}.
\end{equation}
(Note that $a\times b$ is a unique anyon.) This is a unimodular complex number, describing the Berry phase obtained by braiding $a$ and $b$. Note that the braiding phase is Abelian as long as one of the anyons involved in the braiding process is Abelian.

\section{Double-layer toric-code theory}
\label{app:d8}

In this appendix, we discuss the structure of the topological order $\mathcal D$ obtained by gauging the layer-exchange $\mathbb Z_2$ symmetry of a double-layer toric code theory. This result is a special case of the general discussion in Appendix \ref{app:dl}. Here, we separately discuss it as a reference for Sec.~\ref{sec:z2}.

As shown in Ref.~\onlinecite{barkeshli14}, the topological order $\mathcal D$ of the gauged double-layer toric code theory is the same as the quantum double $D(\mathbb D_8)$, where $\mathbb D_8$ is the order-4 dihedral group. A brief review of the general structure of a quantum double $D(G)$ can be found in Sec.~\ref{sec:ex:s3}. Anyons in $D(G)$ are labeled by $([g],\phi)$, where $[g]$ is the conjugacy class of $g$ and $\phi$ is a representation of the centralizer $Z_g=\{h|gh=hg\}$. The group $\mathbb D_8$ is generated by two generators $a$ and $r$ satisfying $a^4=1$, $r^2=1$ and $rar=a^{-1}$. Its eight elements belong to five conjugacy classes: $[1]=\{1\}$, $[a]=\{a, a^3\}$, $[a^2]=\{a^2\}$, $[r]=\{r, ra^2\}$, and $[ra]=\{ra, ra^3\}$. The centralizer groups of $[1]$ and $[a^2]$ are both $\mathbb D_8$ itself, which has five irreducible representations, denoted by $A$ through $E$, respectively. $A$ is the trivial representation. $B$, $C$ and $D$ are 1D representations that act on the generators as the following: $B(a)=+1$ and $B(r)=-1$, $C(a)=-1$ and $C(r)=+1$, and $D(a)=D(r)=-1$, respectively. $E$ is a 2D representation with the character $\tr E(1)=2$, $\tr E(a^2)=-2$ and $\tr E(a)=\tr E(r)=\tr E(ra)=0$. The centralizer group of $[r]$ is a $\mathbb Z_2\times\mathbb Z_2$ group, generated by $r$ and $a^2$. It has four 1D irreducible representations, which we denote by $a^2_\pm r_\pm$, where the two $\pm$ signs denote the image of $a^2$ and $r$, respectively. Similarly, the centralizer group of $[ra]$ is also $\mathbb Z_2\times\mathbb Z_2$, generated by $ra$ and $ra^3$, whose irreducible representations are denoted by $ra_\pm ra^3_\pm$. The centralizer group of $[a]$ is a $\mathbb Z_4$ group generated by $a$. It has four irreducible representations, denoted by the image of $a$, which takes value of $+1$, $i$, $-1$ and $-i$. Hence, there are in total 22 anyons. We relabel them using the notation introduced in Sec.~\ref{sec:z2cl}, as the following
\begin{equation}
\begin{array}{ll}
  ([1], A) = \mathds1 = (\mathds1,\mathds1)^+, \quad\quad\quad  &   ([a^2], A) = (m,m)^+,\\
  ([1], B) = (\mathds1,\mathds1)^-, &   ([a^2], B) = (m,m)^-,\\
  ([1], C) = (e,e)^+, &   ([a^2], C) = (\psi,\psi)^+,\\
  ([1], D) = (e,e)^-, &   ([a^2], D) = (\psi,\psi)^-,\\
  ([1], E) = [e,\1], &   ([a^2], E) = [\psi,m],\\
\end{array}\nonumber
\end{equation}
\begin{equation}
\begin{array}{lll}
  ([ra], ra_+ra^3_+) = [m,\1],\quad\quad\quad &  ([r], a^2_+r_+) = X_{\mathds1}^+, \quad\quad\quad&   ([a], +1) = X_m^+,\\
  ([ra], ra_+ra^3_-) = [e,m], &  ([r], a^2_+r_-) = X_{\mathds1}^-, &   ([a], -1) = X_m^-,\\
  ([ra], ra_-ra^3_+) = [\psi,\1],&   ([r], a^2_-r_+) = X_e^+,&   ([a], +i) = X_\psi^+,\\
  ([ra], ra_-ra^3_-) = [\psi,e];&   ([r], a^2_-r_-) = X_e^-;&   ([a], -i) = X_\psi^-.\\
\end{array}\nonumber
\end{equation}
With this identification, the properties of $\mathbb D_8$ claimed in Sec.~\ref{sec:z2cl} can be derived, using Eqs.~\eqref{eq:theta-dyon} and \eqref{eq:S-dyon}.

\section{Gauging the layer exchanging $\Z2$ symmetry of a general double-layer topological order}
\label{app:dl}

The UMTC that describes a general (decoupled) doule-layer system take the form of $\B = \C \otimes \C$, where $\C$ denotes the UMTC that describes a single-layer system. By gauging the $\Z2$ layer exchange symmetry of the UMTC $\B$, we can obtain the UMTC $\D$ that describes the gauged double-layer system. The main goal of this section is to obtain the fusion rules, the $S$ matrix and the $T$ matrix of the UMTC $\D$ from those of the single layer theory $\C$.

Let's clarify the notation first. For the convenience of this section, we will use a slightly different notation from the main text. For the single-layer theory $\C$, we label its anyon types as $a,b,c...$ and, in particular, the trivial anyon as $\1$.
We denote the $S$ matrix of the single-layer theory $\C$ as $S^\C$. The topological spin and the quantum dimension of the anyon $a$ are denoted as $\theta_a$ and $d_a$ respectively. The fusion multiplicity for anyons $a$ and $b$ fusing into the anyon $c$ is denoted as $^\C \! N^{ab}_c$. The total quantum dimension and the chiral central charge of the UMTC $\C$ are denoted as $D_\C$ and $c_\C$.
Given the data of the single-layer theory $\C$, it is easy to write down the topological data of the double-layer theory $\B =\C \otimes \C$. The anyon types of $\B$ are given by the pair $(a,b)$, where $a$, $b$ label the anyon types in the single-layer theory $\C$. The $S$ matrix element of $\B$ is given by $S^{\B} = S^{\C}\otimes S^{\C}$. In terms of $S$ matrix elements, we have $S^{\B}_{(a,b),(c,d)} = S^{\C}_{ac} S^{\C}_{bd}$. The topological spin and the quantum dimension of the anyon $(a,b)$ are given by $\theta_{(a,b)} = \theta_a \theta_b$ and $d_{(a,b)} = d_a d_b$ respectively. The fusion multiplicity for anyons $(a,b)$ and $(c,d)$ fusing into the anyon $(e,f)$ is given by $^\B \! N^{(a,b)(c,d)}_{(e,f)} = ~^\C \! N^{ac}_e  ~^\C \! N^{bd}_f$. The total quantum dimension and the chiral central charge of the UMTC $\B$ are simply given by $D_\B = D_\C^2$ and $c_\B =2c_\C$. For gauged double-layer theory $\D$, we label its anyon types as $\alpha, \beta,...$. The $S$ matrix is denoted as $S^\D$. The fusion multiplicity for anyons $\alpha$ and $\beta$ fusing into the anyon $\gamma$ is denoted as $^\D \! N^{\alpha\beta}_\gamma$. The chiral central charge of $\D$ is identical to that of the double layer theory $\B$, namely $c_\D =c_\B= 2c_\C $. In the following, we will derive the topological data of the UMTC $\D$ using that of $\C$ and $\B$. In fact, in addition to the topological data of $\C$ and $\B$, we also need to choose an element of $H^3(\Z2, U(1)) = \Z2$ when we gauge the $\Z2$ symmetry to obtain the UMTC $\D$. In the following, we will mostly concentrate on the case with $c_\C = 0 $ and the choice of the trivial element in
$H^3(\Z2, U(1))$, which is directly relevant to our discussion in the main text. Nevertheless, the approach we will use can be directly generalized to the most general situation.

\subsection{Anyon contents and fusion rules of $\D$}
In this subsection, we will obtain the anyon contents of $\D$ and further calculate their fusion rules. The approach to identify the anyon types of $\D$ follows Ref. \onlinecite{Dijkgraaf1989}.

When we gauge the $\mathbb{Z}_2$ layer exchange symmetry in $\B$, the anyon $(a,b)$ and $(b, a)$ form a ``doublet" when $a \neq b$. In the gauged double-layer system $\D$, we denote this doublet as $[a,b]$ (with the implicit assumption that $a \neq b$ and the identification that $[a,b] =[ b,a]$). The anyon type $(a,a)$ of $\B$, which is invariant under the $\mathbb{Z}_2$ layer exchange symmetry, will be lifted to two types of anyon $(a,a)^+$ and $(a,a)^-$ in $\D$ after gauging. The $\pm$ signs indicate the value of $\mathbb{Z}_2$ charges (under the layer exchange) in $(a,a)^\pm$, where $+$ means no charge and $-$ means a non-trivial $\mathbb{Z}_2$ charge. In particular,  $(\1,\1)^+$ represents the trivial anyon in $\D$, while $(\1,\1)^-$ represents the pure $\mathbb{Z}_2$ charge in $\D$. The collection of anyons $(a,a)^\pm$ and $[a,b]$ forms the untwisted sector $\D_0$ of $\D$, namely the collection of the anyons in $\D$ that does not carry $\Z2$ flux. The collection of anyons carrying $\Z2$ fluxes are denoted as the twist sector $\D_1$. The anyons in the twist sector are labeled by $X_a^{\pm}$, which represent a $\mathbb{Z}_2$ gauge flux decorated by an anyon $a$ in the single-layer theory $\C$ and a $\mathbb{Z}_2$ charge ``$\pm$". Unlike the case of $(a,a)^\pm$, the $\pm$ in $X_a^{\pm}$ does not canonically correspond to the absolute value of $\mathbb{Z}_2$ charges. Rather, it is a relative notion in the sense that $X_a^+$ and $X_a^-$ differ by a $\Z2$ charge. Before we study their fusion rules, we quickly summarize anyon contents of $\D$:
\begin{align}
(a,a)^+,~~(a,a)^-,~~[a,b],~~X^+_a~~\text{and}~~ X^-_a,
\end{align}
where the first three types belong the untwisted sector $\D_0$ and the last two belong to the twist sector $\D_1$. The total number of anyon types in $\D$ is $\frac{|\C|(|\C|+7)}{2}$, where $|\C|$ is the number of anyon types in the single layer UMTC $\C$.

We first look at the fusion rules from the $\mathbb{Z}_2$ charge perspective. The ``pure" $ \Z2$ charge is given by the anyon $(\1,\1)^-$. Therefore, we have the fusion rules:
\begin{align}
& (a,a)^\pm \times (\1,\1)^- = (a,a)^\mp,
\label{Eq:[aa] Fusion With Z2 Charge}
\\
& [a,b] \times (\1,\1)^- = [a,b],
\label{Eq:[ab] Fusion With Z2 Charge}
\\
& X_a^\pm \times (\1,\1)^- = X_a^\mp.
\label{Eq:X Fusion With Z2 Charge}
\end{align}

Similar to the $\mathbb{Z}_2$ charge, the $\mathbb{Z}_2$ flux also provide a constraint on the fusion rules. The sector labels of the anyons serve as a $\Z2$-grading which the fusion rules respect.

Now, we study the fusion rules of $\D$ from the anyon condensation perspective. As we discussed, the theory $\D$ is the result of gauging the layer exchange $\mathbb{Z}_2$ symmetry of $\B$. Conversely, the theory $\B$ can be obtained from $\D$ via condensing the anyon $(\1,\1)^-$, namely the $\mathbb{Z}_2$ charge. Throughout this section, we will only make use of the condensation of the anyon $(\1,\1)^-$ for the purpose of obtaining the data of the UMTC $\D$. One should not confuse it with other types anyon condensations studied in the main text. The condensation of anyon $(\1,\1)^-$ leads to the restriction map $r$ that satisfies
\begin{align}
&r((a,a)^+)  = (a,a),
\label{Eq:Condensate Id [aa]+}
\\
& r((a,a)^-) = (a,a) ,
\label{Eq:Condensate Id [aa]-}
\\
&r([a,b]) =  (a,b) + (b,a)
\label{Eq:Condensate Id [ab]}
\end{align}
for the deconfined anyons in the condensate. One notice that anyons in the untwisted sector $\D_0$ are all deconfined. On contrary, all the anyons in the twist sector $\D_1=\{X_a^\pm\}$ are confined due to the non-trivial braiding statistics between the $\Z2$ charge $(\1,\1)^-$ and the $\Z2$ flux carried by the anyons in $\D_1$. From the restriction map given above, we can immediately obtained the quantum dimensions of the anyons $(a,a)^\pm$ and $[a,b]$:
\begin{align}
& d_{(a,a)^+} = d_{(a,a)^-} = d_a^2,
\\
& d_{[a,b]} = 2 d_a d_b.
\end{align}
The restriction map restricted to the deconfined anyons can be encoded in by an $|\D|\times |\B|$ matrix $n_{\alpha,x}$ (with $\alpha\in \D$ and $x \in \B$) with all the non-vanishing elements given by
\begin{align}
& n_{(a,a)^\pm , (a,a) } = 1,
\\
& n_{[a,b] , (a,b) } = n_{[a,b]  ,(b,a) } = 1.
\end{align}
For the deconfined anyons, the restriction map has to be consistent with the fusion rules of $\B$ and $\D$:
\begin{align}
r(\alpha) \times r(\beta) = r (\alpha \times \beta),
\end{align}
where $\alpha,\beta \in \D$. Here the fusion product $\alpha\times \beta$ should be understood as the fusion product in $\D$, while $r(\alpha) \times r(\beta)$ should be understood as the fusion product in $\B$. In terms of the fusion multiplicity, we can write it as
\begin{align}
\sum_{x,y\in \B} n_{\alpha x} n_{\beta y} ~^\B \! N^{xy}_z = \sum_{\gamma\in \D}~^\D \! N^{\alpha\beta}_\gamma n_{\gamma z},
\end{align}
where $x,y,z \in \B$ and $\alpha,\beta, \gamma \in \D$.

The consistency relation $r(\alpha) \times r(\beta) = r (\alpha \times \beta)$ can be utilized to derive the fusion rules of the UMTC $\D$ from that of the UMTC $\B$ which is already known. We first focus on fusion rules in the untwisted sector $\D_0$. For two anyons $\alpha$ and $\beta$ in the UMTC $\D_0$, we can map them to $r(\alpha), r(\beta) \in \B$, obtain $r(\alpha \times \beta)= r(\alpha) \times r(\beta)$ using the fusion rules of $\B$ and reverse engineer $\alpha \times \beta$ from $r(\alpha \times \beta)$. In the $\D_0$ sector, we have

\begin{align}
& r((a,a)^\pm) \times r((b,b)^\pm)
=
(a,a) \times (b,b)
= \sum_{e,f \in \C} ~^\C\! N^{ab}_{e} ~^\C\! N^{ab}_{f} (e,f)
\label{Eq:Fusion Lifting [aa]1}
\\
& r((a,a)^\pm) \times r((b,b)^\mp)
=
(a,a) \times (b,b)
= \sum_{e,f \in \C} ~^\C\! N^{ab}_{e} ~^\C\! N^{ab}_{f} (e,f)
\label{Eq:Fusion Lifting [aa]2}
\\
& r([a,b]) \times r((c,c)^\pm)
=
\Big( (a,b)+(b,a) \Big) \times (c,c)
=\sum_{e,f \in \C} (~^\C\! N^{ac}_{e} ~^\C\! N^{bc}_{f} + ~^\C\! N^{bc}_{e} ~^\C\! N^{ac}_{f} ) (e,f)
\label{Eq:Fusion Lifting [ab]1}
\\
& r([a,b]) \times r([c,d])
=
\Big( (a,b) + (b,a) \Big) \times \Big((c,d) + (d,c)\Big)
\nonumber \\
& ~~~~~~~~~~~~~~~~~~~~~~~
=\sum_{e, f \in \C} \Big(
~^\C\! N^{ac}_{e} ~^\C\! N^{bd}_{f} +
~^\C\! N^{bc}_{e} ~^\C\! N^{ad}_{f} +
~^\C\! N^{ad}_{e} ~^\C\! N^{bc}_{f} +
~^\C\! N^{bd}_{e} ~^\C\! N^{ac}_{f}
\Big) (e,f).
\label{Eq:Fusion Lifting [ab]2}
\end{align}
On the right hand sides of these equations, the combination $(e,f)+(f,e)$ can be naturally identify as $r([e,f])$ when $e \neq f$. The anyon $(e,e)$ can be identified as either $r((e,e)^+)$ or $r((e,e)^-)$ depending on the fusion of $\Z2$ charges in the UMTC $\D$. To be more specific, when we consider $(a,a)^+ \times (b,b)^+ $ for example, both $(a,a)^+ $ and $(b,b)^+ $ have fixed $\Z2$ charges. The $\Z2$ charge of their fusion products (when applicable) should be the sum of their $\Z2$ charges. The same analysis applies to all the cases of  $(a,a)^\pm \times (b,b)^\pm $ and  $(a,a)^\pm \times (b,b)^\mp $. Combining Eq. \eqref{Eq:Fusion Lifting [aa]1}, Eq. \eqref{Eq:Fusion Lifting [aa]2} and the analysis on the $\Z2$ charges, we can conclude
\begin{align}
& (a,a)^\pm \times (b,b)^\pm
=
\sum_{[e,f] \in \D} ~^\C\! N^{ab}_{e} ~^\C\! N^{ab}_{f} [e,f]
+
\sum_{(e,e)^+ \in \D} ~^\C\! N^{ab}_{e} ~^\C\! N^{ab}_{e} (e,e)^+,
\label{Eq:Fusion [aa]1}
\\
& (a,a)^\pm \times (b,b)^\mp
=
\sum_{[e,f] \in \D} ~^\C\! N^{ab}_{e} ~^\C\! N^{ab}_{f} [e,f]
+
\sum_{(e,e)^- \in \D} ~^\C\! N^{ab}_{e} ~^\C\! N^{ab}_{e} (e,e)^- .
\label{Eq:Fusion [aa]2}
\end{align}
Although $[a,b]$ and $[c,d]$ do not carry fixed $\Z2$ charges, we still need to make sure that the fusion product is consistent with Eq. \ref{Eq:[ab] Fusion With Z2 Charge}. Together with Eq. \ref{Eq:Fusion Lifting [ab]1} and \ref{Eq:Fusion Lifting [ab]2}, we can conclude
\begin{align}
 [a,b] \times (c,c)^\pm
= & \sum_{[e,f] \in \D}  (
~^\C\! N^{ac}_{e} ~^\C\! N^{bc}_{f} + ~^\C\! N^{bc}_{e} ~^\C\! N^{ac}_{f})[e,f]
\nonumber \\
&
+
\sum_{[e,e]^+ \in \D}
~^\C\! N^{ac}_{e} ~^\C\! N^{bc}_{e}
(e,e)^+
+
\sum_{[e,e]^- \in \D}
~^\C\! N^{ac}_{e} ~^\C\! N^{bc}_{e}
(e,e)^-,
\\
 [a,b] \times [c,d]
= & \sum_{[e,f] \in \D} (
~^\C\! N^{ac}_{e} ~^\C\! N^{bd}_{f} + ~^\C\! N^{bc}_{e} ~^\C\! N^{ad}_{f} + ~^\C\! N^{ad}_{e} ~^\C\! N^{bc}_{f} + ~^\C\! N^{bd}_{e} ~^\C\! N^{ac}_{f} ) [e,f]
\nonumber \\
&
+
\sum_{(e,e)^+ \in \D} (
~^\C\! N^{ac}_{e} ~^\C\! N^{bd}_{e} + ~^\C\! N^{bc}_{e} ~^\C\! N^{ad}_{e}) (e,e)^+
\nonumber \\
&
+
\sum_{(e,e)^- \in \D} (
~^\C\! N^{ac}_{e} ~^\C\! N^{bd}_{e} + ~^\C\! N^{bc}_{e} ~^\C\! N^{ad}_{e}) (e,e)^-.
\end{align}

Now, we can move on to the twist sector $\D_1$. First of all, by the additive property of the $\Z2$ fluxes, we know that $X^\pm_a \times X^\pm_b$ are entirely contained in the untwisted sector $\D_0$. Upon the condensation of $(\1,\1)^-$, the twist sector $\D_1$ are totally confined. We should view the confined anyon $X^\pm_a$ as the ``$\Z2$ genons" in the UMTC $\B$\cite{Barkeshli2013}, which are the end points of layer exchange branch cuts in the double layer system. This picture allows us to map the double-layer theory $\B$ in the presence of $\Z2$ genons to the single-layer theory $\C$ on Riemann surfaces with higher genus. Using this mapping, we can obtain that
\begin{align}
& r(X^\pm_a \times X^\pm_b ) = (0,a)\times (0,b) \times \left( \sum_{c \in \C} (c,\bar{c})  \right)
=
\sum_{c, e, f\in \C}
~^\C\! N^{a\bar{c}}_e ~^\C\! N^{eb}_f ( c,f),
\\
& r(X^\pm_a \times X^\mp_b ) = (0,a)\times (0,b) \times \left( \sum_{c \in \C} (c,\bar{c})  \right)
=
\sum_{c, e, f\in \C}
~^\C\! N^{a\bar{c}}_e ~^\C\! N^{eb}_f ( c,f).
\end{align}
At this point, we would like to reverse engineer $X^\pm_a \times X^\pm_b $ from $r(X^\pm_a \times X^\pm_b )$. We again encounter the ambiguity that $(c,c)$ can be viewed either as $r((c,c)^+)$ or $r((c,c)^-)$. This ambiguity cannot be resolved at this point. This is because, unlike the case of $(a,a)^\pm \times (b,b)^\pm $, the ``$\pm$" sign of the anyon $X_a^\pm$ does not canonically correspond to the absolute value of $\Z2$ charges. Now, we can only write
\begin{align}
& X^\pm_a \times X^\pm_b
=
\sum_{[c,f]\in \D} \sum_{e\in \C}
~^\C\! N^{a\bar{c}}_e ~^\C\! N^{eb}_f [ c,f]
+
\sum_{c\in \C} \sum_{e\in \C}
~^\C\! N^{a\bar{c}}_e ~^\C\! N^{eb}_c (c,c)^{p(a,b,c,e)}
\label{Eq:Fusion XX1}
\\
& X^\pm_a \times X^\mp_b
=
\sum_{[c,f]\in \D} \sum_{e\in \C}
~^\C\! N^{a\bar{c}}_e ~^\C\! N^{eb}_f [ c,f]
+
\sum_{c\in \C} \sum_{e\in \C}
~^\C\! N^{a\bar{c}}_e ~^\C\! N^{eb}_c (c,c)^{-p(a,b,c,e)},
\label{Eq:Fusion XX2}
\end{align}
where the function $p(a,b,c,e)$ depends on $a,b,c,e\in \C$, takes values $\pm$ and is yet to be determined. Once we obtain the $S$ matrix of the UMTC $\D$, we can calculate this function via the Verlinde formula. Regardless of the specific form of the function $p$, the quantum dimension of the anyon $X^\pm_a$ is given by
\begin{align}
X^\pm_a = D_\C d_a
\end{align}
Having derived the quantum dimensions of all the anyons in $\D$, we obtain the total quantum dimension
\begin{align}
D_\D =2 D_\C ^2.
\end{align}

For the fusion product between the untwisted sector $\D_0$ and the twist sector $\D_1$, we can also use the genon picture and write down the fusion rules
\begin{align}
& X_a^\pm \times (b,b)^\pm = \sum_{d,e \in \C}
~^\C\! N^{ab}_e ~^\C\! N^{be}_d
X^{q(a,b,d,e)}_d,
\label{Eq:Fusion_Xabb1}
\\
& X_a^\pm \times (b,b)^\mp = \sum_{d,e \in \C}
~^\C\! N^{ab}_e ~^\C\! N^{be}_d
X^{-q(a,b,d,e)}_d,
\label{Eq:Fusion_Xabb2}
\\
& X_a^\pm \times [b,c] ~ = \sum_{d, e \in \C}
~^\C\! N^{ab}_e ~^\C\! N^{ce}_d
\Big( X^+_d + X^-_d \Big),
\label{Eq:Fusion_Xabc}
\end{align}
where the function $q(a,b,d,e)$ depends on $a,b,d,e \in \C$ and takes values $\pm$. This function in Eqs. \eqref{Eq:Fusion_Xabb1} and \eqref{Eq:Fusion_Xabb2} is yet to be determined by the $S$ matrix via the Verlinde formula. In contrast, Eq.~\eqref{Eq:Fusion_Xabc} is completely determined by the genon picture and by the consistency with Eq.~\eqref{Eq:[ab] Fusion With Z2 Charge}.

\begin{figure}
\includegraphics[scale=0.4]{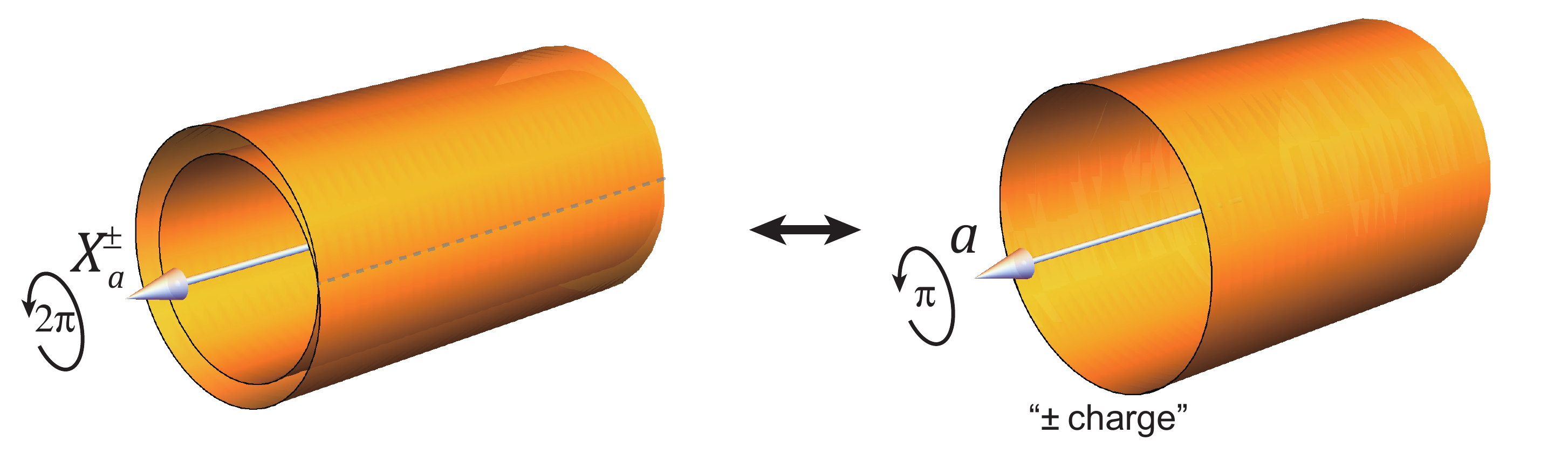}
\caption{The left panel depicts the double-layer cylinder geometry with a layer-exchanging branch cut (dashed line). The white arrow represents the anyon flux $X_a^\pm$ threading through the cylinder. The topological spin $\theta_{X_a^\pm}$ of $X_a^\pm$ is equal to the Berry phase accumulated when one end of the cylinder is rotated by $2\pi$. The right panel depicts a single-layer cylinder geometry carrying the topological order described by $\C$ that is topologically equivalent to the left panel. The anyon flux threading through the single-layer cylinder is given by $a$.}
\label{Fig:TwoCylinderTwists}
\end{figure}

\subsection{$T$ matrix $T^\D$ of $\D$}
The $T$ matrix $T^\D$ of the UMTC $\D$ captures the topological spins of the anyons in $\D$. For the anyons in the untwisted sector $\D_0$, since they are all deconfined in condensation of $(\1,\1)^-$ that yields $\B$, their topological spins should be the same as their image under the restriction map $r$ in the UMTC $\B$. Hence, we have
\begin{align}
& \theta_{(a,a)^\pm} = \theta_a^2,
\\
& \theta_{[a,b]} = \theta_a \theta_b.
\end{align}

Now we study the twist sector $\D_1$. The topological spin of $X_{a}^\pm$ can be measured using the ``momentum polarization" method \cite{Tu2013}. That is to say that we consider the topological order described by $\D$ on a cylinder geometry with the anyon flux of $X_{a}^\pm$ threading the cylinder. The topological spin $\theta_{X_a^\pm}$ of $X_a^\pm$ is directly given by the Berry phase accumulated when one end of the cylinder is rotated by $2\pi$ (together with central charge correction if $c_\D \neq 0$). The genon picture allows us to map the cylinder geometry of the theory $\D$ with the anyon flux of $X_{a}^\pm$ to a double-layer cylinder geometry with a layer-exchanging branch cut (see the left panel of Fig.~\ref{Fig:TwoCylinderTwists}). In the presence of the branch cut, this double-layer cylinder geometry is topologically equivalent to a single-layer cylinder on which a single copy of UMTC $\C$ resides. The type of anyon flux $a$ thread the single-layer cylinder is directly given by the anyon type $X_{a}^\pm$ in the theory $\D$. Another observation is that a $2\pi$ rotation in the double-layer cylinder geometry is equivalent to a $\pi$ rotation in the effective single-layer geometry. The equivalence between the double-layer geometry and single-layer geometry guarantees that
\begin{align}
\theta_{X_{a}^\pm} = \pm \theta_a ^{\frac{1}{2}}.
\end{align}
We can see from this expression that the $\pm$ $\Z2$ charge assigned to $X_{a}^\pm$ generally doesn't have an absolute meaning (because of the ambiguity of the square root $\theta_a ^{\frac{1}{2}}$). In the presence of non-trivial chiral central charge $c_\C$ or the non-trivial element $\omega \in H^3(\Z2, U(1))$, there will be an extra factor of $\omega^\frac{1}{2} e^{i2\pi(2-\frac{1}{2})\frac{c_\C}{24}}$ in the expression of $\theta_{X_{a}^\pm}$. Here, we have implicitly used $\omega=1$ to denote the trivial element of $ H^3(\Z2, U(1))$ and $\omega=-1$ for non-trivial one.

\subsection{$S$ matrix $S^\D$ of $\D$}
In this subsection, we will derive the $S$ matrix $S^\D$ of $\D$. We will separate $S^\D$ into several blocks and use different technique to obtain them.

For the anyons of $\D$ that are deconfined when the anyon $(\1,\1)^-$ condenses, we can utilize the consistency condition between the restriction map $r$ (encoded by the $\D \times \B$ matrix $n$) and the $S$ matrices $S^\D$ and $S^B$:
\begin{align}
S^\D n = n S^\B.
\end{align}
In addition, since the anyon $(\1,\1)^-$ braids trivially with the anyons $(a,a)^\pm$ and $[a,b]$, the following identities hold
\begin{align}
S^\D_{(a,a)^+,(b,b)^+} = S^\D_{(a,a)^+,(b,b)^-} & = S^\D_{(a,a)^-,(b,b)^+} = S^\D_{(a,a)^-,(b,b)^-},
\\
S^\D_{[a,b]^+,(c,c)^+} &  = S^\D_{[a,b]^+,(c,c)^-}.
\end{align}
By studying different components of the matrix identity $S^\D n = n S^\B$ together with the relations given above, we can directly obtain
\begin{align}
& S^\D_{(a,a)^\pm,(b,b)^\pm} = S^\D_{(a,a)^\pm,(b,b)^\mp} = \frac{1}{2} ( S^\C_{ab} )^2,
\\
& S^\D_{[a,b],(c,c)^\pm} = S^\C_{ac} S^\C_{bc} ,
\\
& S^\D_{[a,b],[c,d]} = S^\C_{ac} S^\C_{bd} +S^\C_{ad} S^\C_{bc}.
\end{align}

\begin{figure}
\includegraphics[scale=0.4]{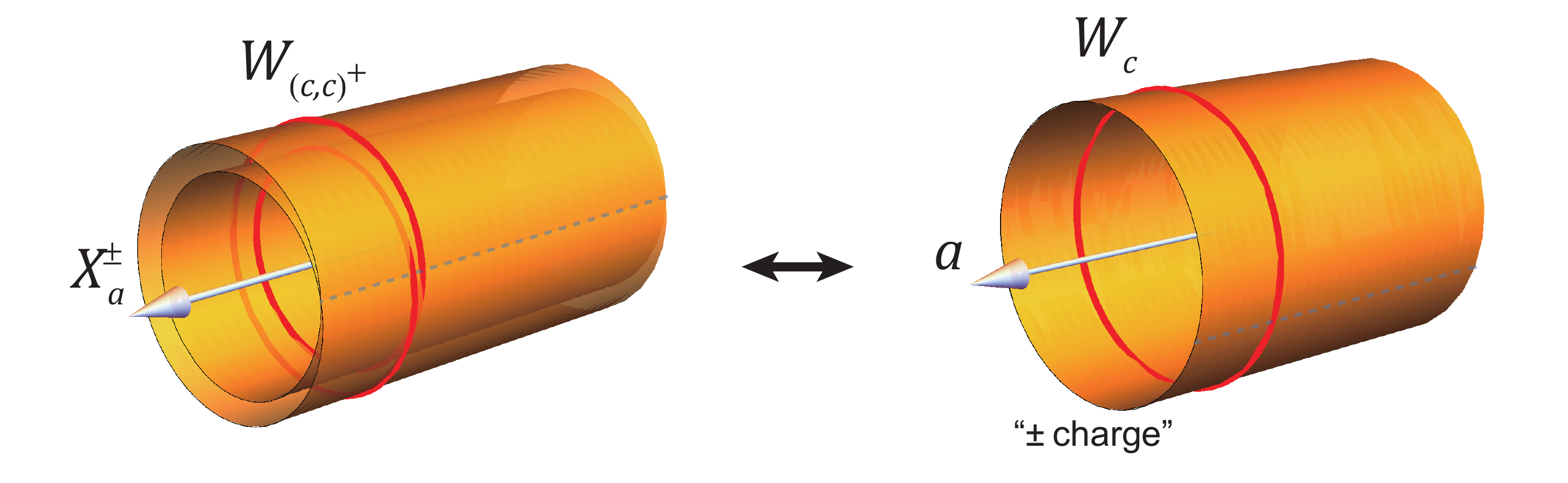}
\caption{The left panel depicts the double-layer cylinder geometry with a layer-exchanging branch cut (dashed line). The white arrow represents the anyon flux $X_a^\pm$ threading through the cylinder. The Wilson line operator $W_{(c,c)^+}$ is indicated by the red line. The right panel depicts a single-layer cylinder geometry carrying the topological order described by $\C$ that is topologically equivalent to the left panel. The anyon flux threading through the single-layer cylinder is given by $a$. The application of the Wilson line operator $W_{(c,c)^+}$ on the left panel is topologically equivalent to applying the Wilson line operator $W_{c}$ on the single-layer cylinder geometry. The Wilson line operator $W_{c}$ brings an anyon $c \in \C$ around the cylinder.}
\label{Fig:TwoCylinderWilsonLine}
\end{figure}

For the matrix elements $S^\D_{X_a^\pm,(c,c)^\pm}$ and $S^\D_{X_a^\pm,(c,c)^\mp}$, the consistency with anyon condensation does not provide a lot of information. We will use a different approach to obtain them. Consider a topological state in the theory $\D$ on a cylinder geometry with $X_a^\pm$ anyon flux threading the cylinder. Such a state is an eigenstate of the Wilson line operator $W_{(c,c)^+}$ which bring an anyon $(c,c)^+$ around the cylinder for any $c\in \C$. The corresponding eigenvalue is $\frac{S^\D_{X_a^+,(c,c)^+}}{S^\D_{X_a^+,(\1,\1)^+}}$. The cylinder geometry with the $X_a^\pm$ anyon flux can be viewed as a double-layer cylinder with a layer exchanging branch cut (left panel of Fig. \ref{Fig:TwoCylinderWilsonLine}), which is topologically equivalent to a single-layer cylinder with the topological order described by $\C$ residing on it and with the anyon flux $a\in \C$ threading through it (right panel of Fig. \ref{Fig:TwoCylinderWilsonLine}). The application of the Wilson line operator $W_{(c,c)^+}$ on the double-layer cylinder with the $X_a^\pm$ anyon flux is topologically equivalent to applying the Wilson line operator $W_{c}$ on the single-layer cylinder geometry. Here $W_{c}$ is the operator that brings the anyon $c\in \C$ around cylinder. The state of the single-layer cylinder with anyon flux $a$ is an eigenstate of $W_c$ with eigenvalue $\frac{S^\C_{ac}}{S^\C_{a\1}}$. Therefore, we have
\begin{align}
\frac{S^\D_{X_a^+,(c,c)^+}}{S^\D_{X_a^+,(\1,\1)^+}} = \frac{S^\C_{ac}}{S^\C_{a\1}}
\end{align}
By further noticing that $S^\D_{X_a^+,(\1,\1)^+} = d_{X_a^+}/D_\D=$ and $S^\C_{a\1} = d_a/ D_\C$, we can conclude that
\begin{align}
S^\D_{X_a^+,(c,c)^+} = \frac{1}{2}S^\C_{ac}.
\end{align}
$S^\D_{X_a^-,(c,c)^+}$ can be obtained by viewing the $X_a^-$ as the fusion product of $X_a^+$ and $(\1,\1)^-$. Since $(\1,\1)^-$ braids trivially with $(c,c)^+$, we have
\begin{align}
S^\D_{X_a^-,(c,c)^+} = S^\D_{X_a^+,(c,c)^+} = \frac{1}{2}S^\C_{ac}.
\end{align}
$S^\D_{X_a^\pm,(c,c)^-}$ can be obtained in a similar fashion. $(c,c)^-$ is the fusion product of $(c,c)^+$ and $(\1,\1)^-$. Due to the braiding statistics between the $\Z2$ charge $(\1,\1)^-$ and the $\Z2$ flux carried by $X_a^\pm$, the matrix $S^\D_{X_a^\pm,(c,c)^-}$ has an extra factor of $-1$:
\begin{align}
S^\D_{X_a^\pm,(c,c)^-} = -S^\D_{X_a^\pm,(c,c)^+} = -\frac{1}{2}S^\C_{ac}.
\end{align}

The elements of the $S$ matrix element can also be expressed in terms of the fusion multiplicities and the topological spins of the anyons. Such a relation will be used to derive the $S$ matrix elements $S^\D_{X_a^\pm,[c,d]}$:
\begin{align}
S^\D_{X_a^\pm,[c,d]}
= \frac{1}{D_\D ~ \theta_{X_a^\pm} \theta_{[c,d]}} \sum_{\alpha \in \D} ~^\D \! N_{\alpha}^{X_a^\pm,[c,d]} d_\alpha \theta_\alpha.
\end{align}
From Eq. \ref{Eq:Fusion_Xabc}, we see that the fusion product of $X_a^\pm$ and $[c,d]$ always contains equal numbers of $X_e^+$ and $X_e^-$, which have their opposite topological spins differ by $-1$, for any $e \in \C$. Therefore, the summation $\sum_{\alpha \in \D} ~^\D \! N_{\alpha}^{X_a^\pm,[c,d]} d_\alpha \theta_\alpha = 0$, which means
\begin{align}
S^\D_{X_a^\pm,[c,d]}  = 0.
\end{align}
The same method can be used to derive $S^\D_{X_a^\pm,X_a^\pm}$ and $S^\D_{X_a^\pm,X_a^\mp}$. For example,
\begin{align}
S^\D_{X_a^+,X_b^+}
= \frac{1}{D_\D~ \theta_{X_a^+}  \theta_{X_b^+}} \sum_{\alpha \in \D } ~^\D \! N_{\alpha}^{X_a^+,X_b^+} d_\alpha \theta_\alpha.
\end{align}
Although Eq. \ref{Eq:Fusion XX1} can only determine the fusion product of $X_a^+$ and $X_b^+$ upto the unknown function $p$ (that takes value $\pm$), the relevant topological spins $\theta_\alpha$ and the quantum dimensions $d_\alpha$ in the fusion product are in fact independent from $p$. Therefore, we have
\begin{align}
S^\D_{X_a^+,X_b^+}
& = \frac{1}{D_\D~ \theta_{X_a^+}  \theta_{X_b^+}} \sum_{\alpha \in \D } ~^\D \! N_{\alpha}^{X_a^+,X_b^+} d_\alpha \theta_\alpha
\nonumber \\
& = \frac{1}{2 D_\C^2  \theta_a^{\frac{1}{2}} \theta_b^{\frac{1}{2}} }
\sum_{d,e,f \in \C}
~^\C \! N_{e}^{a \bar{d}} ~^\C \! N_{f}^{e b} d_d d_{f} \theta_d \theta_f,
\nonumber \\
&= \frac{1}{2    \theta_a^{\frac{1}{2}} \theta_b^{\frac{1}{2}} }
\sum_{e\in \C}
\left[
\left( \frac{1}{D_\C} \sum_{d\in \C} ~^\C \! N_{d}^{a \bar{e}} d_d \theta_d \right)
\left( \frac{1}{D_\C} \sum_{f \in \C} ~^\C \! N_{f}^{e b} d_f \theta_f
\right)
\right]
\nonumber \\
& = \frac{1}{2    \theta_a^{\frac{1}{2}} \theta_b^{\frac{1}{2}} }
\sum_{e\in \C}
\big[
\left( \theta_a \theta_{\bar{e}} S^\C_{a\bar{e}} \right)
\left( \theta_e \theta_{b} S^\C_{eb} \right)
\big]
\nonumber \\
& = \frac{1}{2} \big( \sqrt{T^{\C}} S^\C T^{\C}  (S^\C)^2 T^{\C} S^\C  \sqrt{T^{\C}} \big)_{ab}.
\end{align}
Here, we've used the fact that $~^\C \! N_{e}^{a \bar{d}}  =~^\C \! N_{d}^{a \bar{e}}$ on the third line and the relations between the $S$-matrix, fusion rule and topological spins in $\C$ in deriving the fourth line. $T^\C$ denotes the $T$ matrix of the UMTC $\C$. Given this result, it is easy to obtain that
\begin{align}
& S^\D_{X_a^\pm,X_b^\pm} =  \frac{1}{2} \big( \sqrt{T^{\C}} S^\C T^{\C}  (S^\C)^2 T^{\C} S^\C  \sqrt{T^{\C}} \big)_{ab}
\\
& S^\D_{X_a^\pm,X_b^\mp} = - \frac{1}{2} \big( \sqrt{T^{\C}} S^\C T^{\C}  (S^\C)^2 T^{\C} S^\C  \sqrt{T^{\C}} \big)_{ab}.
\end{align}
These expressions of the $S^\D_{X_a^\pm,X_b^\pm} $ and $S^\D_{X_a^\pm,X_b^\mp}$ apply to the case with no chiral central charge $c_\C$ and with the choice of trivial element in $H^3(\Z2, U(1))$. In the most general case, there will be an extra factor of $\omega e^{-i2\pi c_\C/8}$ in $S^\D_{X_a^\pm,X_b^\pm} $ and $S^\D_{X_a^\pm,X_b^\mp}$. As a reminder, $\omega=1$ and $\omega=-1$ represent the trivial and non-trivial element in $H^3(\Z2, U(1))$ respectively. Given that the $S$ matrix is an symmetric matrix, we've now obtained all the matrix element of $S^\D$. The result (for the case with $c_\C = 0$ and $\omega=1$) is summarized in Table. \ref{Table:tS matrix}.

\begin{table}
\caption{The matrix elements $S^\D_{\alpha \beta}$ of the $S$ matrix of $\D$ (with no chiral central charge, i.e. $c_\C =0$ and with the trivial element $\omega=1 \in H^3(\Z2,U(1))$).}
\label{Table:tS matrix}
\begin{tabular}{|c||c|c|c|c|c|}
\hline
\diagbox[height=2em]{~$\alpha$}{$\beta$}  & $(c,c)^+ $ & $(c,c)^-$ & $[c,d]$ & $X_c^+$ & $X_c^-$ \\
\hline \hline
$(a,a)^+$     &       $\frac{1}{2} ( S^\C_{ac} )^2$       &  $\frac{1}{2} ( S^\C_{ac} )^2$    & $S^\C_{ac} S^\C_{ad}$  & $\frac{1}{2} S^\C_{ac} $ &   $\frac{1}{2} S^\C_{ac} $ \\
\hline
$(a,a)^-$       &       $\frac{1}{2} ( S^\C_{ac} )^2$        &      $\frac{1}{2} ( S^\C_{ac} )^2$  & $S^\C_{ac} S^\C_{ad}$  & $-\frac{1}{2} S^\C_{ac} $ & $-\frac{1}{2} S^\C_{ac} $ \\
\hline
$[a,b]$      &      $S^\C_{ac} S^\C_{bc} $     &  $S^\C_{ac} S^\C_{bc} $    &  $S^\C_{ac} S^\C_{bd} + S^\C_{bc} S^\C_{ad}$ & 0 & 0\\
\hline
$X_a^+$    &    $\frac{1}{2} S^\C_{ac} $        &  $-\frac{1}{2} S^\C_{ac} $    &  0 & $\frac{1}{2} \big( \sqrt{T^{\C}} S^\C T^{\C}  (S^\C)^2 T^{\C} S^\C  \sqrt{T^{\C}} \big)_{ac}$ & $-\frac{1}{2} \big( \sqrt{T^{\C}} S^\C T^{\C}  (S^\C)^2 T^{\C} S^\C  \sqrt{T^{\C}} \big)_{ac}$\\
\hline
$X_a^-$     &      $\frac{1}{2} S^\C_{ac} $       &  $-\frac{1}{2} S^\C_{ac} $   & 0  & $-\frac{1}{2} \big( \sqrt{T^{\C}} S^\C T^{\C}  (S^\C)^2 T^{\C} S^\C  \sqrt{T^{\C}} \big)_{ac}$  & $\frac{1}{2} \big( \sqrt{T^{\C}} S^\C T^{\C}  (S^\C)^2 T^{\C} S^\C  \sqrt{T^{\C}} \big)_{ac}$\\
\hline
\end{tabular}
\end{table}

\section{Derivation of Eq.~\eqref{eq:wxw}}
\label{app:derive_wxw}

In this appendix, we provide a detailed derivation of Eq.~\eqref{eq:wxw}. As discussed in the main text, this result follows the constraint \eqref{eq:nnN=Nn} that the restriction map, describing the anyon condensation, commutes with anyon fusion. In particular, we shall consider the fusion between two symmetry defects $X_{\mathds1}^+$. In this derivation, we want to keep track of the anyon charges in $\mathcal C$ through the fusion and anyon-condensation processes, without worrying about the $\mathbb Z_2$-symmetry charges, which do not appear in Eq.~\eqref{eq:wxw}. Technically, this can be achieved by considering another anyon condensation $r^\prime$, which condenses, or forgets, the $\mathbb Z_2$ charges in $\mathcal T$.
For example, $r^\prime$ further maps both $a^\pm\in\mathcal T$ to $a$, and both defects $x_a^\pm\in\mathcal T$ to $x_a$.
Hence, the composed restricting map $\tilde r=r^\prime\circ r$ follows rules similar to Eqs.~(\ref{eq:rab}-\ref{eq:raa-}), but forgets the $\mathbb Z_2$ charges of the outcomes.

In Eq.~\eqref{eq:lx1}, $w_a$ are defined as the lifting coefficients of $x_{\mathds1}^+$: $w_a\equiv n_{X_a^+,x_{\mathds1}^+}$. Here, we argue that the same coefficients also describe the restriction map of $X_{\mathds1}^+$:
\begin{equation}
  \label{eq:rX1}
  \tilde r(X_{\mathds1}^+) = \sum_{a\in\mathcal C}w_ax_a.
\end{equation}
As discussed in Sec.~\ref{sec:app2dl}, restriction map $\tilde r$ of the defect $X_{\mathds1}^+$ only contains defects $x_b$. Thus, we can write this as a general expression,
\begin{equation}
  \label{eq:rX1gen}
  \tilde r(X_{\mathds1}^+) = \sum_{a\in\mathcal C}
  \tilde n_{X_{\mathds1}^+,x_a}x_a.
\end{equation}
We now derive the coefficient $\tilde n_{X_{\mathds1}^+,x_a}$ of the map $\tilde r$, using the constraint \eqref{eq:nnN=Nn} that the restriction map $\tilde r$ commutes with anyon fusion. In particular, consider the fusion $X_{\mathds1}^+\times[a,\mathds1] = X_a^++X_a^-$ [see Eq.~\eqref{eq:Xabc}]. Constraint \eqref{eq:nnN=Nn} implies that $\tilde r(X_{\mathds1}^+)\times \tilde r([a,\1]) = \tilde r(X_a^+)+\tilde r(X_a^-)$.
The definition $w_a\equiv n_{X_a^+,x_{\mathds1}^+}$ implies that, on the right-hand side, $\tilde r(X_a^+)=w_ax_{\mathds1}+\cdots$. At the same time, according to Eq.~\eqref{eq:lx1-}, $\tilde r(X_a^-)=w_ax_{\mathds1}+\cdots$. Furthermore, Eq.~\eqref{eq:rab} implies that $\tilde r([a,\mathds1])=2a$. Compiling these results, we get
\begin{equation*}
  \left(\sum_{b\in\mathcal C}\tilde n_{X_{\mathds1}^+,x_b}x_b\right) \times 2a=2w_ax_{\mathds1}+\cdots
\end{equation*}
Since $x_{\mathds1}$ can only be generated through the fusion $x_{\bar a}\times a=x_{\mathds1}+\cdots$, we conclude that $\tilde n_{X_{\mathds1}^+,x_a^+}=w_{\bar a}$. Furthermore, using the property that $n_{a,b}=n_{\bar a,\bar b}$, we have $w_{\bar a}=n_{X_{\bar a}^+, x_\1}=n_{\overline{ X_{\bar a}^+}, \overline{x_\1}}=n_{X_a^+,x_\1^+}=w_a$. Hence, $\tilde n_{X_{\mathds1}^+,x_a}=w_{\bar a}=w_a$.
Plugging this in Eq.~\eqref{eq:rX1gen} gives Eq.~\eqref{eq:rX1}.

Now, we are ready to prove Eq.~\eqref{eq:wxw}, using Eq.~\eqref{eq:rX1} and the constraint \eqref{eq:nnN=Nn}. We consider the following fusion rule, which is a special case of Eq.~\eqref{Eq:Fusion XX1},
\begin{equation}
  \label{eq:X1+xX1+}
  X_{\mathds1}^+\times X_{\mathds1}^+
  = \sideset{}{'}\sum_{a\neq\bar a}[a,\bar a]
  + \sum_{a=\bar a}(a,a)^{p(\1,\1,a,a)},
\end{equation}
where $p(\1,\1,a,a)=\pm1$ is a $\mathbb Z_2$ symmetry charge that plays no role in the derivation below, because it will be forgotten after $\tilde r$ is applied. Constraint~\eqref{eq:nnN=Nn} implies that
\begin{equation}
  \label{eq:X1+xX1+:r}
  \tilde r\left(X_{\mathds1}^+\right)\times
  \tilde r\left(X_{\mathds1}^+\right)
  = \sideset{}{'}\sum_{a\neq\bar a}\tilde r\left([a,\bar a]\right)
  + \sum_{a=\bar a}\tilde r\left((a,a)^{p(\1,\1,a,a)}\right).
\end{equation}
Applying Eq.~\eqref{eq:rX1} to the left-hand side, we get
\begin{equation}
  \label{eq:X1+xX1+:LHS}
  \tilde r\left(X_{\mathds1}^+\right)\times
  \tilde r\left(X_{\mathds1}^+\right)
  =\sum_{ab}w_aw_bx_a\times x_b
  =\sum_{abc}w_aw_bN^{ab}_cc.
\end{equation}
Applying Eqs.~(\ref{eq:raa}-\ref{eq:raa-}) to the right-hand side, we get
\begin{equation}
  \label{eq:X1+xX1+:RHS}
  \sideset{}{'}\sum_{a\neq\bar a}\tilde r\left([a,\bar a]\right)
  + \sum_{a=\bar a}\tilde r\left((a,a)^{p(\1,\1,a,a)}\right)
  =\sideset{}{'}\sum_{a\neq\bar a,c}2N^{a\bar\rho_m(a)}_cc
  + \sum_{a=\bar a,c}N^{a\bar\rho_m(a)}_cc
  =\sum_{ac}N^{a\bar\rho_m(a)}_cc.
\end{equation}
Comparing Eqs.~\eqref{eq:X1+xX1+:LHS} and \eqref{eq:X1+xX1+:RHS}, we arrive at the result of Eq.~\eqref{eq:wxw2}, or equivalently Eq.~\eqref{eq:wxw}.

\section{Simplification of Eq.~\eqref{eq:pa=Sw}}
\label{app:simplification}

In this appendix, we derive a simplified form --- Eq.~\eqref{eq:piu} --- of Eq.~\eqref{eq:pa=Sw}, when $a$ is an Abelian anyon. In this case, the $S$ matrix entry $S_{a,b}$ is proportional to an Abelian braiding phase,
\begin{equation}
S_{a,b}=\frac{d_b}{D_{\mathcal C}}M_{a,b}^\ast,
\end{equation}
where $b$ can be any (non-Abelian) anyon, and $M_{a,b}$ is a braiding phase factor. Correspondingly, Eq.~\eqref{eq:pa=Sw} becomes
\begin{equation}
\mu(a) = \sum_{b\in\textswab w}\frac{d_bw_b}{D_{\mathcal C}}M_{a,b}^\ast.
\end{equation}
Here, we assume that $\{w_b\}$ satisfies Eq.~\eqref{eq:wxw2}. Using Eq.~\eqref{eq:Dw}, we see that the right-hand side of this equation is a weighted sum of phases, with the wight being $\frac{d_bw_b}{D_{\mathcal C}}$ and the total weights equal to unity. Therefore, the absolute value of the sum is bounded by one:
\[\left|\sum_{b\in\textswab w}\frac{d_bw_b}{D_{\mathcal C}}M_{a,b}^\ast\right|
\leq \sum_{b\in\textswab w}\frac{d_bw_b}{D_{\mathcal C}}|M_{a,b}^\ast|=1,\]
The bound is saturated if and only if all phases $M_{a,b}^\ast$ are the same. For $a=\bar\rho_m(a)$, the left-hand side $\mu(a)=\pm1$ indeed saturates this bound. Therefore, we conclude that
\begin{equation}
  \label{eq:piu}
  M_{a,b}^\ast = \mu(a) = \pm1,\quad\forall b\in\textswab w.
\end{equation}
This relation is extensively used for Abelian topological order in Ref.~\onlinecite{WangLevinIndicator} for deriving anomaly indicators associated with the time-reversal symmetry.

\section{Relations between constraints}
\label{app:concon}



In this appendix, we show that some parts of the constraints on $\{\mu(a)\}$ can be derived from those on $\{w_a\}$. More explicitly, assuming that $\{\mu(a)\}$ are computed  from $\{w_a\}$ through Eq.~\eqref{eq:pa=Sw}, we show that the constraints Eqs.~\eqref{eq:wxw} and Eq.~\eqref{eq:spinconstraint} on $\{w_a\}$ can lead to the following constraints on $\{\mu(x)\}$: (i) $\mu(x)=\pm1$ if $x=\bar\rho_m(x)$ and $\mu(x)=0$ otherwise; (2) the Abelian case of Eq.~\eqref{eq:pifusion} and (3) the Abelian case of Eq.~\eqref{eq:bbarrhob}. We do not know how to show the non-Abelian case of Eqs.~\eqref{eq:pifusion} and \eqref{eq:bbarrhob}.

First, let us show that $\mu(x)=\pm1$ if $x=\bar\rho_m(x)$ and $\mu(x)=0$ otherwise. That is, $\mu(x)^2=\delta_{x,\bar\rho_m(x)}$ for any $x\in \mathcal{C}$. To do that, we use the Verlinde formula
\begin{equation}
N^{ab}_c = \sum_x\frac{S_{ax}S_{bx}S_{\bar cx}}{S_{1x}}
\end{equation}
Note that $\{w_a\}$ satisfies Eq.~\eqref{eq:wxw2}. Multiplying Eq.~(\ref{eq:wxw2}) with $S_{x\bar{c}}^\dag $ and summing over $c$, we have the left-hand side of Eq.~\eqref{eq:wxw2} equals
\begin{align}
\sum_c S_{x\bar{c}}^\dag \sum_{a,b} w_aw_bN^{ab}_c & =  \sum_{a,b,c,y} S_{x\bar{c}}^\dag w_a w_b \frac{S_{ay}S_{by}S_{\bar cy}}{S_{1y}}  \nonumber \\
 & =  \sum_{c,y} S_{x\bar{c}}^\dag S_{\bar c y} \frac{\mu(y)^2}{S_{1y}} \nonumber\\
 & = \frac{\mu(x)^2}{S_{1x}}
\end{align}
At the same time, the  right-hand side of Eq.~\eqref{eq:wxw2} becomes
\begin{align}
\sum_c S_{x\bar{c}}^\dag \sum_a N^{a\bar\rho_m(a)}_c  & = \sum_{a,c,y} S_{x\bar c}^\dag \frac{S_{ay}S_{\bar\rho_m(a)y}S_{\bar cy}}{S_{1y}} \nonumber \\
 & = \sum_a \frac{S_{ax}S_{\bar\rho_m(a)x}}{S_{1x}} \nonumber\\
 & = \sum_a \frac{S_{ax}S_{a\bar\rho_m(x)}^*}{S_{1x}} =\frac{\delta_{x,\bar\rho_m(x)}}{S_{1x}}
\end{align}
Combining the above two equations, we find that
\begin{equation}
\label{eq:musquare}
\mu(x)^2 = \delta_{x,\bar\rho_m(x)}
\end{equation}
which is exactly what we want to show. Since $d_a$ and $w_a$ are non-negative, we have $\mu(\mathds 1) = \sum_a d_a w_a/D_{\mathcal{C}} = 1$. This agrees with Eq.~\eqref{eq:Dw}.

Second, we use Eq.~\eqref{eq:musquare} to show some additional properties of $\{w_a\}$, which are helpful in the main text as well as for showing other constraints on $\mu(x)$. Since Eq.~\eqref{eq:musquare} comes after Eq.~\eqref{eq:wxw2}, these properties are also consequence of Eq.~\eqref{eq:wxw2}. More explicitly, we show that
\begin{equation}
\label{eq:wproperty}
w_a=w_{\bar a} = w_{\rho_m(a)} = w_{a\times v\times\bar\rho_m(v)}
\end{equation}
where $v$ is any Abelian anyon. First of all, we notice that Eq.~\eqref{eq:musquare} implies that $\mu(x)$ is real. Accordingly, we have
\begin{align}
\mu(\bar x) =\mu(\bar x)^* = \sum_a S_{\bar x,a}^*w_a =  \sum_a S_{ x,a}w_a = \mu(x)
\end{align}
where we have used the property $S_{\bar x,a}^* = S_{x,a}$. With this, we have
\begin{equation}
w_{\bar a} = \sum_x S_{\bar a x}\mu(x) = \sum_x S_{a \bar x}\mu(\bar x) = w_a
\end{equation}
Also, it is not hard to see that
\begin{equation}
\mu(\rho_m(x)) = \mu(\bar x) =  \mu(x)
\end{equation}
which holds for both $x=\bar\rho_m(x)$ and $x\neq\bar\rho_m(x)$. Then, we find
\begin{equation}
w_{\rho_m(a)} = \sum_x S_{\rho_m(a) x}\mu(x) = \sum_x S^*_{a \rho_m(x)}\mu( x) =  \sum_x S^*_{a x}\mu(\rho_m(x)) = w_a^*=w_a
\end{equation}
where in the second equality we have used the fact that $S_{\rho_m(a) x}=S_{a\rho_m(x)}^*$. Finally, consider an Abelian anyon $v$. Then, we have
\begin{align}
w_{a\times v\times \bar\rho_m(v)} = \sum_{x=\bar\rho_m(x)} S_{a\times v\times\bar \rho_m(v), x} \mu(x) = \sum_{x=\rho_m(\bar x)} S_{a,x} \mu(x) = w_a \label{eq4}
\end{align}
where we have used the relation that $S_{a\times u, x} = S_{a, x}M^*_{u,x}$ for any Abelian anyon $u$.


Third, if both $x$ and $y$ are Abelian and are invariant under $\bar\rho_m$, we have $\mu(x)\mu(y)=\mu(x\times y)$, i.e., the Abelian case of the constraint Eq.~\eqref{eq:pifusion}.  This constraint immediately follows from Eq.~\eqref{eq:piu} and the fact that $M_{x,b}M_{y,b}=M_{x\times y, b}$ for Abelian anyons $x$ and $y$. Note that the proof of Eq.~\eqref{eq:piu} only uses properties of the $S$ matrix and that $\textswab{w}$ satisfies Eq.~\eqref{eq:wxw}. We are not able to show the general case of Eq.~\eqref{eq:pifusion}.

Finally, we show the Abelian case of the constraint Eq.~\eqref{eq:bbarrhob}. To do that, we use the constraint Eq.~\eqref{eq:spinconstraint}, i.e., $\theta_a=\eta$ are the same for all  $a\in\textswab w$.  As shown in Appendix \ref{app:simplification}, for Abelian anyon $x$, $\mu(x)=M_{x,a}^*$ for any $a\in\textswab w$. Also, we have $w_a=w_{a\times v\times\bar\rho_m(v)}$ for Abelian anyon $v$ accordingly to Eq.~\eqref{eq:wproperty}. Then, we have
\begin{align}
\mu(v\times\bar\rho_m(v)) &  = M_{v\times\bar\rho_m(v),a}^* =\frac{\theta_a\theta_{v\times\bar\rho_m(v)}}{\theta_{a\times v\times\bar\rho_m(v)}}=\theta_{v\times\bar\rho_m(v)}.
\end{align}
which is exactly Eq.~\eqref{eq:bbarrhob}.


\bibliography{set,quantum_double,extranotes}

\begin{thebibliography}{99}%
\makeatletter
\providecommand \@ifxundefined [1]{%
 \@ifx{#1\undefined}
}%
\providecommand \@ifnum [1]{%
 \ifnum #1\expandafter \@firstoftwo
 \else \expandafter \@secondoftwo
 \fi
}%
\providecommand \@ifx [1]{%
 \ifx #1\expandafter \@firstoftwo
 \else \expandafter \@secondoftwo
 \fi
}%
\providecommand \natexlab [1]{#1}%
\providecommand \enquote  [1]{``#1''}%
\providecommand \bibnamefont  [1]{#1}%
\providecommand \bibfnamefont [1]{#1}%
\providecommand \citenamefont [1]{#1}%
\providecommand \href@noop [0]{\@secondoftwo}%
\providecommand \href [0]{\begingroup \@sanitize@url \@href}%
\providecommand \@href[1]{\@@startlink{#1}\@@href}%
\providecommand \@@href[1]{\endgroup#1\@@endlink}%
\providecommand \@sanitize@url [0]{\catcode `\\12\catcode `\$12\catcode
  `\&12\catcode `\#12\catcode `\^12\catcode `\_12\catcode `\%12\relax}%
\providecommand \@@startlink[1]{}%
\providecommand \@@endlink[0]{}%
\providecommand \url  [0]{\begingroup\@sanitize@url \@url }%
\providecommand \@url [1]{\endgroup\@href {#1}{\urlprefix }}%
\providecommand \urlprefix  [0]{URL }%
\providecommand \Eprint [0]{\href }%
\providecommand \doibase [0]{http://dx.doi.org/}%
\providecommand \selectlanguage [0]{\@gobble}%
\providecommand \bibinfo  [0]{\@secondoftwo}%
\providecommand \bibfield  [0]{\@secondoftwo}%
\providecommand \translation [1]{[#1]}%
\providecommand \BibitemOpen [0]{}%
\providecommand \bibitemStop [0]{}%
\providecommand \bibitemNoStop [0]{.\EOS\space}%
\providecommand \EOS [0]{\spacefactor3000\relax}%
\providecommand \BibitemShut  [1]{\csname bibitem#1\endcsname}%
\let\auto@bib@innerbib\@empty
\bibitem [{\citenamefont {Wen}(1990)}]{WenTo1990}%
  \BibitemOpen
  \bibfield  {author} {\bibinfo {author} {\bibfnamefont {X.~G.}\ \bibnamefont
  {Wen}},\ }\href {\doibase 10.1142/s0217979290000139} {\bibfield  {journal}
  {\bibinfo  {journal} {Int. J. Mod. Phys. B}\ }\textbf {\bibinfo {volume}
  {04}},\ \bibinfo {pages} {239} (\bibinfo {year} {1990})}\BibitemShut
  {NoStop}%
\bibitem [{\citenamefont {Wen}(2015)}]{WenNSR2016}%
  \BibitemOpen
  \bibfield  {author} {\bibinfo {author} {\bibfnamefont {X.-G.}\ \bibnamefont
  {Wen}},\ }\href {\doibase 10.1093/nsr/nwv077} {\bibfield  {journal} {\bibinfo
   {journal} {Natl. Sci. Rev.}\ }\textbf {\bibinfo {volume} {3}},\ \bibinfo
  {pages} {68} (\bibinfo {year} {2015})}\BibitemShut {NoStop}%
\bibitem [{\citenamefont {Levin}\ and\ \citenamefont {Wen}(2006)}]{levin06}%
  \BibitemOpen
  \bibfield  {author} {\bibinfo {author} {\bibfnamefont {M.}~\bibnamefont
  {Levin}}\ and\ \bibinfo {author} {\bibfnamefont {X.-G.}\ \bibnamefont
  {Wen}},\ }\href {\doibase 10.1103/PhysRevLett.96.110405} {\bibfield
  {journal} {\bibinfo  {journal} {Phys. Rev. Lett.}\ }\textbf {\bibinfo
  {volume} {96}},\ \bibinfo {pages} {110405} (\bibinfo {year}
  {2006})}\BibitemShut {NoStop}%
\bibitem [{\citenamefont {Kitaev}\ and\ \citenamefont
  {Preskill}(2006)}]{kitaev06b}%
  \BibitemOpen
  \bibfield  {author} {\bibinfo {author} {\bibfnamefont {A.}~\bibnamefont
  {Kitaev}}\ and\ \bibinfo {author} {\bibfnamefont {J.}~\bibnamefont
  {Preskill}},\ }\href {\doibase 10.1103/PhysRevLett.96.110404} {\bibfield
  {journal} {\bibinfo  {journal} {Phys. Rev. Lett.}\ }\textbf {\bibinfo
  {volume} {96}},\ \bibinfo {pages} {110404} (\bibinfo {year}
  {2006})}\BibitemShut {NoStop}%
\bibitem [{\citenamefont {Haldane}(1983)}]{haldane83}%
  \BibitemOpen
  \bibfield  {author} {\bibinfo {author} {\bibfnamefont {F.~D.~M.}\
  \bibnamefont {Haldane}},\ }\href {\doibase 10.1103/PhysRevLett.50.1153}
  {\bibfield  {journal} {\bibinfo  {journal} {Phys. Rev. Lett.}\ }\textbf
  {\bibinfo {volume} {50}},\ \bibinfo {pages} {1153} (\bibinfo {year}
  {1983})}\BibitemShut {NoStop}%
\bibitem [{\citenamefont {Gu}\ and\ \citenamefont {Wen}(2009)}]{gu09}%
  \BibitemOpen
  \bibfield  {author} {\bibinfo {author} {\bibfnamefont {Z.-C.}\ \bibnamefont
  {Gu}}\ and\ \bibinfo {author} {\bibfnamefont {X.-G.}\ \bibnamefont {Wen}},\
  }\href {\doibase 10.1103/PhysRevB.80.155131} {\bibfield  {journal} {\bibinfo
  {journal} {Phys. Rev. B}\ }\textbf {\bibinfo {volume} {80}},\ \bibinfo
  {pages} {155131} (\bibinfo {year} {2009})}\BibitemShut {NoStop}%
\bibitem [{\citenamefont {Pollmann}\ \emph {et~al.}(2010)\citenamefont
  {Pollmann}, \citenamefont {Turner}, \citenamefont {Berg},\ and\ \citenamefont
  {Oshikawa}}]{pollmann10}%
  \BibitemOpen
  \bibfield  {author} {\bibinfo {author} {\bibfnamefont {F.}~\bibnamefont
  {Pollmann}}, \bibinfo {author} {\bibfnamefont {A.~M.}\ \bibnamefont
  {Turner}}, \bibinfo {author} {\bibfnamefont {E.}~\bibnamefont {Berg}}, \ and\
  \bibinfo {author} {\bibfnamefont {M.}~\bibnamefont {Oshikawa}},\ }\href
  {\doibase 10.1103/PhysRevB.81.064439} {\bibfield  {journal} {\bibinfo
  {journal} {Phys. Rev. B}\ }\textbf {\bibinfo {volume} {81}},\ \bibinfo
  {pages} {064439} (\bibinfo {year} {2010})}\BibitemShut {NoStop}%
\bibitem [{\citenamefont {Fidkowski}\ and\ \citenamefont
  {Kitaev}(2011)}]{fidkowski11}%
  \BibitemOpen
  \bibfield  {author} {\bibinfo {author} {\bibfnamefont {L.}~\bibnamefont
  {Fidkowski}}\ and\ \bibinfo {author} {\bibfnamefont {A.}~\bibnamefont
  {Kitaev}},\ }\href {\doibase 10.1103/PhysRevB.83.075103} {\bibfield
  {journal} {\bibinfo  {journal} {Phys. Rev. B}\ }\textbf {\bibinfo {volume}
  {83}},\ \bibinfo {pages} {075103} (\bibinfo {year} {2011})}\BibitemShut
  {NoStop}%
\bibitem [{\citenamefont {Chen}\ \emph
  {et~al.}(2011{\natexlab{a}})\citenamefont {Chen}, \citenamefont {Gu},\ and\
  \citenamefont {Wen}}]{chen11a}%
  \BibitemOpen
  \bibfield  {author} {\bibinfo {author} {\bibfnamefont {X.}~\bibnamefont
  {Chen}}, \bibinfo {author} {\bibfnamefont {Z.-C.}\ \bibnamefont {Gu}}, \ and\
  \bibinfo {author} {\bibfnamefont {X.-G.}\ \bibnamefont {Wen}},\ }\href
  {\doibase 10.1103/PhysRevB.84.235128} {\bibfield  {journal} {\bibinfo
  {journal} {Phys. Rev. B}\ }\textbf {\bibinfo {volume} {84}},\ \bibinfo
  {pages} {235128} (\bibinfo {year} {2011}{\natexlab{a}})}\BibitemShut
  {NoStop}%
\bibitem [{\citenamefont {Chen}\ \emph
  {et~al.}(2011{\natexlab{b}})\citenamefont {Chen}, \citenamefont {Gu},\ and\
  \citenamefont {Wen}}]{chen11b}%
  \BibitemOpen
  \bibfield  {author} {\bibinfo {author} {\bibfnamefont {X.}~\bibnamefont
  {Chen}}, \bibinfo {author} {\bibfnamefont {Z.-C.}\ \bibnamefont {Gu}}, \ and\
  \bibinfo {author} {\bibfnamefont {X.-G.}\ \bibnamefont {Wen}},\ }\href
  {\doibase 10.1103/PhysRevB.83.035107} {\bibfield  {journal} {\bibinfo
  {journal} {Phys. Rev. B}\ }\textbf {\bibinfo {volume} {83}},\ \bibinfo
  {pages} {035107} (\bibinfo {year} {2011}{\natexlab{b}})}\BibitemShut
  {NoStop}%
\bibitem [{\citenamefont {Schuch}\ \emph {et~al.}(2011)\citenamefont {Schuch},
  \citenamefont {P\'erez-Garc\'{\i}a},\ and\ \citenamefont {Cirac}}]{schuch11}%
  \BibitemOpen
  \bibfield  {author} {\bibinfo {author} {\bibfnamefont {N.}~\bibnamefont
  {Schuch}}, \bibinfo {author} {\bibfnamefont {D.}~\bibnamefont
  {P\'erez-Garc\'{\i}a}}, \ and\ \bibinfo {author} {\bibfnamefont
  {I.}~\bibnamefont {Cirac}},\ }\href {\doibase 10.1103/PhysRevB.84.165139}
  {\bibfield  {journal} {\bibinfo  {journal} {Phys. Rev. B}\ }\textbf {\bibinfo
  {volume} {84}},\ \bibinfo {pages} {165139} (\bibinfo {year}
  {2011})}\BibitemShut {NoStop}%
\bibitem [{\citenamefont {Chen}\ \emph {et~al.}(2012)\citenamefont {Chen},
  \citenamefont {Gu}, \citenamefont {Liu},\ and\ \citenamefont
  {Wen}}]{XieScience2012}%
  \BibitemOpen
  \bibfield  {author} {\bibinfo {author} {\bibfnamefont {X.}~\bibnamefont
  {Chen}}, \bibinfo {author} {\bibfnamefont {Z.-C.}\ \bibnamefont {Gu}},
  \bibinfo {author} {\bibfnamefont {Z.-X.}\ \bibnamefont {Liu}}, \ and\
  \bibinfo {author} {\bibfnamefont {X.-G.}\ \bibnamefont {Wen}},\ }\href
  {\doibase 10.1126/science.1227224} {\bibfield  {journal} {\bibinfo  {journal}
  {Science}\ }\textbf {\bibinfo {volume} {338}},\ \bibinfo {pages} {1604}
  (\bibinfo {year} {2012})}\BibitemShut {NoStop}%
\bibitem [{\citenamefont {Chen}\ \emph {et~al.}(2013)\citenamefont {Chen},
  \citenamefont {Gu}, \citenamefont {Liu},\ and\ \citenamefont {Wen}}]{chen13}%
  \BibitemOpen
  \bibfield  {author} {\bibinfo {author} {\bibfnamefont {X.}~\bibnamefont
  {Chen}}, \bibinfo {author} {\bibfnamefont {Z.-C.}\ \bibnamefont {Gu}},
  \bibinfo {author} {\bibfnamefont {Z.-X.}\ \bibnamefont {Liu}}, \ and\
  \bibinfo {author} {\bibfnamefont {X.-G.}\ \bibnamefont {Wen}},\ }\href
  {\doibase 10.1103/PhysRevB.87.155114} {\bibfield  {journal} {\bibinfo
  {journal} {Phys. Rev. B}\ }\textbf {\bibinfo {volume} {87}},\ \bibinfo
  {pages} {155114} (\bibinfo {year} {2013})}\BibitemShut {NoStop}%
\bibitem [{\citenamefont {Hasan}\ and\ \citenamefont {Kane}(2010)}]{hasan10}%
  \BibitemOpen
  \bibfield  {author} {\bibinfo {author} {\bibfnamefont {M.~Z.}\ \bibnamefont
  {Hasan}}\ and\ \bibinfo {author} {\bibfnamefont {C.~L.}\ \bibnamefont
  {Kane}},\ }\href {\doibase 10.1103/RevModPhys.82.3045} {\bibfield  {journal}
  {\bibinfo  {journal} {Rev. Mod. Phys.}\ }\textbf {\bibinfo {volume} {82}},\
  \bibinfo {pages} {3045} (\bibinfo {year} {2010})}\BibitemShut {NoStop}%
\bibitem [{\citenamefont {Qi}\ and\ \citenamefont {Zhang}(2011)}]{qi11}%
  \BibitemOpen
  \bibfield  {author} {\bibinfo {author} {\bibfnamefont {X.-L.}\ \bibnamefont
  {Qi}}\ and\ \bibinfo {author} {\bibfnamefont {S.-C.}\ \bibnamefont {Zhang}},\
  }\href {\doibase 10.1103/RevModPhys.83.1057} {\bibfield  {journal} {\bibinfo
  {journal} {Rev. Mod. Phys.}\ }\textbf {\bibinfo {volume} {83}},\ \bibinfo
  {pages} {1057} (\bibinfo {year} {2011})}\BibitemShut {NoStop}%
\bibitem [{\citenamefont {Wen}(2002)}]{wen12}%
  \BibitemOpen
  \bibfield  {author} {\bibinfo {author} {\bibfnamefont {X.-G.}\ \bibnamefont
  {Wen}},\ }\href {\doibase 10.1103/PhysRevB.65.165113} {\bibfield  {journal}
  {\bibinfo  {journal} {Phys. Rev. B}\ }\textbf {\bibinfo {volume} {65}},\
  \bibinfo {pages} {165113} (\bibinfo {year} {2002})}\BibitemShut {NoStop}%
\bibitem [{\citenamefont {Kitaev}(2006)}]{kitaev06}%
  \BibitemOpen
  \bibfield  {author} {\bibinfo {author} {\bibfnamefont {A.}~\bibnamefont
  {Kitaev}},\ }\href {\doibase http://dx.doi.org/10.1016/j.aop.2005.10.005}
  {\bibfield  {journal} {\bibinfo  {journal} {Annals of Physics}\ }\textbf
  {\bibinfo {volume} {321}},\ \bibinfo {pages} {2} (\bibinfo {year}
  {2006})}\BibitemShut {NoStop}%
\bibitem [{\citenamefont {Mesaros}\ and\ \citenamefont
  {Ran}(2013)}]{mesaros13}%
  \BibitemOpen
  \bibfield  {author} {\bibinfo {author} {\bibfnamefont {A.}~\bibnamefont
  {Mesaros}}\ and\ \bibinfo {author} {\bibfnamefont {Y.}~\bibnamefont {Ran}},\
  }\href {\doibase 10.1103/PhysRevB.87.155115} {\bibfield  {journal} {\bibinfo
  {journal} {Phys. Rev. B}\ }\textbf {\bibinfo {volume} {87}},\ \bibinfo
  {pages} {155115} (\bibinfo {year} {2013})}\BibitemShut {NoStop}%
\bibitem [{\citenamefont {Essin}\ and\ \citenamefont
  {Hermele}(2013)}]{essin13}%
  \BibitemOpen
  \bibfield  {author} {\bibinfo {author} {\bibfnamefont {A.~M.}\ \bibnamefont
  {Essin}}\ and\ \bibinfo {author} {\bibfnamefont {M.}~\bibnamefont
  {Hermele}},\ }\href {\doibase 10.1103/PhysRevB.87.104406} {\bibfield
  {journal} {\bibinfo  {journal} {Phys. Rev. B}\ }\textbf {\bibinfo {volume}
  {87}},\ \bibinfo {pages} {104406} (\bibinfo {year} {2013})}\BibitemShut
  {NoStop}%
\bibitem [{\citenamefont {{Barkeshli}}\ \emph {et~al.}(2014)\citenamefont
  {{Barkeshli}}, \citenamefont {{Bonderson}}, \citenamefont {{Cheng}},\ and\
  \citenamefont {{Wang}}}]{barkeshli14}%
  \BibitemOpen
  \bibfield  {author} {\bibinfo {author} {\bibfnamefont {M.}~\bibnamefont
  {{Barkeshli}}}, \bibinfo {author} {\bibfnamefont {P.}~\bibnamefont
  {{Bonderson}}}, \bibinfo {author} {\bibfnamefont {M.}~\bibnamefont
  {{Cheng}}}, \ and\ \bibinfo {author} {\bibfnamefont {Z.}~\bibnamefont
  {{Wang}}},\ }\href@noop {} {\bibfield  {journal} {\bibinfo  {journal} {ArXiv
  e-prints}\ } (\bibinfo {year} {2014})},\ \Eprint
  {http://arxiv.org/abs/1410.4540} {arXiv:1410.4540} \BibitemShut {NoStop}%
\bibitem [{\citenamefont {Tarantino}\ \emph {et~al.}(2016)\citenamefont
  {Tarantino}, \citenamefont {Lindner},\ and\ \citenamefont
  {Fidkowski}}]{tarantino16}%
  \BibitemOpen
  \bibfield  {author} {\bibinfo {author} {\bibfnamefont {N.}~\bibnamefont
  {Tarantino}}, \bibinfo {author} {\bibfnamefont {N.~H.}\ \bibnamefont
  {Lindner}}, \ and\ \bibinfo {author} {\bibfnamefont {L.}~\bibnamefont
  {Fidkowski}},\ }\href {http://stacks.iop.org/1367-2630/18/i=3/a=035006}
  {\bibfield  {journal} {\bibinfo  {journal} {New Journal of Physics}\ }\textbf
  {\bibinfo {volume} {18}},\ \bibinfo {pages} {035006} (\bibinfo {year}
  {2016})}\BibitemShut {NoStop}%
\bibitem [{\citenamefont {Heinrich}\ \emph {et~al.}(2016)\citenamefont
  {Heinrich}, \citenamefont {Burnell}, \citenamefont {Fidkowski},\ and\
  \citenamefont {Levin}}]{heinrich16}%
  \BibitemOpen
  \bibfield  {author} {\bibinfo {author} {\bibfnamefont {C.}~\bibnamefont
  {Heinrich}}, \bibinfo {author} {\bibfnamefont {F.}~\bibnamefont {Burnell}},
  \bibinfo {author} {\bibfnamefont {L.}~\bibnamefont {Fidkowski}}, \ and\
  \bibinfo {author} {\bibfnamefont {M.}~\bibnamefont {Levin}},\ }\href
  {\doibase 10.1103/PhysRevB.94.235136} {\bibfield  {journal} {\bibinfo
  {journal} {Phys. Rev. B}\ }\textbf {\bibinfo {volume} {94}},\ \bibinfo
  {pages} {235136} (\bibinfo {year} {2016})}\BibitemShut {NoStop}%
\bibitem [{\citenamefont {Cheng}\ \emph {et~al.}(2017)\citenamefont {Cheng},
  \citenamefont {Gu}, \citenamefont {Jiang},\ and\ \citenamefont
  {Qi}}]{cheng16}%
  \BibitemOpen
  \bibfield  {author} {\bibinfo {author} {\bibfnamefont {M.}~\bibnamefont
  {Cheng}}, \bibinfo {author} {\bibfnamefont {Z.-C.}\ \bibnamefont {Gu}},
  \bibinfo {author} {\bibfnamefont {S.}~\bibnamefont {Jiang}}, \ and\ \bibinfo
  {author} {\bibfnamefont {Y.}~\bibnamefont {Qi}},\ }\href {\doibase
  10.1103/PhysRevB.96.115107} {\bibfield  {journal} {\bibinfo  {journal} {Phys.
  Rev. B}\ }\textbf {\bibinfo {volume} {96}},\ \bibinfo {pages} {115107}
  (\bibinfo {year} {2017})}\BibitemShut {NoStop}%
\bibitem [{\citenamefont {Wang}\ and\ \citenamefont
  {Levin}(2017)}]{WangLevinIndicator}%
  \BibitemOpen
  \bibfield  {author} {\bibinfo {author} {\bibfnamefont {C.}~\bibnamefont
  {Wang}}\ and\ \bibinfo {author} {\bibfnamefont {M.}~\bibnamefont {Levin}},\
  }\href {\doibase 10.1103/PhysRevLett.119.136801} {\bibfield  {journal}
  {\bibinfo  {journal} {Phys. Rev. Lett.}\ }\textbf {\bibinfo {volume} {119}},\
  \bibinfo {pages} {136801} (\bibinfo {year} {2017})}\BibitemShut {NoStop}%
\bibitem [{\citenamefont {{Barkeshli}}\ \emph {et~al.}()\citenamefont
  {{Barkeshli}}, \citenamefont {Bonderson}, \citenamefont {Cheng},
  \citenamefont {Jian},\ and\ \citenamefont {Walker}}]{BarkeshliTRSET2016X}%
  \BibitemOpen
  \bibfield  {author} {\bibinfo {author} {\bibfnamefont {M.}~\bibnamefont
  {{Barkeshli}}}, \bibinfo {author} {\bibfnamefont {P.}~\bibnamefont
  {Bonderson}}, \bibinfo {author} {\bibfnamefont {M.}~\bibnamefont {Cheng}},
  \bibinfo {author} {\bibfnamefont {C.-M.}\ \bibnamefont {Jian}}, \ and\
  \bibinfo {author} {\bibfnamefont {K.}~\bibnamefont {Walker}},\ }\href@noop {}
  {\ }\Eprint {http://arxiv.org/abs/1612.07792} {arXiv:1612.07792
  [cond-mat.str-el]} \BibitemShut {NoStop}%
\bibitem [{\citenamefont {{Tachikawa}}\ and\ \citenamefont
  {{Yonekura}}(2016)}]{tackikawa16a}%
  \BibitemOpen
  \bibfield  {author} {\bibinfo {author} {\bibfnamefont {Y.}~\bibnamefont
  {{Tachikawa}}}\ and\ \bibinfo {author} {\bibfnamefont {K.}~\bibnamefont
  {{Yonekura}}},\ }\href@noop {} {\bibfield  {journal} {\bibinfo  {journal}
  {ArXiv e-prints}\ } (\bibinfo {year} {2016})},\ \Eprint
  {http://arxiv.org/abs/1610.07010} {arXiv:1610.07010} \BibitemShut {NoStop}%
\bibitem [{\citenamefont {Tachikawa}\ and\ \citenamefont
  {Yonekura}(2017)}]{tackikawa16b}%
  \BibitemOpen
  \bibfield  {author} {\bibinfo {author} {\bibfnamefont {Y.}~\bibnamefont
  {Tachikawa}}\ and\ \bibinfo {author} {\bibfnamefont {K.}~\bibnamefont
  {Yonekura}},\ }\href {\doibase 10.1103/PhysRevLett.119.111603} {\bibfield
  {journal} {\bibinfo  {journal} {Phys. Rev. Lett.}\ }\textbf {\bibinfo
  {volume} {119}},\ \bibinfo {pages} {111603} (\bibinfo {year}
  {2017})}\BibitemShut {NoStop}%
\bibitem [{\citenamefont {Vishwanath}\ and\ \citenamefont
  {Senthil}(2013)}]{vishwanath13}%
  \BibitemOpen
  \bibfield  {author} {\bibinfo {author} {\bibfnamefont {A.}~\bibnamefont
  {Vishwanath}}\ and\ \bibinfo {author} {\bibfnamefont {T.}~\bibnamefont
  {Senthil}},\ }\href {\doibase 10.1103/PhysRevX.3.011016} {\bibfield
  {journal} {\bibinfo  {journal} {Phys. Rev. X}\ }\textbf {\bibinfo {volume}
  {3}},\ \bibinfo {pages} {011016} (\bibinfo {year} {2013})}\BibitemShut
  {NoStop}%
\bibitem [{\citenamefont {{Wang}}\ \emph {et~al.}(2014)\citenamefont {{Wang}},
  \citenamefont {{Potter}},\ and\ \citenamefont {{Senthil}}}]{wangc-science}%
  \BibitemOpen
  \bibfield  {author} {\bibinfo {author} {\bibfnamefont {C.}~\bibnamefont
  {{Wang}}}, \bibinfo {author} {\bibfnamefont {A.~C.}\ \bibnamefont
  {{Potter}}}, \ and\ \bibinfo {author} {\bibfnamefont {T.}~\bibnamefont
  {{Senthil}}},\ }\href {\doibase 10.1126/science.1243326} {\bibfield
  {journal} {\bibinfo  {journal} {Science}\ }\textbf {\bibinfo {volume}
  {343}},\ \bibinfo {pages} {629} (\bibinfo {year} {2014})}\BibitemShut
  {NoStop}%
\bibitem [{\citenamefont {Metlitski}\ \emph {et~al.}(2015)\citenamefont
  {Metlitski}, \citenamefont {Kane},\ and\ \citenamefont
  {Fisher}}]{metlitski15}%
  \BibitemOpen
  \bibfield  {author} {\bibinfo {author} {\bibfnamefont {M.~A.}\ \bibnamefont
  {Metlitski}}, \bibinfo {author} {\bibfnamefont {C.~L.}\ \bibnamefont {Kane}},
  \ and\ \bibinfo {author} {\bibfnamefont {M.~P.~A.}\ \bibnamefont {Fisher}},\
  }\href {\doibase 10.1103/PhysRevB.92.125111} {\bibfield  {journal} {\bibinfo
  {journal} {Phys. Rev. B}\ }\textbf {\bibinfo {volume} {92}},\ \bibinfo
  {pages} {125111} (\bibinfo {year} {2015})}\BibitemShut {NoStop}%
\bibitem [{\citenamefont {Wang}\ \emph {et~al.}(2013)\citenamefont {Wang},
  \citenamefont {Potter},\ and\ \citenamefont {Senthil}}]{wangc13b}%
  \BibitemOpen
  \bibfield  {author} {\bibinfo {author} {\bibfnamefont {C.}~\bibnamefont
  {Wang}}, \bibinfo {author} {\bibfnamefont {A.~C.}\ \bibnamefont {Potter}}, \
  and\ \bibinfo {author} {\bibfnamefont {T.}~\bibnamefont {Senthil}},\ }\href
  {\doibase 10.1103/PhysRevB.88.115137} {\bibfield  {journal} {\bibinfo
  {journal} {Phys. Rev. B}\ }\textbf {\bibinfo {volume} {88}},\ \bibinfo
  {pages} {115137} (\bibinfo {year} {2013})}\BibitemShut {NoStop}%
\bibitem [{\citenamefont {Chen}\ \emph {et~al.}(2014)\citenamefont {Chen},
  \citenamefont {Fidkowski},\ and\ \citenamefont {Vishwanath}}]{chen14a}%
  \BibitemOpen
  \bibfield  {author} {\bibinfo {author} {\bibfnamefont {X.}~\bibnamefont
  {Chen}}, \bibinfo {author} {\bibfnamefont {L.}~\bibnamefont {Fidkowski}}, \
  and\ \bibinfo {author} {\bibfnamefont {A.}~\bibnamefont {Vishwanath}},\
  }\href {\doibase 10.1103/PhysRevB.89.165132} {\bibfield  {journal} {\bibinfo
  {journal} {Phys. Rev. B}\ }\textbf {\bibinfo {volume} {89}},\ \bibinfo
  {pages} {165132} (\bibinfo {year} {2014})}\BibitemShut {NoStop}%
\bibitem [{\citenamefont {Bonderson}\ \emph {et~al.}(2013)\citenamefont
  {Bonderson}, \citenamefont {Nayak},\ and\ \citenamefont {Qi}}]{bonderson13}%
  \BibitemOpen
  \bibfield  {author} {\bibinfo {author} {\bibfnamefont {P.}~\bibnamefont
  {Bonderson}}, \bibinfo {author} {\bibfnamefont {C.}~\bibnamefont {Nayak}}, \
  and\ \bibinfo {author} {\bibfnamefont {X.-L.}\ \bibnamefont {Qi}},\ }\href
  {http://stacks.iop.org/1742-5468/2013/i=09/a=P09016} {\bibfield  {journal}
  {\bibinfo  {journal} {Journal of Statistical Mechanics: Theory and
  Experiment}\ }\textbf {\bibinfo {volume} {2013}},\ \bibinfo {pages} {P09016}
  (\bibinfo {year} {2013})}\BibitemShut {NoStop}%
\bibitem [{\citenamefont {Fidkowski}\ \emph {et~al.}(2013)\citenamefont
  {Fidkowski}, \citenamefont {Chen},\ and\ \citenamefont
  {Vishwanath}}]{fidkowski13}%
  \BibitemOpen
  \bibfield  {author} {\bibinfo {author} {\bibfnamefont {L.}~\bibnamefont
  {Fidkowski}}, \bibinfo {author} {\bibfnamefont {X.}~\bibnamefont {Chen}}, \
  and\ \bibinfo {author} {\bibfnamefont {A.}~\bibnamefont {Vishwanath}},\
  }\href {\doibase 10.1103/PhysRevX.3.041016} {\bibfield  {journal} {\bibinfo
  {journal} {Phys. Rev. X}\ }\textbf {\bibinfo {volume} {3}},\ \bibinfo {pages}
  {041016} (\bibinfo {year} {2013})}\BibitemShut {NoStop}%
\bibitem [{\citenamefont {{Metlitski}}\ \emph {et~al.}()\citenamefont
  {{Metlitski}}, \citenamefont {{Fidkowski}}, \citenamefont {{Chen}},\ and\
  \citenamefont {{Vishwanath}}}]{metlitski14}%
  \BibitemOpen
  \bibfield  {author} {\bibinfo {author} {\bibfnamefont {M.~A.}\ \bibnamefont
  {{Metlitski}}}, \bibinfo {author} {\bibfnamefont {L.}~\bibnamefont
  {{Fidkowski}}}, \bibinfo {author} {\bibfnamefont {X.}~\bibnamefont {{Chen}}},
  \ and\ \bibinfo {author} {\bibfnamefont {A.}~\bibnamefont {{Vishwanath}}},\
  }\href@noop {} {\ }\Eprint {http://arxiv.org/abs/1406.3032} {arXiv:1406.3032}
  \BibitemShut {NoStop}%
\bibitem [{\citenamefont {Chen}\ \emph {et~al.}(2015)\citenamefont {Chen},
  \citenamefont {Burnell}, \citenamefont {Vishwanath},\ and\ \citenamefont
  {Fidkowski}}]{chen14}%
  \BibitemOpen
  \bibfield  {author} {\bibinfo {author} {\bibfnamefont {X.}~\bibnamefont
  {Chen}}, \bibinfo {author} {\bibfnamefont {F.~J.}\ \bibnamefont {Burnell}},
  \bibinfo {author} {\bibfnamefont {A.}~\bibnamefont {Vishwanath}}, \ and\
  \bibinfo {author} {\bibfnamefont {L.}~\bibnamefont {Fidkowski}},\ }\href
  {\doibase 10.1103/PhysRevX.5.041013} {\bibfield  {journal} {\bibinfo
  {journal} {Phys. Rev. X}\ }\textbf {\bibinfo {volume} {5}},\ \bibinfo {pages}
  {041013} (\bibinfo {year} {2015})}\BibitemShut {NoStop}%
\bibitem [{\citenamefont {{Kapustin}}\ and\ \citenamefont
  {{Thorngren}}(2014)}]{kapustin14}%
  \BibitemOpen
  \bibfield  {author} {\bibinfo {author} {\bibfnamefont {A.}~\bibnamefont
  {{Kapustin}}}\ and\ \bibinfo {author} {\bibfnamefont {R.}~\bibnamefont
  {{Thorngren}}},\ }\href@noop {} {\bibfield  {journal} {\bibinfo  {journal}
  {ArXiv e-prints}\ } (\bibinfo {year} {2014})},\ \Eprint
  {http://arxiv.org/abs/1404.3230} {arXiv:1404.3230} \BibitemShut {NoStop}%
\bibitem [{\citenamefont {Cho}\ \emph {et~al.}(2014)\citenamefont {Cho},
  \citenamefont {Teo},\ and\ \citenamefont {Ryu}}]{cho14}%
  \BibitemOpen
  \bibfield  {author} {\bibinfo {author} {\bibfnamefont {G.~Y.}\ \bibnamefont
  {Cho}}, \bibinfo {author} {\bibfnamefont {J.~C.~Y.}\ \bibnamefont {Teo}}, \
  and\ \bibinfo {author} {\bibfnamefont {S.}~\bibnamefont {Ryu}},\ }\href
  {\doibase 10.1103/PhysRevB.89.235103} {\bibfield  {journal} {\bibinfo
  {journal} {Phys. Rev. B}\ }\textbf {\bibinfo {volume} {89}},\ \bibinfo
  {pages} {235103} (\bibinfo {year} {2014})}\BibitemShut {NoStop}%
\bibitem [{\citenamefont {Wang}\ \emph {et~al.}(2015)\citenamefont {Wang},
  \citenamefont {Santos},\ and\ \citenamefont {Wen}}]{wangj15}%
  \BibitemOpen
  \bibfield  {author} {\bibinfo {author} {\bibfnamefont {J.~C.}\ \bibnamefont
  {Wang}}, \bibinfo {author} {\bibfnamefont {L.~H.}\ \bibnamefont {Santos}}, \
  and\ \bibinfo {author} {\bibfnamefont {X.-G.}\ \bibnamefont {Wen}},\ }\href
  {\doibase 10.1103/PhysRevB.91.195134} {\bibfield  {journal} {\bibinfo
  {journal} {Phys. Rev. B}\ }\textbf {\bibinfo {volume} {91}},\ \bibinfo
  {pages} {195134} (\bibinfo {year} {2015})}\BibitemShut {NoStop}%
\bibitem [{\citenamefont {Wang}\ \emph {et~al.}(2016)\citenamefont {Wang},
  \citenamefont {Lin},\ and\ \citenamefont {Levin}}]{bbc}%
  \BibitemOpen
  \bibfield  {author} {\bibinfo {author} {\bibfnamefont {C.}~\bibnamefont
  {Wang}}, \bibinfo {author} {\bibfnamefont {C.-H.}\ \bibnamefont {Lin}}, \
  and\ \bibinfo {author} {\bibfnamefont {M.}~\bibnamefont {Levin}},\ }\href
  {\doibase 10.1103/PhysRevX.6.021015} {\bibfield  {journal} {\bibinfo
  {journal} {Phys. Rev. X}\ }\textbf {\bibinfo {volume} {6}},\ \bibinfo {pages}
  {021015} (\bibinfo {year} {2016})}\BibitemShut {NoStop}%
\bibitem [{\citenamefont {Metlitski}\ \emph {et~al.}(2013)\citenamefont
  {Metlitski}, \citenamefont {Kane},\ and\ \citenamefont
  {Fisher}}]{metlitski13}%
  \BibitemOpen
  \bibfield  {author} {\bibinfo {author} {\bibfnamefont {M.~A.}\ \bibnamefont
  {Metlitski}}, \bibinfo {author} {\bibfnamefont {C.~L.}\ \bibnamefont {Kane}},
  \ and\ \bibinfo {author} {\bibfnamefont {M.~P.~A.}\ \bibnamefont {Fisher}},\
  }\href {\doibase 10.1103/PhysRevB.88.035131} {\bibfield  {journal} {\bibinfo
  {journal} {Phys. Rev. B}\ }\textbf {\bibinfo {volume} {88}},\ \bibinfo
  {pages} {035131} (\bibinfo {year} {2013})}\BibitemShut {NoStop}%
\bibitem [{\citenamefont {Wang}\ and\ \citenamefont {Senthil}(2013)}]{wangc13}%
  \BibitemOpen
  \bibfield  {author} {\bibinfo {author} {\bibfnamefont {C.}~\bibnamefont
  {Wang}}\ and\ \bibinfo {author} {\bibfnamefont {T.}~\bibnamefont {Senthil}},\
  }\href {\doibase 10.1103/PhysRevB.87.235122} {\bibfield  {journal} {\bibinfo
  {journal} {Phys. Rev. B}\ }\textbf {\bibinfo {volume} {87}},\ \bibinfo
  {pages} {235122} (\bibinfo {year} {2013})}\BibitemShut {NoStop}%
\bibitem [{\citenamefont {Qi}\ and\ \citenamefont
  {Fu}(2015{\natexlab{a}})}]{qi15}%
  \BibitemOpen
  \bibfield  {author} {\bibinfo {author} {\bibfnamefont {Y.}~\bibnamefont
  {Qi}}\ and\ \bibinfo {author} {\bibfnamefont {L.}~\bibnamefont {Fu}},\ }\href
  {\doibase 10.1103/PhysRevLett.115.236801} {\bibfield  {journal} {\bibinfo
  {journal} {Phys. Rev. Lett.}\ }\textbf {\bibinfo {volume} {115}},\ \bibinfo
  {pages} {236801} (\bibinfo {year} {2015}{\natexlab{a}})}\BibitemShut
  {NoStop}%
\bibitem [{\citenamefont {Hermele}\ and\ \citenamefont
  {Chen}(2016)}]{HermeleFFAT}%
  \BibitemOpen
  \bibfield  {author} {\bibinfo {author} {\bibfnamefont {M.}~\bibnamefont
  {Hermele}}\ and\ \bibinfo {author} {\bibfnamefont {X.}~\bibnamefont {Chen}},\
  }\href {\doibase 10.1103/PhysRevX.6.041006} {\bibfield  {journal} {\bibinfo
  {journal} {Phys. Rev. X}\ }\textbf {\bibinfo {volume} {6}},\ \bibinfo {pages}
  {041006} (\bibinfo {year} {2016})}\BibitemShut {NoStop}%
\bibitem [{\citenamefont {{Senthil}}(2015)}]{senthil15}%
  \BibitemOpen
  \bibfield  {author} {\bibinfo {author} {\bibfnamefont {T.}~\bibnamefont
  {{Senthil}}},\ }\href {\doibase 10.1146/annurev-conmatphys-031214-014740}
  {\bibfield  {journal} {\bibinfo  {journal} {Annual Review of Condensed Matter
  Physics}\ }\textbf {\bibinfo {volume} {6}},\ \bibinfo {pages} {299} (\bibinfo
  {year} {2015})}\BibitemShut {NoStop}%
\bibitem [{\citenamefont {Burnell}\ \emph {et~al.}(2014)\citenamefont
  {Burnell}, \citenamefont {Chen}, \citenamefont {Fidkowski},\ and\
  \citenamefont {Vishwanath}}]{burnell14}%
  \BibitemOpen
  \bibfield  {author} {\bibinfo {author} {\bibfnamefont {F.~J.}\ \bibnamefont
  {Burnell}}, \bibinfo {author} {\bibfnamefont {X.}~\bibnamefont {Chen}},
  \bibinfo {author} {\bibfnamefont {L.}~\bibnamefont {Fidkowski}}, \ and\
  \bibinfo {author} {\bibfnamefont {A.}~\bibnamefont {Vishwanath}},\ }\href
  {\doibase 10.1103/PhysRevB.90.245122} {\bibfield  {journal} {\bibinfo
  {journal} {Phys. Rev. B}\ }\textbf {\bibinfo {volume} {90}},\ \bibinfo
  {pages} {245122} (\bibinfo {year} {2014})}\BibitemShut {NoStop}%
\bibitem [{\citenamefont {Bernevig}\ \emph {et~al.}(2006)\citenamefont
  {Bernevig}, \citenamefont {Hughes},\ and\ \citenamefont
  {Zhang}}]{Bernevig2006}%
  \BibitemOpen
  \bibfield  {author} {\bibinfo {author} {\bibfnamefont {B.~A.}\ \bibnamefont
  {Bernevig}}, \bibinfo {author} {\bibfnamefont {T.}~\bibnamefont {Hughes}}, \
  and\ \bibinfo {author} {\bibfnamefont {S.}~\bibnamefont {Zhang}},\
  }\href@noop {} {\bibfield  {journal} {\bibinfo  {journal} {Science}\ }\textbf
  {\bibinfo {volume} {314}},\ \bibinfo {pages} {1757} (\bibinfo {year}
  {2006})},\ \Eprint {http://arxiv.org/abs/cond-mat/0611399} {cond-mat/0611399}
  \BibitemShut {NoStop}%
\bibitem [{\citenamefont {Bernevig}\ and\ \citenamefont
  {Zhang}(2006)}]{Bernevig_PRL2006}%
  \BibitemOpen
  \bibfield  {author} {\bibinfo {author} {\bibfnamefont {B.~A.}\ \bibnamefont
  {Bernevig}}\ and\ \bibinfo {author} {\bibfnamefont {S.-C.}\ \bibnamefont
  {Zhang}},\ }\href {\doibase 10.1103/PhysRevLett.96.106802} {\bibfield
  {journal} {\bibinfo  {journal} {Phys. Rev. Lett.}\ }\textbf {\bibinfo
  {volume} {96}},\ \bibinfo {pages} {106802} (\bibinfo {year}
  {2006})}\BibitemShut {NoStop}%
\bibitem [{\citenamefont {Kane}\ and\ \citenamefont
  {Mele}(2005{\natexlab{a}})}]{Kane2005a}%
  \BibitemOpen
  \bibfield  {author} {\bibinfo {author} {\bibfnamefont {C.~L.}\ \bibnamefont
  {Kane}}\ and\ \bibinfo {author} {\bibfnamefont {E.~J.}\ \bibnamefont
  {Mele}},\ }\href {\doibase 10.1103/PhysRevLett.95.146802} {\bibfield
  {journal} {\bibinfo  {journal} {Phys. Rev. Lett.}\ }\textbf {\bibinfo
  {volume} {95}},\ \bibinfo {pages} {146802} (\bibinfo {year}
  {2005}{\natexlab{a}})},\ \Eprint {http://arxiv.org/abs/cond-mat/0506581}
  {cond-mat/0506581} \BibitemShut {NoStop}%
\bibitem [{\citenamefont {Kane}\ and\ \citenamefont
  {Mele}(2005{\natexlab{b}})}]{Kane2005b}%
  \BibitemOpen
  \bibfield  {author} {\bibinfo {author} {\bibfnamefont {C.~L.}\ \bibnamefont
  {Kane}}\ and\ \bibinfo {author} {\bibfnamefont {E.~J.}\ \bibnamefont
  {Mele}},\ }\href@noop {} {\bibfield  {journal} {\bibinfo  {journal} {Phys.
  Rev. Lett.}\ }\textbf {\bibinfo {volume} {95}},\ \bibinfo {pages} {226801}
  (\bibinfo {year} {2005}{\natexlab{b}})},\ \Eprint
  {http://arxiv.org/abs/cond-mat/0411737} {cond-mat/0411737} \BibitemShut
  {NoStop}%
\bibitem [{\citenamefont {Fu}\ \emph {et~al.}(2007)\citenamefont {Fu},
  \citenamefont {Kane},\ and\ \citenamefont {Mele}}]{Fu_PRL07}%
  \BibitemOpen
  \bibfield  {author} {\bibinfo {author} {\bibfnamefont {L.}~\bibnamefont
  {Fu}}, \bibinfo {author} {\bibfnamefont {C.~L.}\ \bibnamefont {Kane}}, \ and\
  \bibinfo {author} {\bibfnamefont {E.~J.}\ \bibnamefont {Mele}},\ }\href
  {\doibase 10.1103/PhysRevLett.98.106803} {\bibfield  {journal} {\bibinfo
  {journal} {Phys. Rev. Lett.}\ }\textbf {\bibinfo {volume} {98}},\ \bibinfo
  {pages} {106803} (\bibinfo {year} {2007})},\ \Eprint
  {http://arxiv.org/abs/cond-mat/0607699} {cond-mat/0607699} \BibitemShut
  {NoStop}%
\bibitem [{\citenamefont {Moore}\ and\ \citenamefont
  {Balents}(2007)}]{Moore_PRB07}%
  \BibitemOpen
  \bibfield  {author} {\bibinfo {author} {\bibfnamefont {J.~E.}\ \bibnamefont
  {Moore}}\ and\ \bibinfo {author} {\bibfnamefont {L.}~\bibnamefont
  {Balents}},\ }\href {\doibase 10.1103/PhysRevB.75.121306} {\bibfield
  {journal} {\bibinfo  {journal} {Phys. Rev. B}\ }\textbf {\bibinfo {volume}
  {75}},\ \bibinfo {pages} {121306} (\bibinfo {year} {2007})},\ \Eprint
  {http://arxiv.org/abs/cond-mat/0607314} {cond-mat/0607314} \BibitemShut
  {NoStop}%
\bibitem [{\citenamefont {Roy}(2009)}]{Roy_PRB2009}%
  \BibitemOpen
  \bibfield  {author} {\bibinfo {author} {\bibfnamefont {R.}~\bibnamefont
  {Roy}},\ }\href {\doibase 10.1103/PhysRevB.79.195322} {\bibfield  {journal}
  {\bibinfo  {journal} {Phys. Rev. B}\ }\textbf {\bibinfo {volume} {79}},\
  \bibinfo {pages} {195322} (\bibinfo {year} {2009})},\ \Eprint
  {http://arxiv.org/abs/cond-mat/0607531} {cond-mat/0607531} \BibitemShut
  {NoStop}%
\bibitem [{\citenamefont {Kitaev}(2009)}]{kitaev2009}%
  \BibitemOpen
  \bibfield  {author} {\bibinfo {author} {\bibfnamefont {A.}~\bibnamefont
  {Kitaev}},\ }\href@noop {} {\bibfield  {journal} {\bibinfo  {journal} {AIP
  Conf. Proc.}\ }\textbf {\bibinfo {volume} {1134}},\ \bibinfo {pages} {22}
  (\bibinfo {year} {2009})},\ \Eprint {http://arxiv.org/abs/arXiv:0901.2686}
  {arXiv:0901.2686} \BibitemShut {NoStop}%
\bibitem [{\citenamefont {Schnyder}\ \emph {et~al.}(2008)\citenamefont
  {Schnyder}, \citenamefont {Ryu}, \citenamefont {Furusaki},\ and\
  \citenamefont {Ludwig}}]{schnyder2008}%
  \BibitemOpen
  \bibfield  {author} {\bibinfo {author} {\bibfnamefont {A.~P.}\ \bibnamefont
  {Schnyder}}, \bibinfo {author} {\bibfnamefont {S.}~\bibnamefont {Ryu}},
  \bibinfo {author} {\bibfnamefont {A.}~\bibnamefont {Furusaki}}, \ and\
  \bibinfo {author} {\bibfnamefont {A.~W.~W.}\ \bibnamefont {Ludwig}},\
  }\href@noop {} {\bibfield  {journal} {\bibinfo  {journal} {Phys. Rev. B}\
  }\textbf {\bibinfo {volume} {78}},\ \bibinfo {pages} {195125} (\bibinfo
  {year} {2008})},\ \Eprint {http://arxiv.org/abs/arXiv:0803.2786}
  {arXiv:0803.2786} \BibitemShut {NoStop}%
\bibitem [{\citenamefont {Wang}\ and\ \citenamefont {Senthil}(2014)}]{wangc14}%
  \BibitemOpen
  \bibfield  {author} {\bibinfo {author} {\bibfnamefont {C.}~\bibnamefont
  {Wang}}\ and\ \bibinfo {author} {\bibfnamefont {T.}~\bibnamefont {Senthil}},\
  }\href {\doibase 10.1103/PhysRevB.89.195124} {\bibfield  {journal} {\bibinfo
  {journal} {Phys. Rev. B}\ }\textbf {\bibinfo {volume} {89}},\ \bibinfo
  {pages} {195124} (\bibinfo {year} {2014})}\BibitemShut {NoStop}%
\bibitem [{\citenamefont {Song}\ \emph {et~al.}(2017)\citenamefont {Song},
  \citenamefont {Huang}, \citenamefont {Fu},\ and\ \citenamefont
  {Hermele}}]{Song2017}%
  \BibitemOpen
  \bibfield  {author} {\bibinfo {author} {\bibfnamefont {H.}~\bibnamefont
  {Song}}, \bibinfo {author} {\bibfnamefont {S.-J.}\ \bibnamefont {Huang}},
  \bibinfo {author} {\bibfnamefont {L.}~\bibnamefont {Fu}}, \ and\ \bibinfo
  {author} {\bibfnamefont {M.}~\bibnamefont {Hermele}},\ }\href {\doibase
  10.1103/PhysRevX.7.011020} {\bibfield  {journal} {\bibinfo  {journal} {Phys.
  Rev. X}\ }\textbf {\bibinfo {volume} {7}},\ \bibinfo {pages} {011020}
  (\bibinfo {year} {2017})}\BibitemShut {NoStop}%
\bibitem [{\citenamefont {Lake}(2016)}]{Lake2016}%
  \BibitemOpen
  \bibfield  {author} {\bibinfo {author} {\bibfnamefont {E.}~\bibnamefont
  {Lake}},\ }\href {\doibase 10.1103/PhysRevB.94.205149} {\bibfield  {journal}
  {\bibinfo  {journal} {Phys. Rev. B}\ }\textbf {\bibinfo {volume} {94}},\
  \bibinfo {pages} {205149} (\bibinfo {year} {2016})}\BibitemShut {NoStop}%
\bibitem [{\citenamefont {Kitaev}()}]{e8}%
  \BibitemOpen
  \bibfield  {author} {\bibinfo {author} {\bibfnamefont {A.}~\bibnamefont
  {Kitaev}},\ }\href@noop {} {}\bibinfo {note}
  {Http://online.kitp.ucsb.edu/online/topomat11 /kitaev/}\BibitemShut {NoStop}%
\bibitem [{\citenamefont {Lu}\ and\ \citenamefont {Vishwanath}(2012)}]{lu12}%
  \BibitemOpen
  \bibfield  {author} {\bibinfo {author} {\bibfnamefont {Y.-M.}\ \bibnamefont
  {Lu}}\ and\ \bibinfo {author} {\bibfnamefont {A.}~\bibnamefont
  {Vishwanath}},\ }\href {\doibase 10.1103/PhysRevB.86.125119} {\bibfield
  {journal} {\bibinfo  {journal} {Phys. Rev. B}\ }\textbf {\bibinfo {volume}
  {86}},\ \bibinfo {pages} {125119} (\bibinfo {year} {2012})}\BibitemShut
  {NoStop}%
\bibitem [{\citenamefont {{Kapustin}}(2014)}]{kapustin14a}%
  \BibitemOpen
  \bibfield  {author} {\bibinfo {author} {\bibfnamefont {A.}~\bibnamefont
  {{Kapustin}}},\ }\href@noop {} {\bibfield  {journal} {\bibinfo  {journal}
  {ArXiv e-prints}\ } (\bibinfo {year} {2014})},\ \Eprint
  {http://arxiv.org/abs/1403.1467} {arXiv:1403.1467} \BibitemShut {NoStop}%
\bibitem [{\citenamefont {{Kitaev}}(2003)}]{kitaev03}%
  \BibitemOpen
  \bibfield  {author} {\bibinfo {author} {\bibfnamefont {A.~Y.}\ \bibnamefont
  {{Kitaev}}},\ }\href {\doibase 10.1016/S0003-4916(02)00018-0} {\bibfield
  {journal} {\bibinfo  {journal} {Annals of Physics}\ }\textbf {\bibinfo
  {volume} {303}},\ \bibinfo {pages} {2} (\bibinfo {year} {2003})}\BibitemShut
  {NoStop}%
\bibitem [{\citenamefont {Zaletel}\ \emph {et~al.}(2017)\citenamefont
  {Zaletel}, \citenamefont {Lu},\ and\ \citenamefont
  {Vishwanath}}]{zaletel15b}%
  \BibitemOpen
  \bibfield  {author} {\bibinfo {author} {\bibfnamefont {M.~P.}\ \bibnamefont
  {Zaletel}}, \bibinfo {author} {\bibfnamefont {Y.-M.}\ \bibnamefont {Lu}}, \
  and\ \bibinfo {author} {\bibfnamefont {A.}~\bibnamefont {Vishwanath}},\
  }\href {\doibase 10.1103/PhysRevB.96.195164} {\bibfield  {journal} {\bibinfo
  {journal} {Phys. Rev. B}\ }\textbf {\bibinfo {volume} {96}},\ \bibinfo
  {pages} {195164} (\bibinfo {year} {2017})}\BibitemShut {NoStop}%
\bibitem [{\citenamefont {Zaletel}\ and\ \citenamefont
  {Vishwanath}(2015)}]{zaletel15}%
  \BibitemOpen
  \bibfield  {author} {\bibinfo {author} {\bibfnamefont {M.~P.}\ \bibnamefont
  {Zaletel}}\ and\ \bibinfo {author} {\bibfnamefont {A.}~\bibnamefont
  {Vishwanath}},\ }\href {\doibase 10.1103/PhysRevLett.114.077201} {\bibfield
  {journal} {\bibinfo  {journal} {Phys. Rev. Lett.}\ }\textbf {\bibinfo
  {volume} {114}},\ \bibinfo {pages} {077201} (\bibinfo {year}
  {2015})}\BibitemShut {NoStop}%
\bibitem [{\citenamefont {Jiang}\ and\ \citenamefont
  {Ran}(2015)}]{SJiangTPS2015X}%
  \BibitemOpen
  \bibfield  {author} {\bibinfo {author} {\bibfnamefont {S.}~\bibnamefont
  {Jiang}}\ and\ \bibinfo {author} {\bibfnamefont {Y.}~\bibnamefont {Ran}},\
  }\href {\doibase 10.1103/PhysRevB.92.104414} {\bibfield  {journal} {\bibinfo
  {journal} {Phys. Rev. B}\ }\textbf {\bibinfo {volume} {92}},\ \bibinfo
  {pages} {104414} (\bibinfo {year} {2015})}\BibitemShut {NoStop}%
\bibitem [{\citenamefont {Song}\ and\ \citenamefont
  {Hermele}(2015)}]{Song2015}%
  \BibitemOpen
  \bibfield  {author} {\bibinfo {author} {\bibfnamefont {H.}~\bibnamefont
  {Song}}\ and\ \bibinfo {author} {\bibfnamefont {M.}~\bibnamefont {Hermele}},\
  }\href {\doibase 10.1103/PhysRevB.91.014405} {\bibfield  {journal} {\bibinfo
  {journal} {Phys. Rev. B}\ }\textbf {\bibinfo {volume} {91}},\ \bibinfo
  {pages} {014405} (\bibinfo {year} {2015})}\BibitemShut {NoStop}%
\bibitem [{\citenamefont {Levin}\ and\ \citenamefont {Gu}(2012)}]{levin12}%
  \BibitemOpen
  \bibfield  {author} {\bibinfo {author} {\bibfnamefont {M.}~\bibnamefont
  {Levin}}\ and\ \bibinfo {author} {\bibfnamefont {Z.-C.}\ \bibnamefont {Gu}},\
  }\href {\doibase 10.1103/PhysRevB.86.115109} {\bibfield  {journal} {\bibinfo
  {journal} {Phys. Rev. B}\ }\textbf {\bibinfo {volume} {86}},\ \bibinfo
  {pages} {115109} (\bibinfo {year} {2012})}\BibitemShut {NoStop}%
\bibitem [{\citenamefont {Teo}\ \emph {et~al.}(2015)\citenamefont {Teo},
  \citenamefont {Hughes},\ and\ \citenamefont {Fradkin}}]{SET2}%
  \BibitemOpen
  \bibfield  {author} {\bibinfo {author} {\bibfnamefont {J.~C.~Y.}\
  \bibnamefont {Teo}}, \bibinfo {author} {\bibfnamefont {T.~L.}\ \bibnamefont
  {Hughes}}, \ and\ \bibinfo {author} {\bibfnamefont {E.}~\bibnamefont
  {Fradkin}},\ }\href@noop {} {\bibfield  {journal} {\bibinfo  {journal} {Ann.
  Phys.}\ }\textbf {\bibinfo {volume} {360}},\ \bibinfo {pages} {349} (\bibinfo
  {year} {2015})}\BibitemShut {NoStop}%
\bibitem [{\citenamefont {Wang}\ and\ \citenamefont {Levin}(2013)}]{wang13}%
  \BibitemOpen
  \bibfield  {author} {\bibinfo {author} {\bibfnamefont {C.}~\bibnamefont
  {Wang}}\ and\ \bibinfo {author} {\bibfnamefont {M.}~\bibnamefont {Levin}},\
  }\href {\doibase 10.1103/PhysRevB.88.245136} {\bibfield  {journal} {\bibinfo
  {journal} {Phys. Rev. B}\ }\textbf {\bibinfo {volume} {88}},\ \bibinfo
  {pages} {245136} (\bibinfo {year} {2013})}\BibitemShut {NoStop}%
\bibitem [{\citenamefont {Lu}\ and\ \citenamefont {Fidkowski}(2014)}]{lu14}%
  \BibitemOpen
  \bibfield  {author} {\bibinfo {author} {\bibfnamefont {Y.-M.}\ \bibnamefont
  {Lu}}\ and\ \bibinfo {author} {\bibfnamefont {L.}~\bibnamefont {Fidkowski}},\
  }\href {\doibase 10.1103/PhysRevB.89.115321} {\bibfield  {journal} {\bibinfo
  {journal} {Phys. Rev. B}\ }\textbf {\bibinfo {volume} {89}},\ \bibinfo
  {pages} {115321} (\bibinfo {year} {2014})}\BibitemShut {NoStop}%
\bibitem [{\citenamefont {{Bravyi}}\ and\ \citenamefont
  {{Kitaev}}(1998)}]{bravyi98}%
  \BibitemOpen
  \bibfield  {author} {\bibinfo {author} {\bibfnamefont {S.~B.}\ \bibnamefont
  {{Bravyi}}}\ and\ \bibinfo {author} {\bibfnamefont {A.~Y.}\ \bibnamefont
  {{Kitaev}}},\ }\href@noop {} {\bibfield  {journal} {\bibinfo  {journal}
  {eprint arXiv:quant-ph/9811052}\ } (\bibinfo {year} {1998})},\ \Eprint
  {http://arxiv.org/abs/quant-ph/9811052} {quant-ph/9811052} \BibitemShut
  {NoStop}%
\bibitem [{\citenamefont {{Beigi}}\ \emph {et~al.}(2011)\citenamefont
  {{Beigi}}, \citenamefont {{Shor}},\ and\ \citenamefont {{Whalen}}}]{beigi11}%
  \BibitemOpen
  \bibfield  {author} {\bibinfo {author} {\bibfnamefont {S.}~\bibnamefont
  {{Beigi}}}, \bibinfo {author} {\bibfnamefont {P.~W.}\ \bibnamefont {{Shor}}},
  \ and\ \bibinfo {author} {\bibfnamefont {D.}~\bibnamefont {{Whalen}}},\
  }\href {\doibase 10.1007/s00220-011-1294-x} {\bibfield  {journal} {\bibinfo
  {journal} {Communications in Mathematical Physics}\ }\textbf {\bibinfo
  {volume} {306}},\ \bibinfo {pages} {663} (\bibinfo {year} {2011})},\ \Eprint
  {http://arxiv.org/abs/1006.5479} {arXiv:1006.5479} \BibitemShut {NoStop}%
\bibitem [{\citenamefont {Levin}\ and\ \citenamefont {Stern}(2009)}]{levin09}%
  \BibitemOpen
  \bibfield  {author} {\bibinfo {author} {\bibfnamefont {M.}~\bibnamefont
  {Levin}}\ and\ \bibinfo {author} {\bibfnamefont {A.}~\bibnamefont {Stern}},\
  }\href {\doibase 10.1103/PhysRevLett.103.196803} {\bibfield  {journal}
  {\bibinfo  {journal} {Phys. Rev. Lett.}\ }\textbf {\bibinfo {volume} {103}},\
  \bibinfo {pages} {196803} (\bibinfo {year} {2009})}\BibitemShut {NoStop}%
\bibitem [{\citenamefont {Levin}\ and\ \citenamefont {Stern}(2012)}]{levin12b}%
  \BibitemOpen
  \bibfield  {author} {\bibinfo {author} {\bibfnamefont {M.}~\bibnamefont
  {Levin}}\ and\ \bibinfo {author} {\bibfnamefont {A.}~\bibnamefont {Stern}},\
  }\href {\doibase 10.1103/PhysRevB.86.115131} {\bibfield  {journal} {\bibinfo
  {journal} {Phys. Rev. B}\ }\textbf {\bibinfo {volume} {86}},\ \bibinfo
  {pages} {115131} (\bibinfo {year} {2012})}\BibitemShut {NoStop}%
\bibitem [{\citenamefont {Levin}(2013)}]{levin13}%
  \BibitemOpen
  \bibfield  {author} {\bibinfo {author} {\bibfnamefont {M.}~\bibnamefont
  {Levin}},\ }\href {\doibase 10.1103/PhysRevX.3.021009} {\bibfield  {journal}
  {\bibinfo  {journal} {Phys. Rev. X}\ }\textbf {\bibinfo {volume} {3}},\
  \bibinfo {pages} {021009} (\bibinfo {year} {2013})}\BibitemShut {NoStop}%
\bibitem [{\citenamefont {{Kitaev}}\ and\ \citenamefont
  {{Kong}}(2012)}]{kitaev11}%
  \BibitemOpen
  \bibfield  {author} {\bibinfo {author} {\bibfnamefont {A.}~\bibnamefont
  {{Kitaev}}}\ and\ \bibinfo {author} {\bibfnamefont {L.}~\bibnamefont
  {{Kong}}},\ }\href {\doibase 10.1007/s00220-012-1500-5} {\bibfield  {journal}
  {\bibinfo  {journal} {Communications in Mathematical Physics}\ }\textbf
  {\bibinfo {volume} {313}},\ \bibinfo {pages} {351} (\bibinfo {year}
  {2012})},\ \Eprint {http://arxiv.org/abs/1104.5047} {arXiv:1104.5047}
  \BibitemShut {NoStop}%
\bibitem [{\citenamefont {Barkeshli}\ \emph
  {et~al.}(2013{\natexlab{a}})\citenamefont {Barkeshli}, \citenamefont {Jian},\
  and\ \citenamefont {Qi}}]{barkeshli13}%
  \BibitemOpen
  \bibfield  {author} {\bibinfo {author} {\bibfnamefont {M.}~\bibnamefont
  {Barkeshli}}, \bibinfo {author} {\bibfnamefont {C.-M.}\ \bibnamefont {Jian}},
  \ and\ \bibinfo {author} {\bibfnamefont {X.-L.}\ \bibnamefont {Qi}},\ }\href
  {\doibase 10.1103/PhysRevB.88.235103} {\bibfield  {journal} {\bibinfo
  {journal} {Phys. Rev. B}\ }\textbf {\bibinfo {volume} {88}},\ \bibinfo
  {pages} {235103} (\bibinfo {year} {2013}{\natexlab{a}})}\BibitemShut
  {NoStop}%
\bibitem [{\citenamefont {Barkeshli}\ \emph
  {et~al.}(2013{\natexlab{b}})\citenamefont {Barkeshli}, \citenamefont {Jian},\
  and\ \citenamefont {Qi}}]{barkeshli13b}%
  \BibitemOpen
  \bibfield  {author} {\bibinfo {author} {\bibfnamefont {M.}~\bibnamefont
  {Barkeshli}}, \bibinfo {author} {\bibfnamefont {C.-M.}\ \bibnamefont {Jian}},
  \ and\ \bibinfo {author} {\bibfnamefont {X.-L.}\ \bibnamefont {Qi}},\ }\href
  {\doibase 10.1103/PhysRevB.88.241103} {\bibfield  {journal} {\bibinfo
  {journal} {Phys. Rev. B}\ }\textbf {\bibinfo {volume} {88}},\ \bibinfo
  {pages} {241103} (\bibinfo {year} {2013}{\natexlab{b}})}\BibitemShut
  {NoStop}%
\bibitem [{\citenamefont {{Fuchs}}\ \emph {et~al.}(2013)\citenamefont
  {{Fuchs}}, \citenamefont {{Schweigert}},\ and\ \citenamefont
  {{Valentino}}}]{Fuchs2013}%
  \BibitemOpen
  \bibfield  {author} {\bibinfo {author} {\bibfnamefont {J.}~\bibnamefont
  {{Fuchs}}}, \bibinfo {author} {\bibfnamefont {C.}~\bibnamefont
  {{Schweigert}}}, \ and\ \bibinfo {author} {\bibfnamefont {A.}~\bibnamefont
  {{Valentino}}},\ }\href {\doibase 10.1007/s00220-013-1723-0} {\bibfield
  {journal} {\bibinfo  {journal} {Communications in Mathematical Physics}\
  }\textbf {\bibinfo {volume} {321}},\ \bibinfo {pages} {543} (\bibinfo {year}
  {2013})},\ \Eprint {http://arxiv.org/abs/1203.4568} {arXiv:1203.4568
  [hep-th]} \BibitemShut {NoStop}%
\bibitem [{\citenamefont {Lan}\ \emph {et~al.}(2015)\citenamefont {Lan},
  \citenamefont {Wang},\ and\ \citenamefont {Wen}}]{TLanAnyCon2015}%
  \BibitemOpen
  \bibfield  {author} {\bibinfo {author} {\bibfnamefont {T.}~\bibnamefont
  {Lan}}, \bibinfo {author} {\bibfnamefont {J.~C.}\ \bibnamefont {Wang}}, \
  and\ \bibinfo {author} {\bibfnamefont {X.-G.}\ \bibnamefont {Wen}},\ }\href
  {\doibase 10.1103/PhysRevLett.114.076402} {\bibfield  {journal} {\bibinfo
  {journal} {Phys. Rev. Lett.}\ }\textbf {\bibinfo {volume} {114}},\ \bibinfo
  {pages} {076402} (\bibinfo {year} {2015})}\BibitemShut {NoStop}%
\bibitem [{\citenamefont {Hung}\ and\ \citenamefont {Wan}(2015)}]{hung15}%
  \BibitemOpen
  \bibfield  {author} {\bibinfo {author} {\bibfnamefont {L.-Y.}\ \bibnamefont
  {Hung}}\ and\ \bibinfo {author} {\bibfnamefont {Y.}~\bibnamefont {Wan}},\
  }\href {\doibase 10.1007/JHEP07(2015)120} {\bibfield  {journal} {\bibinfo
  {journal} {J. High Energ. Phys.}\ }\textbf {\bibinfo {volume} {2015}},\
  \bibinfo {pages} {120} (\bibinfo {year} {2015})}\BibitemShut {NoStop}%
\bibitem [{\citenamefont {{Wan}}\ and\ \citenamefont {{Wang}}(2017)}]{wan17}%
  \BibitemOpen
  \bibfield  {author} {\bibinfo {author} {\bibfnamefont {Y.}~\bibnamefont
  {{Wan}}}\ and\ \bibinfo {author} {\bibfnamefont {C.}~\bibnamefont {{Wang}}},\
  }\href {\doibase 10.1007/JHEP03(2017)172} {\bibfield  {journal} {\bibinfo
  {journal} {Journal of High Energy Physics}\ }\textbf {\bibinfo {volume}
  {3}},\ \bibinfo {eid} {172} (\bibinfo {year} {2017})},\ \Eprint
  {http://arxiv.org/abs/1607.01388} {arXiv:1607.01388} \BibitemShut {NoStop}%
\bibitem [{\citenamefont {Hu}\ \emph {et~al.}(2018)\citenamefont {Hu},
  \citenamefont {Luo}, \citenamefont {Pankovich}, \citenamefont {Wan},\ and\
  \citenamefont {Wu}}]{hu17}%
  \BibitemOpen
  \bibfield  {author} {\bibinfo {author} {\bibfnamefont {Y.}~\bibnamefont
  {Hu}}, \bibinfo {author} {\bibfnamefont {Z.-X.}\ \bibnamefont {Luo}},
  \bibinfo {author} {\bibfnamefont {R.}~\bibnamefont {Pankovich}}, \bibinfo
  {author} {\bibfnamefont {Y.}~\bibnamefont {Wan}}, \ and\ \bibinfo {author}
  {\bibfnamefont {Y.-S.}\ \bibnamefont {Wu}},\ }\href {\doibase
  10.1007/JHEP01(2018)134} {\bibfield  {journal} {\bibinfo  {journal} {J. High
  Energ. Phys.}\ }\textbf {\bibinfo {volume} {2018}},\ \bibinfo {pages} {134}
  (\bibinfo {year} {2018})}\BibitemShut {NoStop}%
\bibitem [{\citenamefont {Bais}\ and\ \citenamefont
  {Slingerland}(2009)}]{bais2009}%
  \BibitemOpen
  \bibfield  {author} {\bibinfo {author} {\bibfnamefont {F.~A.}\ \bibnamefont
  {Bais}}\ and\ \bibinfo {author} {\bibfnamefont {J.~K.}\ \bibnamefont
  {Slingerland}},\ }\href {\doibase 10.1103/PhysRevB.79.045316} {\bibfield
  {journal} {\bibinfo  {journal} {Phys. Rev. B}\ }\textbf {\bibinfo {volume}
  {79}},\ \bibinfo {pages} {045316} (\bibinfo {year} {2009})},\ \Eprint
  {http://arxiv.org/abs/arXiv:0808.0627} {arXiv:0808.0627} \BibitemShut
  {NoStop}%
\bibitem [{\citenamefont {Eli\"ens}\ \emph {et~al.}(2014)\citenamefont
  {Eli\"ens}, \citenamefont {Romers},\ and\ \citenamefont {Bais}}]{eliens14}%
  \BibitemOpen
  \bibfield  {author} {\bibinfo {author} {\bibfnamefont {I.~S.}\ \bibnamefont
  {Eli\"ens}}, \bibinfo {author} {\bibfnamefont {J.~C.}\ \bibnamefont
  {Romers}}, \ and\ \bibinfo {author} {\bibfnamefont {F.~A.}\ \bibnamefont
  {Bais}},\ }\href {\doibase 10.1103/PhysRevB.90.195130} {\bibfield  {journal}
  {\bibinfo  {journal} {Phys. Rev. B}\ }\textbf {\bibinfo {volume} {90}},\
  \bibinfo {pages} {195130} (\bibinfo {year} {2014})}\BibitemShut {NoStop}%
\bibitem [{\citenamefont {Neupert}\ \emph {et~al.}(2016)\citenamefont
  {Neupert}, \citenamefont {He}, \citenamefont {von Keyserlingk}, \citenamefont
  {Sierra},\ and\ \citenamefont {Bernevig}}]{NeupertAnyCon2016}%
  \BibitemOpen
  \bibfield  {author} {\bibinfo {author} {\bibfnamefont {T.}~\bibnamefont
  {Neupert}}, \bibinfo {author} {\bibfnamefont {H.}~\bibnamefont {He}},
  \bibinfo {author} {\bibfnamefont {C.}~\bibnamefont {von Keyserlingk}},
  \bibinfo {author} {\bibfnamefont {G.}~\bibnamefont {Sierra}}, \ and\ \bibinfo
  {author} {\bibfnamefont {B.~A.}\ \bibnamefont {Bernevig}},\ }\href {\doibase
  10.1103/PhysRevB.93.115103} {\bibfield  {journal} {\bibinfo  {journal} {Phys.
  Rev. B}\ }\textbf {\bibinfo {volume} {93}},\ \bibinfo {pages} {115103}
  (\bibinfo {year} {2016})}\BibitemShut {NoStop}%
\bibitem [{\citenamefont {Wang}\ and\ \citenamefont {Levin}(2015)}]{wangcj15}%
  \BibitemOpen
  \bibfield  {author} {\bibinfo {author} {\bibfnamefont {C.}~\bibnamefont
  {Wang}}\ and\ \bibinfo {author} {\bibfnamefont {M.}~\bibnamefont {Levin}},\
  }\href {\doibase 10.1103/PhysRevB.91.165119} {\bibfield  {journal} {\bibinfo
  {journal} {Phys. Rev. B}\ }\textbf {\bibinfo {volume} {91}},\ \bibinfo
  {pages} {165119} (\bibinfo {year} {2015})}\BibitemShut {NoStop}%
\bibitem [{\citenamefont {Lu}\ \emph {et~al.}(2017)\citenamefont {Lu},
  \citenamefont {Cho},\ and\ \citenamefont {Vishwanath}}]{LuBFU}%
  \BibitemOpen
  \bibfield  {author} {\bibinfo {author} {\bibfnamefont {Y.-M.}\ \bibnamefont
  {Lu}}, \bibinfo {author} {\bibfnamefont {G.~Y.}\ \bibnamefont {Cho}}, \ and\
  \bibinfo {author} {\bibfnamefont {A.}~\bibnamefont {Vishwanath}},\ }\href
  {\doibase 10.1103/PhysRevB.96.205150} {\bibfield  {journal} {\bibinfo
  {journal} {Phys. Rev. B}\ }\textbf {\bibinfo {volume} {96}},\ \bibinfo
  {pages} {205150} (\bibinfo {year} {2017})}\BibitemShut {NoStop}%
\bibitem [{\citenamefont {Qi}\ and\ \citenamefont
  {Fu}(2015{\natexlab{b}})}]{QiCSF}%
  \BibitemOpen
  \bibfield  {author} {\bibinfo {author} {\bibfnamefont {Y.}~\bibnamefont
  {Qi}}\ and\ \bibinfo {author} {\bibfnamefont {L.}~\bibnamefont {Fu}},\ }\href
  {\doibase 10.1103/PhysRevB.91.100401} {\bibfield  {journal} {\bibinfo
  {journal} {Phys. Rev. B}\ }\textbf {\bibinfo {volume} {91}},\ \bibinfo
  {pages} {100401(R)} (\bibinfo {year} {2015}{\natexlab{b}})}\BibitemShut
  {NoStop}%
\bibitem [{\citenamefont {Cincio}\ and\ \citenamefont {Qi}()}]{CincioCSL2016}%
  \BibitemOpen
  \bibfield  {author} {\bibinfo {author} {\bibfnamefont {L.}~\bibnamefont
  {Cincio}}\ and\ \bibinfo {author} {\bibfnamefont {Y.}~\bibnamefont {Qi}},\
  }\href@noop {} {\ }\Eprint {http://arxiv.org/abs/1511.02226}
  {arXiv:1511.02226 [cond-mat.str-el]} \BibitemShut {NoStop}%
\bibitem [{\citenamefont {{Davydov}}(2014)}]{davydov13}%
  \BibitemOpen
  \bibfield  {author} {\bibinfo {author} {\bibfnamefont {A.}~\bibnamefont
  {{Davydov}}},\ }\href {\doibase 10.1063/1.4895764} {\bibfield  {journal}
  {\bibinfo  {journal} {Journal of Mathematical Physics}\ }\textbf {\bibinfo
  {volume} {55}},\ \bibinfo {eid} {092305} (\bibinfo {year} {2014})},\ \Eprint
  {http://arxiv.org/abs/1312.7466} {arXiv:1312.7466} \BibitemShut {NoStop}%
\bibitem [{\citenamefont {Bakalov}\ and\ \citenamefont
  {Kirillov}(2001)}]{Bakalov2001}%
  \BibitemOpen
  \bibfield  {author} {\bibinfo {author} {\bibfnamefont {B.}~\bibnamefont
  {Bakalov}}\ and\ \bibinfo {author} {\bibfnamefont {A.}~\bibnamefont
  {Kirillov}, \bibfnamefont {Jr}},\ }\href@noop {} {\emph {\bibinfo {title}
  {Lectures on Tensor Categories and Modular Functors}}},\ \bibinfo {series}
  {University Lecture Series}, Vol.~\bibinfo {volume} {21}\ (\bibinfo
  {publisher} {American Mathematical Society},\ \bibinfo {address} {Providence,
  RI},\ \bibinfo {year} {2001})\BibitemShut {NoStop}%
\bibitem [{\citenamefont {Levin}\ and\ \citenamefont {Wen}(2005)}]{levin05}%
  \BibitemOpen
  \bibfield  {author} {\bibinfo {author} {\bibfnamefont {M.~A.}\ \bibnamefont
  {Levin}}\ and\ \bibinfo {author} {\bibfnamefont {X.-G.}\ \bibnamefont
  {Wen}},\ }\href {\doibase 10.1103/PhysRevB.71.045110} {\bibfield  {journal}
  {\bibinfo  {journal} {Phys. Rev. B}\ }\textbf {\bibinfo {volume} {71}},\
  \bibinfo {pages} {045110} (\bibinfo {year} {2005})}\BibitemShut {NoStop}%
\bibitem [{\citenamefont {{Metlitski}}(2015)}]{metlitski15b}%
  \BibitemOpen
  \bibfield  {author} {\bibinfo {author} {\bibfnamefont {M.~A.}\ \bibnamefont
  {{Metlitski}}},\ }\href@noop {} {\bibfield  {journal} {\bibinfo  {journal}
  {ArXiv e-prints}\ } (\bibinfo {year} {2015})},\ \Eprint
  {http://arxiv.org/abs/1510.05663} {arXiv:1510.05663} \BibitemShut {NoStop}%
\bibitem [{\citenamefont {Fidkowski}\ and\ \citenamefont
  {Vishwanath}(2017)}]{fidkowski15}%
  \BibitemOpen
  \bibfield  {author} {\bibinfo {author} {\bibfnamefont {L.}~\bibnamefont
  {Fidkowski}}\ and\ \bibinfo {author} {\bibfnamefont {A.}~\bibnamefont
  {Vishwanath}},\ }\href {\doibase 10.1103/PhysRevB.96.045131} {\bibfield
  {journal} {\bibinfo  {journal} {Phys. Rev. B}\ }\textbf {\bibinfo {volume}
  {96}},\ \bibinfo {pages} {045131} (\bibinfo {year} {2017})}\BibitemShut
  {NoStop}%
\bibitem [{\citenamefont {Barkeshli}\ and\ \citenamefont
  {Cheng}(2018)}]{barkeshli17}%
  \BibitemOpen
  \bibfield  {author} {\bibinfo {author} {\bibfnamefont {M.}~\bibnamefont
  {Barkeshli}}\ and\ \bibinfo {author} {\bibfnamefont {M.}~\bibnamefont
  {Cheng}},\ }\href {\doibase 10.1103/PhysRevB.98.115129} {\bibfield  {journal}
  {\bibinfo  {journal} {Phys. Rev. B}\ }\textbf {\bibinfo {volume} {98}},\
  \bibinfo {pages} {115129} (\bibinfo {year} {2018})}\BibitemShut {NoStop}%
\bibitem [{\citenamefont {Dijkgraaf}\ \emph {et~al.}(1989)\citenamefont
  {Dijkgraaf}, \citenamefont {Vafa}, \citenamefont {Verlinde},\ and\
  \citenamefont {Verlinde}}]{Dijkgraaf1989}%
  \BibitemOpen
  \bibfield  {author} {\bibinfo {author} {\bibfnamefont {R.}~\bibnamefont
  {Dijkgraaf}}, \bibinfo {author} {\bibfnamefont {C.}~\bibnamefont {Vafa}},
  \bibinfo {author} {\bibfnamefont {E.}~\bibnamefont {Verlinde}}, \ and\
  \bibinfo {author} {\bibfnamefont {H.}~\bibnamefont {Verlinde}},\ }\href
  {\doibase 10.1007/BF01238812} {\bibfield  {journal} {\bibinfo  {journal}
  {Communications in Mathematical Physics}\ }\textbf {\bibinfo {volume}
  {123}},\ \bibinfo {pages} {485} (\bibinfo {year} {1989})}\BibitemShut
  {NoStop}%
\bibitem [{\citenamefont {Barkeshli}\ \emph
  {et~al.}(2013{\natexlab{c}})\citenamefont {Barkeshli}, \citenamefont {Jian},\
  and\ \citenamefont {Qi}}]{Barkeshli2013}%
  \BibitemOpen
  \bibfield  {author} {\bibinfo {author} {\bibfnamefont {M.}~\bibnamefont
  {Barkeshli}}, \bibinfo {author} {\bibfnamefont {C.-M.}\ \bibnamefont {Jian}},
  \ and\ \bibinfo {author} {\bibfnamefont {X.-L.}\ \bibnamefont {Qi}},\ }\href
  {\doibase 10.1103/PhysRevB.87.045130} {\bibfield  {journal} {\bibinfo
  {journal} {Phys. Rev. B}\ }\textbf {\bibinfo {volume} {87}},\ \bibinfo
  {pages} {045130} (\bibinfo {year} {2013}{\natexlab{c}})}\BibitemShut
  {NoStop}%
\bibitem [{\citenamefont {{Tu}}\ \emph {et~al.}(2013)\citenamefont {{Tu}},
  \citenamefont {{Zhang}},\ and\ \citenamefont {{Qi}}}]{Tu2013}%
  \BibitemOpen
  \bibfield  {author} {\bibinfo {author} {\bibfnamefont {H.-H.}\ \bibnamefont
  {{Tu}}}, \bibinfo {author} {\bibfnamefont {Y.}~\bibnamefont {{Zhang}}}, \
  and\ \bibinfo {author} {\bibfnamefont {X.-L.}\ \bibnamefont {{Qi}}},\ }\href
  {\doibase 10.1103/PhysRevB.88.195412} {\bibfield  {journal} {\bibinfo
  {journal} {\prb}\ }\textbf {\bibinfo {volume} {88}},\ \bibinfo {eid} {195412}
  (\bibinfo {year} {2013})},\ \Eprint {http://arxiv.org/abs/1212.6951}
  {arXiv:1212.6951 [cond-mat.str-el]} \BibitemShut {NoStop}%
\end{thebibliography}%

\end{document}